%% file: ANA-SUSY-2019-22-PAPER.tex
\pdfoutput=1
\documentclass[cernpreprint, atlasdraft=false, orcidlogo=true, UKenglish, texmf, orcidlogo]{atlasdoc}
 
\usepackage[backend=biber,abbrevref=false]{atlaspackage}
\usepackage[backref=false]{atlasbiblatex}
 
\usepackage[BSM,snippets]{atlasphysics}
\usepackage{xcolor}

\graphicspath{{logos/}{figures/}}
 
\usepackage{ANA-SUSY-2019-22-PAPER-defs}

\AtlasTitle{Search for direct production of winos and higgsinos in events with two same-charge leptons or three leptons in $pp$ collision data at $\sqrt{s}=13$ TeV with the ATLAS detector
}

\AtlasAbstract{
A search for supersymmetry targeting the direct production of winos and higgsinos is conducted in final states with either two  leptons ($e$ or $\mu$) with the same electric charge, or three leptons. The analysis uses 139 fb$^{-1}$ of $pp$ collision data at $\sqrt{s}=13$ TeV collected with the ATLAS detector during Run 2 of the Large Hadron Collider. No significant excess over the Standard Model expectation is observed. Simplified and complete models with and without $R$-parity conservation are considered. In topologies with intermediate states including either $Wh$ or $WZ$ pairs, wino masses up to 525 GeV and 250 GeV are excluded, respectively, for a bino of vanishing mass. Higgsino masses smaller than 440 GeV are excluded in a natural $R$-parity-violating model with bilinear terms. Upper limits on the production cross section of generic events beyond the Standard Model as low as 40 ab are obtained in signal regions optimised for these models and also for an $R$-parity-violating scenario with baryon-number-violating higgsino decays into top quarks and jets. The analysis  significantly improves sensitivity to supersymmetric models and other processes beyond the Standard Model that may contribute to the considered final states.}
 
\author{The ATLAS Collaboration}
 
\AtlasRefCode{ANA-SUSY-2019-22}
 
\PreprintIdNumber{CERN-EP-2023-063}

\arXivId{2305.09322}

\AtlasJournalRef{JHEP 11 (2023) 150}
\AtlasDOI{10.1007/JHEP11(2023)150}

\hypersetup{pdftitle={ATLAS document},pdfauthor={The ATLAS Collaboration}}
 
\begin{document}
 
\maketitle
 
\tableofcontents

\section{Introduction}
\label{sec:intro}

Experimental searches for signals of physics beyond the Standard Model (SM) at colliders have long exploited the signature of a pair of isolated light leptons (electrons or muons) with same-sign (SS) electric charges. In the SM, the production of such lepton pairs is rare and originates mainly from pairs of weak-boson decays. In proton--proton (\pp) collisions at $\sqs = 13$~\tev, the inclusive cross section of same-sign lepton pair production is of the order of one~pb~\cite{Alwall:2014hca,ParticleDataGroup:2022pth}, i.e.\ it is suppressed by more than three orders of magnitude relative to the production of opposite-sign lepton pairs. On the other hand, heavy particles beyond the SM (BSM) could decay into multiple massive SM bosons or top quarks, which subsequently decay into jets and same-sign leptons, thus involving a relatively low SM background. Examples of such BSM states include supersymmetric (SUSY) particles~\cite{Barnett:1993ea,Guchait:1994zk}, SS top-quark pairs~\cite{Bai:2008sk,Berger:2011ua}, scalar gluons (sgluons)~\cite{Plehn:2008ae,Calvet:2012rk}, heavy scalar bosons of extended Higgs sectors~\cite{Gaemers:1984sj,Branco:2011iw}, Majorana heavy neutrinos~\cite{Almeida:1997em,Atre:2009rg}, and vector-like top quarks~\cite{Contino:2008hi}.
 
At the Large Hadron Collider (LHC)~\cite{Evans:2008zzb}, the ATLAS~\cite{PERF-2007-01} and CMS~\cite{CMS-CMS-00-001} experiments have extensively probed possible SM extensions in the same-sign dilepton channel. Among these theoretical proposals, SUSY~\cite{Golfand:1971iw,Volkov:1973ix,Wess:1974tw,Wess:1974jb,Ferrara:1974pu,Salam:1974ig,Martin:1997ns} remains a compelling framework as it provides solutions to the gauge hierarchy problem~\cite{Sakai:1981gr,Dimopoulos:1981yj,Ibanez:1981yh,Dimopoulos:1981zb} without the need for large fine-tuning of fundamental parameters~\cite{Barbieri:1987fn,deCarlos:1993yy}, offers gauge coupling unification~\cite{Sakai:1981gr,Dimopoulos:1981yj,Ibanez:1981yh,Dimopoulos:1981zb}, and contains weakly interacting particles that can contribute to the dark matter~\cite{Goldberg:1983nd,Ellis:1983ew}.
 
Charginos, $\susy{\chi}^{\pm}_{1,2}$, and neutralinos, $\susy{\chi}^0_{1,2,3,4}$, collectively referred to as `electroweakinos', are the ordered mass eigenstates formed from the linear superposition of the higgsinos, winos, and binos, which are the SUSY partners of the Higgs and electroweak gauge bosons, respectively. A discrete multiplicative symmetry, \Rp~\cite{Farrar:1978xj}, is often introduced in SUSY models to avoid rapid proton decay. In \Rp-conserving (RPC) models, the lightest supersymmetric particle (LSP) is stable and is required to be neutral and colourless to evade observation as a dark matter candidate~\cite{Hooper:2002nq}. It would therefore also be invisible in a hadron collider experiment, only manifested through large missing transverse momentum, \met. Models of \Rp-violating (RPV) SUSY~\cite{Dreiner:1997uz} are also well-motivated, while introducing more parameters to constrain. In RPV SUSY, a \ninoone LSP would decay into SM particles and, due to its Majorana nature, it may give rise to SS-lepton final states.
 
In this article, the search described in Ref.~\cite{SUSY-2018-09} is extended to more signal models using the full data set of \pp\ collisions at  $\sqs = 13$~\tev\ recorded by the ATLAS detector during \RunTwo of the LHC, corresponding to an integrated luminosity of 139~\ifb. The selection is based on final states with two SS leptons or three leptons accompanied by large \met and a number of hadronic jets, possibly containing $b$-hadrons and tagged as `\bjets'. This search provides the first ATLAS result from a two-SS-lepton selection targeting direct chargino and neutralino production. Such production may be dominant at the LHC according to naturalness considerations~\cite{Barbieri:1987fn,deCarlos:1993yy}, which suggest that the lightest electroweakinos have masses near the electroweak scale while the superpartners of the gluon and quarks can be heavier than a few \TeV, evading their so far direct detection in searches of strongly produced SUSY.  This search covers so-far unconstrained kinematic regions, not yet excluded by previous three-lepton analyses. The smaller background faced by SS-lepton analyses allows looser kinematic requirements to be imposed, e.g.\ on \met\  or on the momenta of jets and leptons, which provides sensitivity to scenarios with small mass splittings between the superpartners~\cite{SUSY-2016-25,CMS-SUS-17-007,SUSY-2018-16,CMS-SUS-18-004}. In addition to directly exploring such scenarios, the analysis aims to provide signal regions orthogonal to others targeting different final states, thus improving the overall sensitivity through future statistical combinations. The event selection is optimised to target four models: (i, ii) simplified models of winos and binos with on-shell \WZ or \Wh boson pairs as intermediate states; (iii) higgsino production with bilinear \Rp-violating (bRPV) terms; and (iv) higgsino production with \Rp-violating decays to top quarks via baryon-number-violating (BNV) \textit{UDD} couplings.
 
All prior searches for SS lepton pairs and several three-lepton searches carried out by ATLAS~\cite{SUSY-2013-09,SUSY-2013-23,SUSY-2015-09,SUSY-2016-14,SUSY-2017-01,SUSY-2018-09,ATLAS:2023afl} and CMS~\cite{,CMS-SUS-16-035,CMS-SUS-19-008,CMS-SUS-19-012} focused on strong production of superpartners, on electroweak SUSY production with low hadronic activity, or on slepton resonant production~\cite{CMS-SUS-17-008}. Other analyses with three-lepton selections focused on direct electroweakino production in events without jets~\cite{CMS-SUS-13-002,SUSY-2013-12,CMS-SUS-13-003,SUSY-2016-24,CMS-SUS-16-003,CMS-SUS-16-041,SUSY-2017-03,SUSY-2018-06,SUSY-2019-09} or with trilepton resonances~\cite{SUSY-2018-36}.
 
Simplified models with \conentwo\ production and \Wh bosons in the decay chain have been explored by ATLAS in fully hadronic~\cite{SUSY-2018-41}, semileptonic~\cite{SUSY-2013-23}, photon~\cite{SUSY-2018-23}, and multilepton~\cite{SUSY-2017-01} final states with large \met. CMS has constrained this scenario by combining a variety of leptonic signatures, including dileptons and $\tau$-leptons~\cite{CMS-SUS-16-039,CMS-SUS-19-012,CMS-SUS-20-001,CMS-SUS-20-003}, and fully hadronic final states~\cite{CMS-SUS-21-002}. Intermediate decays to \WZ\ bosons have been probed previously in ATLAS assuming the presence of boosted hadronically decaying bosons~\cite{SUSY-2018-41}, two~\cite{SUSY-2018-05} or three leptons~\cite{SUSY-2019-09} in the final state. CMS has investigated this channel in searches for multileptons~\cite{CMS-SUS-16-039}, two SS leptons or three leptons~\cite{CMS-SUS-19-012}, soft leptons~\cite{CMS-SUS-17-007,CMS-SUS-18-004} and jets~\cite{CMS-SUS-21-002}.
 
ATLAS has set limits on bRPV models assuming strong superpartner production~\cite{Mitsou:2015kpa}. Minimal Supergravity~\cite{Chamseddine:1982jx,Barbieri:1982eh,Kane:1993td} with bilinear terms has been constrained in events with one lepton~\cite{SUSY-2011-08,SUSY-2013-20}, one $\tau$~\cite{SUSY-2013-10},  or two SS leptons~\cite{SUSY-2013-09}, and in their combination~\cite{SUSY-2014-06}. A reinterpretation of a SS-lepton analysis~\cite{SUSY-2013-09} set bounds~\cite{ATLAS-CONF-2015-018} in a `natural' bRPV scenario~\cite{Barbieri:1987fn,deCarlos:1993yy} within the phenomenological Minimal Supersymmetric Standard Model (pMSSM)~\cite{Djouadi:1998di,Berger:2008cq}.
 
Baryonic \textit{UDD} operators have been probed by the ATLAS~\cite{SUSY-2013-07,SUSY-2015-05,SUSY-2016-11,SUSY-2016-09,SUSY-2016-22,SUSY-2018-38,SUSY-2018-41} and CMS~\cite{CMS-SUS-11-013,CMS-EXO-12-049,CMS-EXO-12-052,CMS-SUS-14-003,CMS-SUS-14-020,CMS-SUS-16-040,CMS-EXO-17-021,CMS-EXO-17-030,CMS-SUS-19-008,CMS-SUS-19-004} experiments in multijet final states and by ATLAS in events with at least one lepton~\cite{SUSY-2019-04}. Models with $\ninoone \to tbs$ have been constrained in gluino and top-squark production in a wide range of $\lambda''_{323}$ couplings, by reinterpreting several ATLAS searches optimised for RPC and RPV SUSY models~\cite{ATLAS-CONF-2018-003}.
 
The paper is structured as follows. \cref{sec:models} is dedicated to the targeted signal models. Details of the ATLAS detector are described in \cref{sec:detector}, with the utilised data set and simulation samples listed in \cref{sec:datamc}.  The object definitions and the event categorisation are discussed in \cref{sec:objects} and \cref{sec:selection}, respectively. The background modelling and validation is given in \cref{sec:background}. Systematic uncertainties are discussed in \cref{sec:systs}, and the results and interpretations are presented in \cref{sec:result} and \cref{sec:interpret}, respectively. The conclusions are summarised in \cref{sec:conclusion}. In addition, the \textit{UDD} RPV model analysis, which provides a relatively small improvement in this search, is discussed in \cref{app:RPVUDD}.


\FloatBarrier
\section{Signal models}
\label{sec:models}

\graphicspath{{figures/models/}}
 
The models targeted in this analysis can be divided into two main scenarios: directly produced wino-like electroweakinos with a bino-like LSP in RPC SUSY, shown in \cref{fig:signals-RPC}, and higgsino-like electroweakinos with RPV terms, shown in \cref{fig:signals-RPV}.
 
\subsection{Wino--bino \conentwo\ production with \Wh or \WZ bosons}
\label{ssec:model-wino}
 
Simplified models~\cite{Alwall:2008ve,Alwall:2008ag,Alves:2011wf} involving the direct production of a lightest chargino, \chinoonepm, and a next-to-lightest neutralino, \ninotwo, are considered.  The \chinoonepm and the \ninotwo are assumed to be mass-degenerate. The \chinoonepm/\ninotwo are wino-like, i.e.\ superpartners of the SU(2)$_\mathrm{L}$ gauge fields, whilst the \ninoone is bino-like, i.e.\ the superpartner of the U(1)$_\mathrm{Y}$ gauge field~\cite{ParticleDataGroup:2022pth}. The \chinoonepm is assumed to decay into an on-shell, leptonically decaying \Wboson and \ninoone. For the \ninotwo, two decay cases are examined: (i) a SM-like Higgs boson and (ii) a leptonically decaying $Z$ boson. The Higgs boson decay is dominant for many choices of MSSM parameters as long as the mass-splitting between the two lightest neutralinos is larger than the Higgs boson mass and the higgsinos are heavier than the winos. All possible decays of the Higgs boson which ultimately result in a single lepton and jets (mostly via intermediate states) are taken into account. This is indicated by the grey-filled dot in the Higgs decay in \cref{fig:C1N2_Wh}. In the case of the leptonically decaying $Z$ boson, this is produced on-shell and leads to the diagram of \cref{fig:C1N2_WZ}.
 
\begin{figure}[!htb]
\centering
\subfloat[]{\includegraphics[width=0.3\textwidth]{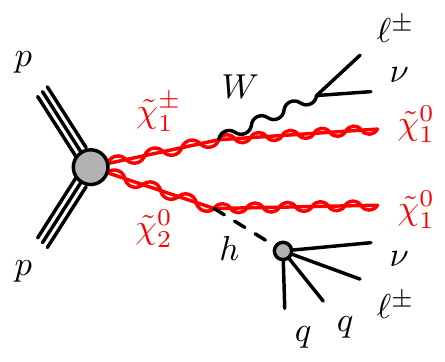}\label{fig:C1N2_Wh}}
\qquad
\subfloat[]{\includegraphics[width=0.3\textwidth]{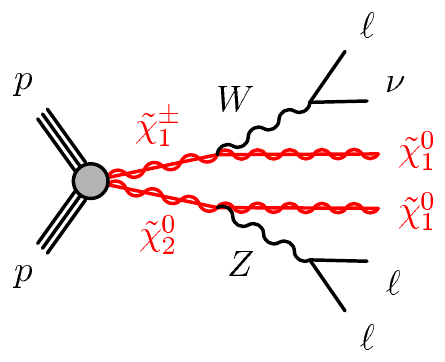}\label{fig:C1N2_WZ}}
\caption{Diagrams of the targeted RPC simplified models with intermediate gauge vector and Higgs boson production.}\label{fig:signals-RPC}
\end{figure}
 
\subsection{Higgsino-like electroweakinos in RPV scenarios}
\label{ssec:model-RPV}
 
The RPV component of the generic superpotential can be written as~\cite{Dreiner:1997uz}:
\begin{equation}
W_{\not  R_p} = \frac{1}{2} \lambda_{ijk} L_{i}L_{j}\bar{E}_{k} + \lambda_{ijk}^{\prime} L_{i}Q_{j}\bar{D}_{k} + \epsilon_i L_{i}H_{2} +
\frac{1}{2}\lambda_{ijk}^{\prime\prime}\bar{U}_{i}\bar{D}_{j}\bar{D}_{k},
\label{eq:rpvandudd:wrpv}
\end{equation}
where $i,j,k=1,2,3$ are generation indices. The $L_i$, $Q_i$ represent the lepton and quark SU(2)$_\mathrm{L}$ doublet superfields, whereas $H_2$ is the Higgs superfield. The $\bar{E}_j$, $\bar{D}_j$, and $\bar{U}_j$ are the charged lepton, down-type quark, and up-type quark SU(2)$_\mathrm{L}$ singlet superfields, respectively. The Yukawa couplings are $\lambda$, $\lambda^{\prime}$, and $\lambda^{\prime\prime}$, whilst $\epsilon$ is a dimensionful mass parameter. Two RPV scenarios are explored, the first from bilinear lepton-number-violating terms $LH_{2}$, and the second from BNV terms \textit{UDD}, in \cref{eq:rpvandudd:wrpv}.
 
RPV SUSY through bilinear terms is strongly motivated by its inherent connection with neutrino physics~\cite{Porod:2000hv,Roy:1996bua,Romao:1999up}. Sneutrino vacuum expectation values (VEVs)  introduce a mixing between neutrinos and neutralinos, leading to a see-saw mechanism that gives mass to one neutrino at tree level, with the other two neutrino masses being induced by loop effects~\cite{Hirsch:2000ef,Diaz:2003as}. The same VEVs are also involved in the decay of the LSP, which is thus constrained by experimental neutrino measurements.
 
The model considered features pair production of light higgsinos, \ninotwo, \chinoonepm and \ninoone, decaying into all possible final states allowed by the bRPV couplings -- it is primarily inspired by naturalness arguments~\cite{Barbieri:1987fn,deCarlos:1993yy}. The dominant production processes are \conenone, \conentwo, \nonentwo, and $\chinoonepm\!\chinoonemp$. The first three processes can lead to a two-SS-lepton or three-lepton final state. The dominant decays are $\chinoonepm\to\Wpm\nu$, $\susy{\chi}^0_{1,2}\to \Wpm\ell^{\mp}, \Wpm\tau^{\mp}$, and $\ninotwo\to\chinoonepm\pimp$.  Higgsino mass splittings of less than 2~\gev~\cite{Biswas:2016ffy} are targeted. A ratio of Higgs doublet VEVs of $\tan\beta=5$ is chosen to primarily favour light leptons, thus suppressing the $\susy{\chi}^0_{1,2}\to\Wpm\tau^{\mp}$ decays, which are preferred at high $\tan\beta\sim50$ with a branching ratio of more than 90\%. At $\tan\beta=5$, the branching ratio for $\susy{\chi}^0_{1,2}\to \Wpm\tau^{\mp}$ drops to less than 50\%, while for $\ninotwo\to\chinoonepm\pimp$ it is ${\sim}20\%$.
 
Example diagrams are given in \cref{fig:C1N1-vWlW-bRPV} and \cref{fig:N1N2-lWlW-bRPV}. The decay modes are partly determined by a fit to neutrino oscillation experimental data~\cite{Thomas:2011kt}, leading to flavour non-universality of lepton decays, with more details given in \cref{sec:datamc}. The bRPV couplings are large enough to ensure prompt higgsino decays. All possible allowed higgsino decays are considered in the analysis.
 
\begin{figure}[!htb]
\centering
\subfloat[]{\includegraphics[width=0.25\textwidth]{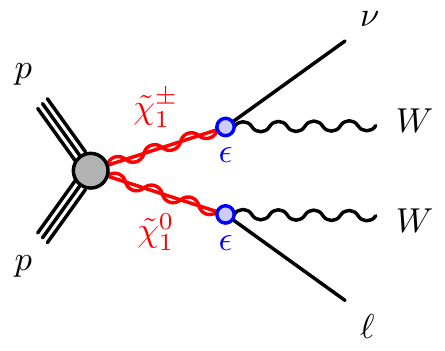}\label{fig:C1N1-vWlW-bRPV}}
\qquad
\subfloat[]{\includegraphics[width=0.25\textwidth]{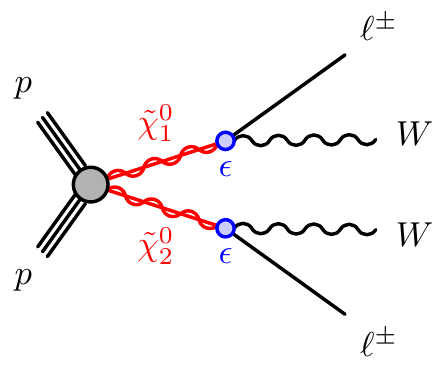}\label{fig:N1N2-lWlW-bRPV}}
\qquad
\subfloat[]{\includegraphics[width=0.25\textwidth]{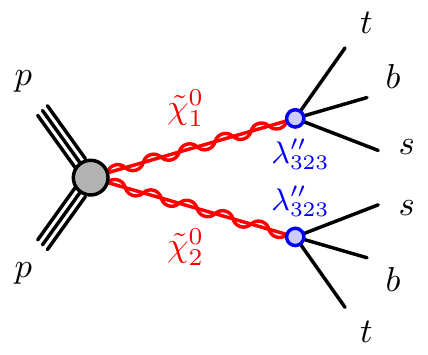}\label{fig:N1N2_RPV}}
\caption{Diagrams of the targeted RPV models. Diagrams \protect\subref{fig:C1N1-vWlW-bRPV} and \protect\subref{fig:N1N2-lWlW-bRPV} serve as examples, since inclusive bRPV production is considered. The \textit{UDD} RPV scenario with BNV terms in diagram \protect\subref{fig:N1N2_RPV} is a simplified model.}\label{fig:signals-RPV}
\end{figure}
 
Besides SUSY with \textit{UDD} terms in \cref{eq:rpvandudd:wrpv}~\cite{Dreiner:1997uz,Barbier:2004ez,Csaki:2011ge}, baryon-number violation is featured in BSM scenarios such as grand unified theories~\cite{Georgi:1974sy} and models with black holes~\cite{Bekenstein:1971hc}. Moreover, in a universe with initially equal amounts of baryonic and anti-baryonic matter, BNV is necessary to describe the observed baryon asymmetry~\cite{Sakharov:1967dj}.
 
In the simplified topology considered, higgsino \nonentwo pairs are produced directly and undergo prompt RPV decays as shown in the diagram of \cref{fig:N1N2_RPV}. The $UDD$-type BNV coupling $\lambda''_{323}$, defined in  \cref{eq:rpvandudd:wrpv}, is chosen to be non-vanishing, as it is predicted to be dominant under the minimal flavour violation hypothesis~\cite{Csaki:2011ge}. Its value is chosen to be $\mathcal{O}(10^{-3})$ to $\mathcal{O}(10^{-2})$, which guarantees prompt decays for electroweakino masses down to \SI{180}{GeV}. The \ninotwo next-to-LSP (NLSP) and the \ninoone LSP are mass degenerate and decay with a 100\% branching ratio into $tbs$, thus possibly leading to a final state with two SS leptons and at least six jets, of which at least four are \bjets. Other electroweakino production modes do not lead to the final states targeted by this search.


\FloatBarrier
\section{ATLAS detector}
\label{sec:detector}

 
The ATLAS detector~\cite{PERF-2007-01} is a multipurpose particle detector with a forward--backward symmetric cylindrical geometry and a near \(4\pi\) coverage in solid angle.\footnote{
ATLAS uses a right-handed coordinate system with its origin at the nominal interaction point (IP) in the centre of the detector and the \(z\)-axis along the beam pipe.
The \(x\)-axis points from the IP to the centre of the LHC ring, and the \(y\)-axis points upwards.
Cylindrical coordinates \((r,\phi)\) are used in the transverse plane, \(\phi\) being the azimuthal angle around the \(z\)-axis.
The pseudorapidity is defined in terms of the polar angle \(\theta\) as \(\eta = -\ln \tan(\theta/2)\).
Angular distance is measured in units of \(\Delta R \equiv \sqrt{(\Delta\eta)^{2} + (\Delta\phi)^{2}}\).}  
It consists of an inner tracking detector surrounded by a thin superconducting solenoid providing a \SI{2}{\tesla} axial magnetic field, electromagnetic and hadronic calorimeters, and a muon spectrometer.
The inner tracking detector (ID) covers the pseudorapidity range \(|\eta| < 2.5\).
It consists of silicon pixel, silicon microstrip, and transition radiation tracking detectors.
An additional layer of silicon pixels, the  insertable B-layer~\cite{ATLAS-TDR-19,PIX-2018-001}, was installed closer to the beamline before Run~2.
Lead/liquid-argon (LAr) sampling calorimeters provide electromagnetic (EM) energy measurements with high granularity.
A steel/scintillator-tile hadron calorimeter covers the central pseudorapidity range (\(|\eta| < 1.7\)).
The endcap and forward regions are instrumented with LAr calorimeters for both the EM and hadronic energy measurements up to \(|\eta| = 4.9\).
The muon spectrometer surrounds the calorimeters and is based on three large superconducting air-core toroidal magnets with eight coils each.
The field integral of the toroids ranges between \num{2.0} and \SI{6.0}{\tesla\metre} across most of the detector.
The muon spectrometer (MS) includes a system of precision chambers for tracking and fast detectors for triggering.
A two-level trigger system is used to select events.
The first-level trigger is implemented in hardware and uses a subset of the detector information to accept events at a rate below \SI{100}{\kHz}.
This is followed by a software-based trigger that reduces the accepted event rate to \SI{1}{\kHz} on average depending on the data-taking conditions.
An extensive software suite~\cite{ATL-SOFT-PUB-2021-001} is used in data simulation, in the reconstruction and analysis of real and simulated data, in detector operations, and in the trigger and data acquisition systems of the experiment.


\FloatBarrier
\section{Data set and simulated event samples}
\label{sec:datamc}

This paper analyses proton--proton collision data collected by the ATLAS detector between 2015 and 2018. In this period, the LHC delivered colliding beams with a peak instantaneous luminosity reaching $2.1\times\qty{E34}{\per\cm\squared\per\second}$, achieved in 2018, and an average number of \pp\ interactions per bunch crossing, $\left<\mu \right>$, of 33.7. After the application of beam, detector, and data-quality criteria~\cite{DAPR-2018-01}, the total integrated luminosity of the data set is 139~\ifb\ with a combined uncertainty of 1.7\%~\cite{ATLAS-CONF-2019-021}, obtained using the LUCID-2 detector~\cite{Avoni:2018iuv} for the primary luminosity measurements.
 
Events with $\met<\SI{250}{GeV}$ were selected using dilepton triggers~\cite{TRIG-2018-05,TRIG-2018-01}, with lepton \pt\ thresholds increasing during the \RunTwo data-taking period to a maximum of \SI{24}{GeV} for triggers requiring two electrons, $\SI{22}{GeV}$ for the leading-\pt\ muon in triggers requiring two muons, and $\SI{17}{GeV}$ (\SI{14}{GeV}) for the electron (muon) in different-flavour dilepton triggers.
For events with $\met>\SI{250}{GeV}$, a logical OR of these triggers and $\met$ triggers~\cite{TRIG-2019-01} was used.
The above strategy was chosen to maximise the trigger efficiency, while selecting events relevant to the targeted final states~\cite{SUSY-2018-09}. The selection thresholds are defined such that the trigger efficiencies are constant throughout the lepton \pt and \met\  range considered in the analysis.
 
Signal and background events produced in \pp\ collisions were simulated with various Monte Carlo (MC) generators.  They include the effect of multiple \pp\ interactions in the same and neighbouring bunch crossings (`pile-up'), which was modelled by overlaying each simulated hard-scattering interaction with simulated inelastic \pp\ events generated by   \PYTHIA[8.186]~\cite{Sjostrand:2007gs,Sjostrand:2014zea} with the \NNPDF[2.3lo] set of parton distribution functions (PDF)~\cite{Ball:2012cx} and a set of tuned parameter values called the A3 tune~\cite{ATL-PHYS-PUB-2016-017}. The simulated events were weighted to reproduce the  $\left<\mu \right>$ distribution observed in the data. The \EVTGEN~\cite{Lange:2001uf} program was used to simulate the $b$- and $c$-flavoured hadron decays.
 
The detector response was simulated using either the full ATLAS detector description~\cite{SOFT-2010-01} based on \GEANT~\cite{Agostinelli:2002hh}, or a fast simulation based on a parameterisation of the performance of the electromagnetic and hadronic calorimeters and \GEANT for the other parts of the detector~\cite{ATL-PHYS-PUB-2010-013}. The generated events are reconstructed in the same manner as the data.
 
\subsection{Signal samples}
\label{sec:signal-mc}
 
The signal samples for the targeted models were generated using \MGNLO[2.2.3]~\cite{Alwall:2014hca,Frederix:2012ps} interfaced to \PYTHIA[8.186] with the A14 tune~\cite{ATL-PHYS-PUB-2014-021} for the modelling of the parton showering (PS)~\cite{Mrenna:2016sih}, hadronisation and underlying event. The matrix element (ME) calculation was performed at tree level, including the emission of up to two additional partons. The PDF set used for the generation was \NNPDF[2.3lo]~\cite{Ball:2012cx}. The ME–PS matching was carried out using the CKKW-L prescription~\cite{Lonnblad:2001iq,Lonnblad:2011xx}, with a matching scale set to one quarter of the pair-produced superpartner mass. For the bRPV model, the RPV parameters (together with the mass spectra and the decay modes) were determined by a fit to neutrino experimental data performed by the \textsc{SPheno}~\cite{Porod:2003um,Porod:2011nf} spectrum calculator produced by the \textsc{Sarah}~\cite{Staub:2015kfa,Vicente:2015zba} package.
 
Signal cross sections were calculated to next-to-leading order (NLO) in the strong coupling constant, adding the resummation of soft gluon emission at next-to-leading-logarithm (NLL) accuracy (NLO+NLL) using \AtlasMC[2.0.1]{Resummino}~\cite{Beenakker:1999xh,Debove:2010kf,Fuks:2012qx,Fuks:2013vua,Fiaschi:2018hgm}. The nominal cross section and its uncertainty were taken from an envelope of cross-section predictions using different PDF sets and factorisation and renormalisation scales~\cite{Borschensky:2014cia,Butterworth:2015oua}. Production cross sections range between $\mathcal{O}(10^{-3}~\mathrm{pb})$ and $\mathcal{O}(1~\mathrm{pb})$.
 
\subsection{Irreducible-background samples}
\label{sec:irreduc-mc}
 
Production of \WZ and \WWSS represents the dominant irreducible background in most signal regions. Samples of fully leptonic, semileptonic and loop-induced \(VV\) (\(V=W,Z\)) processes and electroweak \(VVjj\) processes were simulated. The associated production of a vector gauge boson with a \ttbar pair, \ttVp, is also an important background. Depending on the targeted signal model, considerable background contributions come from \ttHp, tribosons and rare top processes, with the last including \tWZ, \tZq and samples with three or four top quarks. Higgs boson production via vector-boson fusion (VBF) and in association with a vector boson (\(VH\)) was also simulated, whereas production via gluon--gluon fusion and decay into two vector bosons was not simulated separately since the events are included in the diboson processes.
 
Samples of diboson final states \(VV\) were simulated with the \SHERPA[2.2.2]~\cite{Bothmann:2019yzt} generator,  including off-shell effects and Higgs boson contributions where appropriate. Fully leptonic final states and semileptonic final states, where one boson decays leptonically and the other hadronically, were generated using MEs at NLO accuracy in QCD for up to one additional parton and at leading-order (LO) accuracy for up to three additional parton emissions. Samples for the loop-induced processes \(gg \to VV\) were generated using LO-accurate MEs for up to one additional parton emission for both the cases of fully leptonic and semileptonic final states. The ME calculations were matched and merged with the \SHERPA PS based on Catani--Seymour (CS) dipole factorisation~\cite{Gleisberg:2008fv,Schumann:2007mg} using the MEPS@NLO prescription~\cite{Hoeche:2011fd,Hoeche:2012yf,Catani:2001cc,Hoeche:2009rj}. The virtual QCD corrections were provided by the \OPENLOOPS library~\cite{Buccioni:2019sur,Cascioli:2011va,Denner:2016kdg}. The \NNPDF[3.0nnlo] set of PDFs was used~\cite{Ball:2014uwa}, along with the dedicated set of tuned PS parameters developed by the \SHERPA authors.
 
Electroweak diboson production in association with two jets, \(VVjj\), was simulated with the \SHERPA[2.2.2] generator. The LO-accurate MEs were matched to the PS based on CS dipole factorisation using the MEPS@LO prescription. Samples were generated using the \NNPDF[3.0nnlo] PDF set, along with the dedicated set of tuned PS parameters developed by the \SHERPA authors.
 
The production of \ttVp\ events was modelled using the \MGNLO[2.3.3]~\cite{Alwall:2014hca} generator at NLO with the \NNPDF[3.0nlo]~\cite{Ball:2014uwa} PDF. The events were interfaced to \PYTHIA[8.210]~\cite{Sjostrand:2014zea}, which used the A14 tune and the \NNPDF[2.3lo]~\cite{Ball:2014uwa} PDF set.
 
Higgs bosons produced in association with a \ttbar pair, \ttHp, were generated using the \POWHEGBOX[v2]~\cite{Frixione:2007nw,Nason:2004rx,Frixione:2007vw,Alioli:2010xd,Hartanto:2015uka} generator at NLO with the \NNPDF[3.0nlo] PDF set. The events were interfaced to \PYTHIA[8.230]~\cite{Sjostrand:2014zea}, which used the A14 tune~\cite{ATL-PHYS-PUB-2014-021} and the \NNPDF[2.3lo] PDF set.
 
Triboson (\(VVV\)) event production was simulated with the \SHERPA[2.2.1]~\cite{Bothmann:2019yzt} generator. MEs accurate to LO in QCD for up to one additional parton emission were matched and merged with the \SHERPA PS based on CS dipole factorisation using the \MEPSatLO prescription. Samples were generated using the \NNPDF[3.0nnlo] PDF set, along with the dedicated set of tuned PS parameters developed by the \SHERPA authors.
 
The production of rare top events was modelled using the \MGNLO[2.3.3] generator, which provides MEs at NLO in the strong coupling constant \alphas, with the \NNPDF[3.1nlo]~\cite{Ball:2014uwa} PDF. The functional form of the renormalisation and factorisation scales was set to 0.25$\times \sum_i \sqrt{m^2_i+p^2_{\text{T},i}}$, where the sum runs over all the particles generated by the ME calculation, following Ref.~\cite{Frederix:2017wme}. Top quarks were decayed at LO using \MADSPIN~\cite{Frixione:2007zp,Artoisenet:2012st} to preserve all spin correlations. The events were interfaced with \PYTHIA[8.230]~\cite{Sjostrand:2014zea} for the PS and hadronisation, using the A14 tune and the \NNPDF[2.3lo] PDF set.
 
Higgs boson production was simulated with \POWHEGBOX[v2]~\cite{Nason:2009ai,Alioli:2010xd,Nason:2004rx,Frixione:2007vw} and interfaced with \PYTHIA[8]~\cite{Sjostrand:2014zea} for the PS and non-perturbative effects. The \POWHEGBOX prediction is accurate to NLO and uses the \PDFforLHC[15nlo] PDF set~\cite{Butterworth:2015oua} and the \AZNLO tune~\cite{STDM-2012-23} of \PYTHIA[8]~\cite{Sjostrand:2014zea}. The loop-induced \(gg\to ZH\) process was generated separately at LO. The MC prediction was normalised to cross sections calculated at next-to-NLO (NNLO) in QCD with NLO electroweak corrections for \(q\bar{q}/qg \to VH\) and at NLO and NLL in QCD for \(gg \to ZH\)~\cite{Ciccolini:2003jy,Brein:2003wg,Brein:2011vx,Altenkamp:2012sx,Denner:2014cla,Brein:2012ne,Harlander:2014wda}. The VBF production sample was normalised to an approximate-NNLO QCD cross section with NLO electroweak corrections~\cite{Ciccolini:2007jr,Ciccolini:2007ec,Bolzoni:2010xr}.  The normalisation of all Higgs boson samples accounts for the decay branching ratio calculated with \textsc{HDECAY}~\cite{Djouadi:1997yw,Spira:1997dg,Djouadi:2006bz} and \PROPHECY~\cite{Bredenstein:2006ha,Bredenstein:2006rh,Bredenstein:2006nk}.
 
\subsection{Reducible-background samples}
\label{sec:reduc-mc}
 
Even though they do not share the same final state as the signal, some SM processes are possible sources of background due to misidentification of leptons or their charges. These \textit{reducible} backgrounds, discussed in detail in \cref{sec:background}, are estimated with data-driven techniques. They include $V$+jets and electroweak VBF \(Vjj\), as well as top-quark pairs and single-top events.
 
The production of \(V+\)jets was simulated with the \SHERPA[2.2.1]~\cite{Bothmann:2019yzt} generator using NLO MEs for up to two partons, and LO MEs for up to four partons, calculated with the Comix~\cite{Gleisberg:2008fv} and \OPENLOOPS~\cite{Buccioni:2019sur,Cascioli:2011va,Denner:2016kdg} libraries. They were matched with the \SHERPA parton shower~\cite{Schumann:2007mg} using the \MEPSatNLO prescription~\cite{Hoeche:2011fd,Hoeche:2012yf,Catani:2001cc,Hoeche:2009rj} and the set of tuned parameters developed by the \SHERPA authors. The \NNPDF[3.0nnlo] set of PDFs~\cite{Ball:2014uwa} was used and the samples were normalised to a NNLO prediction~\cite{Anastasiou:2003ds}.
 
Electroweak VBF \(Vjj\) production leading to \(\ell\ell jj\), \(\ell\nu jj\) and \(\nu\nu jj\) final states was simulated with \SHERPA[2.2.11]~\cite{Bothmann:2019yzt} using LO MEs with up to one additional parton emission. The MEs were merged with the \SHERPA PS~\cite{Schumann:2007mg} following the \MEPSatLO prescription~\cite{Catani:2001cc} and using the set of tuned parameters developed by the \SHERPA authors. The \NNPDF[3.0nnlo] set of PDFs~\cite{Ball:2014uwa} was employed. The samples were produced in the VBF approximation, which avoids overlap  with semileptonic diboson topologies by requiring a \(t\)-channel colour-singlet exchange~\cite{Campanario:2018ppz}. The starting conditions of the CS shower were set according to the large-\(N_c\) amplitudes supplied by Comix~\cite{Buckley:2021gfw} to achieve  the correct VBF-appropriate radiation pattern.
 
The production of \ttbar events was modelled using the \POWHEGBOX[v2]~\cite{Frixione:2007nw,Nason:2004rx,Frixione:2007vw,Alioli:2010xd} generator at NLO with the \NNPDF[3.0nlo]~\cite{Ball:2014uwa} PDF set and the \hdamp parameter\footnote{The
\hdamp parameter is a resummation damping factor and one of the
parameters that controls the matching of \POWHEG MEs to
the PS and thus effectively regulates the
high-\pT radiation against which the \ttbar system recoils.} set to 1.5\,\mtop~\cite{ATL-PHYS-PUB-2016-020}.  The events were interfaced to \PYTHIA[8.230]~\cite{Sjostrand:2014zea} to model the PS, hadronisation, and underlying event, with parameter values set according to the A14 tune~\cite{ATL-PHYS-PUB-2014-021} and using the \NNPDF[2.3lo] set of PDFs~\cite{Ball:2012cx}.
 
The associated production of a top quark and a \(W\) boson (\(tW\)) and production of single-top in the \(s\)-channel (\(t\)-channel) were modelled using the \POWHEGBOX[v2]~\cite{Alioli:2009je,Nason:2004rx,Frixione:2007vw,Alioli:2010xd,Frederix:2012dh} generator at NLO in QCD in the five-flavour (four-flavour) scheme with the \NNPDF[3.0nlo]~\cite{Ball:2014uwa} PDF set. For \(tW\) production, the diagram removal scheme~\cite{Frixione:2008yi} was used to remove interference and overlap with \ttbar production. The events were interfaced with \PYTHIA[8.230]~\cite{Sjostrand:2014zea}, which used the A14 tune~\cite{ATL-PHYS-PUB-2014-021} and the \NNPDF[2.3lo] PDF set.


\FloatBarrier
\section{Object identification and reconstruction}
\label{sec:objects}

Leptons and jets selected for analysis are categorised as `baseline' (BL) or `signal' (Sig) according to various quality and kinematic selection criteria. The baseline objects are used in the computation of the missing-transverse-momentum vector \pTmiss and its magnitude \met, defined below, and to resolve ambiguities between closely spaced analysis objects.
 
Each electron candidate is reconstructed from a cluster of energy deposits in the EM calorimeter matched to an ID track. Baseline electrons are required to satisfy the {\ttfamily Loose} identification~\cite{EGAM-2018-01} and to have $\pt > \SI{10}{GeV}$ and $|\eta|< 2.47$, excluding the barrel-to-endcap transition region $1.37<|\eta|<1.52$ in the EM calorimeter.
The electron track's transverse impact parameter $d_0$, measured from the beamline with uncertainty $\sigma(d_0)$, must satisfy $|d_0/\sigma(d_0)|<5$, and its longitudinal impact parameter $z_0$, the $z$-distance from the primary vertex\footnote{The primary vertex is defined as the vertex with the largest sum of track $\pt^{2}$.} to the point where $d_0$ is measured, must satisfy $|z_0 \sin(\theta)|<\SI{0.5}{mm}$.
Baseline electrons that satisfy the tighter {\ttfamily Medium} identification~\cite{EGAM-2018-01} and satisfy both a track-based and a calorimeter-based isolation criterion are selected as signal electrons.
Track-based isolation requires the summed scalar $\pt$ of nearby ID tracks not to exceed 6\% of the electron \pt. Similarly to the isolation variables defined in Ref.~\cite{PERF-2015-10}, these nearby tracks must lie within in a cone of $\pt(e)$-dependent size $\Delta R = \sqrt{(\Delta\eta)^2+(\Delta\phi)^2} = \mathrm{min}\left\{0.2, \SI{10}{GeV}/\pt(e)\right\}$ around the electron, and must be associated with the primary vertex to limit sensitivity to pile-up.
Calorimeter-based isolation requires the sum of the transverse energies of the calorimeter energy clusters in a cone of $\Delta R = 0.2$ around the electron (excluding its own energy) to be less than 6\% of the electron's energy.
Only signal electrons with $|\eta|<2.0$ are considered, since this suppresses contributions from electrons having misidentified charge, and these are further rejected by exploiting information related to the electron track reconstruction and its compatibility with the primary vertex and the electron's energy cluster~\cite{EGAM-2018-01}.
 
Muon candidates are reconstructed~\cite{PERF-2015-10} in the region $|\eta|<2.5$ from MS tracks matching ID tracks.
Baseline muons satisfy $\pt>\SI{10}{GeV}$, $|\eta|<2.5$, $|z_0 \sin(\theta)|<\SI{0.5}{mm}$ and a set of {\ttfamily Medium} requirements~\cite{MUON-2018-03} on the quality of the tracks.
Signal muons are defined as baseline muons that also satisfy the requirement $|d_0/\sigma(d_0)|<3$ and pass track-based isolation requirements that are robust against pile-up and similar to those for electrons, but with the maximal cone size increased to $0.3$.
 
Jets are reconstructed from particle-flow energy deposits using the anti-$k_t$ algorithm~\cite{Cacciari:2008gp} with four-momentum recombination and distance parameter $R = 0.4$. The reconstructed jets are then calibrated by the application of a jet energy scale derived from 13~\TeV\ data and simulation~\cite{PERF-2016-04}. Jets with $\pt > 20$~\GeV\ and $|\eta| < 4.5$ are used as baseline jets in the analysis and are also used in computing the \met.
Signal jets are selected as jets satisfying the requirements of $\pt>\SI{20}{GeV}$ and $|\eta|<2.8$. To suppress jets originating from pile-up, additional track-based criteria are applied by using the {\ttfamily Tight} working point of the jet vertex tagger~\cite{PERF-2014-03, PERF-2016-04}.
 
Signal jets containing $b$-hadrons, referred to as \bjets, are identified ($b$-tagged) by the {\ttfamily DL1r} algorithm~\cite{FTAG-2018-01,ATL-PHYS-PUB-2017-013} via a multivariate discriminant combining information from the impact parameters of displaced tracks with topological properties of secondary and tertiary decay vertices reconstructed within the jet. The chosen working point has a \bjet tagging efficiency of 70\% and rejection factors of 6 and 134 for charm-jets and light-flavour jets, respectively. Additionally, the selected \bjets must satisfy $|\eta| < 2.5$.
 
To avoid the double counting of analysis baseline objects, a procedure to remove reconstruction ambiguities is applied as follows:
\begin{itemize}
\item Electron candidates within $\Delta R^{\prime} = \sqrt{(\Delta y)^2+(\Delta\phi)^2} = 0.01$ of a muon are removed.\footnote{The quantity $y = (1/2)[(E+p_z)/(E-p_z)]$ denotes the rapidity of an object.} Softer electron candidates are removed if they are within $\Delta R^{\prime} = 0.05$ of other electron candidates.
\item Jet candidates within $\Delta R^{\prime} = 0.2$ of an electron candidate are removed unless the jet candidate is a \bjet with $\pt< \SI{100}{GeV}$. Jets with fewer than three tracks that lie within $\Delta R^{\prime} = 0.4$ of a muon candidate are removed.
\item Subsequently, electrons and muons within $\Delta R^{\prime}=\operatorname{min}\left\{0.4, 0.1+\SI{9.6}{GeV}/\pt(\ell)\right\}$ of a jet candidate are removed to reject non-prompt or fake leptons originating from hadron decays.
\end{itemize}
 
The \pTmiss is defined as the negative vector sum of the transverse momenta of all identified objects
(baseline electrons, photons~\cite{EGAM-2018-01}, muons and jets) and an additional soft term.
The soft term is constructed from all tracks associated with the primary vertex but not with leptons or jets.
In this way, the magnitude of the \pTmiss, \met, is adjusted for the best calibration of the identified objects listed above, while maintaining approximate pile-up independence in the soft term~\cite{PERF-2016-07,ATLAS-CONF-2018-023}.
Overlaps between objects in the \met\ calculation are resolved as described in Ref.~\cite{PERF-2016-07}.


\FloatBarrier
\section{Analysis strategy and event selection}
\label{sec:selection}

\graphicspath{{figures/}}
 
After a basic event-cleaning procedure is applied, events are required to have a primary vertex with at least two associated tracks with $\pt > 500~\mev$. Jets likely to have been produced by beam-induced backgrounds, cosmic rays or detector noise are removed and other jet quality criteria are imposed~\cite{ATLAS-CONF-2015-029}. Events with at least one muon with low momentum resolution are rejected.
 
Events with at least two signal leptons, with the leading lepton satisfying $\pt>\SI{20}{GeV}$, are selected.
In addition, there must be either at least one pair of leptons with identical electric charges among the ensemble of signal leptons or exactly three leptons. The presence of at least one jet is also required in most signal regions (SRs) in order to improve the selection of signal events and to specifically target compressed-spectra regions. To distinguish between hypothetical SUSY signal processes and SM backgrounds, sets of SRs are optimised for the SUSY models defined in \cref{sec:models}. Each of these SRs, described in \cref{tab:SR_Wh,tab:SR_WZ,tab:SR_bRPV}, is kept orthogonal to those in other ATLAS analyses~\cite{SUSY-2019-09} to facilitate future statistical combinations. Several kinematic variables are deployed to maximise the sensitivities to the targeted signals.

\begin{table}[!tb]
\centering
\caption{Signal region definitions designed for the \Wh model. The variables are defined in the text.}
\begin{threeparttable}[hb]
\begin{tabular}{l|c|c|c|c|c|c}
\hline \hline
& \multicolumn{3}{c|}{} & \multicolumn{3}{c}{} \\ [1pt]
& \multicolumn{3}{c|}{\SRhmttWh} & \multicolumn{3}{c}{\SRlmttWh} \\
& \multicolumn{3}{c|}{} & \multicolumn{3}{c}{} \\ [1pt]
& $e^{\pm}e^{\pm}$ & $e^{\pm}\mu^{\pm}$ & $\mu^{\pm}\mu^{\pm}$ & $e^{\pm}e^{\pm}$ & $e^{\pm}\mu^{\pm}$ & $\mu^{\pm}\mu^{\pm}$ \\
\hline\hline
$N_{\mathrm{BL}}(\ell)$		&	\multicolumn{6}{c}{$=2$} \\
$N_{\mathrm{Sig}}(\ell)$	&	\multicolumn{6}{c}{$=2$} \\
Charge$(\ell)$				&	\multicolumn{6}{c}{same-sign} \\
$\pt(\ell)$					&	\multicolumn{6}{c}{$\geq\SI{25}{GeV}$} \\
\nJ	($\pt > 25~\gev$)		                &	\multicolumn{6}{c}{$\geq1$} \\
\nb		                    &	\multicolumn{6}{c}{$=0$} \\
\mjj      & \multicolumn{6}{c}{$<\SI{350}{GeV}$} \\
\hline \hline
\mttwo    & \multicolumn{3}{c|}{$\geq\SI{80}{GeV}$} & \multicolumn{3}{c}{$<\SI{80}{GeV}$} \\
\minmt    & \multicolumn{3}{c|}{--}                  & \multicolumn{3}{c}{$\geq\SI{100}{GeV}$} \\
\metsig   & \multicolumn{3}{c|}{$\geq 7$}           & \multicolumn{3}{c}{$\geq 6$} \\
\met	  & \multicolumn{3}{c|}{$\geq\SI{75}{GeV}$} & \multicolumn{3}{c}{$\geq\SI{50}{GeV}$} \\
\hline \hline
\multirow{3}{*}{\met binning [\gev]\tnote{a}} & \multicolumn{3}{c|}{\SRhmttWh-1: $\in[75,125)$} & \multicolumn{3}{c}{} \\
& \multicolumn{3}{c|}{\SRhmttWh-2: $\in[125,175)$}     & \multicolumn{3}{c}{--} \\
& \multicolumn{3}{c|}{\SRhmttWh-3: $\in[175,+\infty)$} & \multicolumn{3}{c}{} \\
\hline
\end{tabular}
\begin{tablenotes}
\footnotesize
\item[a]{The \met binning applies separately to each flavour channel of \SRhmttWh.}
\end{tablenotes}
\end{threeparttable}
\label{tab:SR_Wh}
\end{table}


\begin{table}[!tb]
\centering
\caption{Signal region definitions designed for the \WZ model. The variables are defined in the text.}
\begin{tabular}{c|c|c}
\hline \hline
& & \\ [1pt]
& \SRhmttWZ & \SRlmttWZ \\
& & \\ [1pt]
\hline \hline
 
$N_{\mathrm{BL}}(\ell)$ & \multicolumn{2}{c}{= 2} \\
 
$N_{\mathrm{Sig}}(\ell)$ & \multicolumn{2}{c}{= 2} \\
 
Charge$(\ell)$ & \multicolumn{2}{c}{same-sign} \\
 
$\pt(\ell)$ & \multicolumn{2}{c}{$\geq \SI{25}{GeV}$} \\
 
\nJ	$(\pt > 25~\gev)$ & \multicolumn{2}{c}{$\geq 1$} \\
 
\nb	 & \multicolumn{2}{c}{= 0} \\
 
\mjj & \multicolumn{2}{c}{$\leq 350$ GeV} \\
 
\hline \hline
\mttwo  & $\geq 100$~\gev & $\leq 100$~\gev \\
\minmt & $\geq 100$~\gev & $\geq 130$~\gev \\
$\met$ & $\geq 100$~\gev & $\geq 140$~\gev \\
$\meff$ & -- & $\leq 600$~\gev \\
$\Delta R(\ell^{\pm}, \ell^{\pm})$ & -- & $\leq 3$ \\
\hline \hline
\multirow{5}{*}{Bins} & \metsig: $\in [0,10)$ & \multirow{5}{*}{--} \\
& Spread($\Phi$) $\geq 2.2$ & \\
\cline{2-2}
& \metsig: $\in [10,13)$                               & \\
\cline{2-2}
& \metsig: $\in [13,+\infty]$ & \\
& $\Delta R(\ell^{\pm}, \ell^{\pm})\geq 1$ & \\
\hline \hline
\end{tabular}
\label{tab:SR_WZ}
\end{table}


\begin{table}[!tb]
\centering
\caption{Signal region definitions designed for the bRPV model. The variables are defined in the text.}
\begin{tabular}{c|c|c}
\hline \hline
& & \\ [1pt]
& \SRtwolbrpv & \SRthreelbrpv \\
& & \\ [1pt]
\hline \hline
$N_{\mathrm{BL}}(\ell)$ & \multicolumn{2}{c}{--} \\
\pt($\ell$) & \multicolumn{2}{c}{$\geq \SI{20}{GeV}$ for (sub)leading leptons}  \\
\nJ $(\pt > 25~\gev)$ & \multicolumn{2}{c}{$\geq 1$} \\
\hline \hline
$N_{\mathrm{Sig}}(\ell)$ & = 2 & = 3 \\
Charge($\ell$) & same-sign & -- \\
\mttwo & $\geq\SI{60}{GeV}$ & $\geq\SI{80}{GeV}$ \\
\met & $\geq \SI{100}{GeV}$ & $\geq \SI{120}{GeV}$ \\
\meff & -- & $\geq \SI{350}{GeV}$ \\
\nb	 & = 0 & -- \\
\nJ $(\pt > 40~\gev)$ & $\geq 4$ & -- \\
$m_{e^{\pm}e^{\mp}}, \; m_{\mu^{\pm}\mu^{\mp}}$ & -- & $ \notin [81,101]$~\gev \\
\hline \hline
\end{tabular}
\label{tab:SR_bRPV}
\end{table}


The `stransverse mass', $\mttwo$, is an event variable used to bound the masses of an unseen pair of particles which are presumed to have decayed semi-invisibly into particles which were seen~\cite{Lester:1999tx,Barr:2003rg}. Therefore, it is defined as a function of the momenta of two visible particles and the \pTmiss of the event:
\begin{equation*}\label{eq:mttwo}
\mttwo = \min_{\qTvec}\left[\max\left(m_{{\mathrm T}, \ell_1}(\mathbf{p}_{\mathrm{T}, \ell_1},\qTvec),m_{{\mathrm T}, \ell_2}(\mathbf{p}_{\mathrm{T}, \ell_2},\pTmiss-\qTvec)\right)\right],
\end{equation*}
where $\mathbf{p}_{\mathrm{T}, \ell_1}$ and $\mathbf{p}_{\mathrm{T}, \ell_2}$ are the transverse momenta of the two leading leptons, and $\qTvec$ is the transverse momentum vector that minimises the larger of the two transverse masses $m_{{\mathrm T}, \ell_1}$ and $m_{{\mathrm T}, \ell_2}$.
These two transverse masses are defined as
\begin{equation*}\label{eq:mtpq}
m_{{\mathrm T}}(\pTvec,\qTvec) = \sqrt{2(\pT\qT-\pTvec\cdot\qTvec)}.
\end{equation*}
In this analysis, the invisible particle mass is always set to zero when calculating the event \mttwo.
 
For the \Wh and \WZ models, \mttwo was used to define two orthogonal sets of signal regions, `high-\mttwo' and `low-\mttwo', to target models with different kinematics.
Exactly two baseline leptons, $N_{\mathrm{BL}}(\ell)=2$, were required for these two models to further suppress the background.
Requiring the invariant mass of the two leading jets, \mjj,\footnote{If the event has only one jet, \mjj was set to zero.} to be less than $\SI{350}{GeV}$ proved to be efficient in reducing the \WWSS background.
The transverse mass of the \pTmiss and each of the two leading leptons was calculated, and the smaller of the two values, \minmt, is used to recover the sensitivity which would otherwise be lost if only high \mttwo were considered. The
\met and its significance, $\mathcal{S}(\met)$~\cite{ATLAS-CONF-2018-038},\footnote{$\mathcal{S}(\met) = \frac{|\met|^{2}}{\sigma^{2}_\mathrm{L}(1-\rho^{2}_\mathrm{LT})}$, with $\sigma^{2}_\mathrm{L}$ the total variance in the longitudinal direction along \pTmiss and $\rho^{2}_\mathrm{LT}$ the correlation between the longitudinal and transverse resolutions of the objects.} which quantifies the robustness of the \met values against object mismeasurements in events without a genuine source of \met, are also used to target the large \met induced by the (invisible) LSP in RPC scenarios.
The angular distance between the two SS leptons, $\Delta R(\ell^{\pm}, \ell^{\pm})$, is used only for the \WZ model since the SS leptons come from two separate decay legs and should not be too far apart when the masses of the SUSY particles are similar.
 
A multi-bin strategy is applied in the `high-\mttwo' SRs, using \met and flavour for the \Wh model and \metsig for the \WZ model, to maximise the sensitivity across the model's phase space. No similar binning is employed in the `low-\mttwo' SRs for the \WZ model, due to the limited number of surviving events.
For the bins defined for \SRhmttWZ, requirements on the Spread($\Phi$)\footnote{The spread of the $\Phi$ angles of the leptons, \met, and jets is used to describe the event topology in the transverse plane. It was defined and used in Ref.~\cite{HDBS-2019-06}.
It is defined as:
$\text{Spread}(\Phi) = \frac{\mathcal{R}(\phi_{\ell 1},~ \phi_{\ell 2},~ \phi_{\met} ) \cdot \mathcal{R}(\phi_{j 1},~ \phi_{j 2}, \dots) }{ \mathcal{R}(\phi_{\ell 1},~ \phi_{\ell 2},~ \phi_{\met},~ \phi_{j 1},~ \phi_{j 2}, \dots ) }$, where $\mathcal{R}$ means the root-mean-square of the inputs.} or $\Delta R(\ell^{\pm}, \ell^{\pm})$ are applied to further improve the sensitivity to the  benchmark model, increasing the significance by up to 20\%. The \Wh SRs are divided into different flavour channels to maximise the power of the analysis.
 
For the bRPV model, large \met is expected due to the presence of a neutrino in the leptonic decay of the \Wboson boson. High jet multiplicity is required in the two-SS-lepton SR to improve the sensitivity to possible hadronic decays of the higgsinos and the \Wboson boson. An \mttwo threshold at \SI{60}{GeV} or \SI{80}{GeV} is found to be helpful because the \met composition differs between the signal and the background sources.
For the two-SS-lepton SR, a \bjet veto is applied to further reduce the \ttbar backgrounds.
For the three-lepton SR, a lower bound on the effective mass \meff, defined as the scalar sum of the \met and the objects' \pt values, is placed at \SI{350}{GeV} and has proven useful in reducing the remaining background after applying the \Zboson-boson veto ($m_{e^{\pm}e^{\mp}}, \; m_{\mu^{\pm}\mu^{\mp}} \notin [81, 101]~\gev$).
 
Within each signal model, the SRs are designed to be orthogonal to allow their statistical combination in the interpretation of the results. In the wino--bino models, this is achieved with the \mttwo variable, while the number of signal leptons, $N_{\mathrm{Sig}}(\ell)$, ensures orthogonality in the bRPV model.
 
While these SRs are designed to maximise the sensitivity to specific benchmark models, a different set of \emph{discovery SRs} were defined to enhance the discovery potential for a variety of BSM scenarios, such as simplified models with electroweak SUSY production which span the compressed to high mass-splitting scenarios. The inclusive \SRhmttWh and \SRlmttWh for the \Wh model and the \SRhmttWZ and \SRlmttWZ for the \WZ model, defined without any \met or \metsig binning or flavour splitting, act as such discovery regions.
 
The product $A\times\epsilon$ of the acceptance $A$ of the selection criteria and the efficiency $\epsilon$  that accounts for the detector effects, ranges from 0.01\% to a few percent  for the SRs defined in \cref{tab:SR_Wh,tab:SR_WZ,tab:SR_bRPV}. For example, \SRhmttWh (\SRlmttWZ) yields an
$A\times\epsilon$ of ${\sim}0.02\%$ (${\sim}2\%$) in the \Wh (\WZ) model for $m(\chinoonepm/\ninotwo)=\SI{200}{GeV}$ and a massless LSP.


\FloatBarrier
\section{Background estimation}
\label{sec:background}

\graphicspath{{figures/}}
 
The treatment of the SM backgrounds is based on their classification as either irreducible backgrounds, from processes with genuine same-sign prompt leptons, or reducible backgrounds, with events entering the SRs because of misidentification of the lepton (`fake/non-prompt') or the lepton charge (`charge-flip').
The `charge-flip' events (referred to in the following as CF events) are caused by the emission of a bremsstrahlung photon which, through interaction with detector material, converts into a pair of secondary electron tracks. One of those tracks happens to match the position of the calorimeter energy cluster better than the original electron track does, and has a charge opposite to that of the prompt electron.
The CF contribution coming from muons is negligible due to the small cross section for interactions with matter.
The `fake/non-prompt' events (referred to in the following as FNP events) are mainly due to heavy-flavour meson decays, converted photons of various origin, light hadrons faking the electron shower, and in-flight decays of kaons or pions to muons.
Lepton candidates reconstructed from these different sources share the properties of being generally not well-isolated and being mostly rejected by the lepton identification and isolation criteria and impact parameter requirements.
 
The dominant irreducible background processes in the SRs defined in this analysis are \WZ, \WWSS for SRs with a \bjet veto (\SRhmttWh, \SRlmttWh, \SRhmttWZ, \SRlmttWZ and \SRtwolbrpv) and \ttVp for the \bjet-agnostic SR (\SRthreelbrpv).
The \WZ and \WWSS contributions to the respective SRs are evaluated by normalising the MC prediction in dedicated control regions (CRs). All other irreducible backgrounds, discussed in~\cref{sec:datamc}, are estimated from MC simulation.   The reducible backgrounds are estimated through data-driven estimation techniques.
 
The background estimates are obtained by performing a profile log-likelihood fit~\cite{Cowan:2010js}, implemented in the \HISTFITTER~\cite{Baak:2014wma} software framework, considering only the CRs and assuming no signal presence. The statistical and systematic uncertainties are implemented as nuisance parameters in the likelihood; Poisson constraints are used to estimate the uncertainties arising from limited numbers of events in the MC samples, whilst Gaussian constraints are used for experimental and theoretical systematic uncertainties. The normalisation factors and nuisance parameters are adjusted by maximising the likelihood. The significance of the difference between the observed and expected yields is calculated with the profile likelihood method~\cite{Cousins:2007bmb}.
 
The validation regions (VRs), which serve solely to validate the background estimation in the SRs, are defined to be orthogonal to, but close to, both the SRs and the CRs. The background prediction as obtained from this background-only fit is compared with data in the VRs to assess the quality of the background modelling.
 
For the dominant backgrounds in SRs optimised for the \Wh model (\SRhmttWh and \SRlmttWh), dedicated CRs are designed for the \WZ (\CRWZWh) process and the \WWSS process (\CRWWWh). The scale factor for each targeted background process  is obtained via a simultaneous fit in the specific control region. VRs with enriched contributions from \WZ (\VRWZWh) or \WWSS (\VRWWWh) are also defined in order to validate the estimates. The CR and VR definitions for the \Wh model are listed in~\cref{tab:Wh_CRVR}.

\begin{table}[!tb]
\centering
\caption{Control region and validation region definitions for evaluating and validating the dominant irreducible backgrounds in SRs defined for the \Wh model. Requirements guaranteeing orthogonality to SRs are in boldface.}
\begin{tabular}{c|cc|cc}
\hline \hline
& & & & \\ [1pt]
& \CRWZWh & \VRWZWh & \CRWWWh & \VRWWWh \\
& & & & \\ [1pt]
\hline \hline
$N_{\mathrm{BL}}(\ell)$ & \multicolumn{2}{c|}{$= 3$} & \multicolumn{2}{c}{$= 2$} \\
$N_{\mathrm{Sig}}(\ell)$ & \multicolumn{4}{c}{= 2} \\
Charge$(\ell)$ & \multicolumn{4}{c}{same-sign} \\
$\pt(\ell)$ & \multicolumn{4}{c}{$\geq \SI{25}{GeV}$} \\
\nb & \multicolumn{4}{c}{= 0} \\
\met & \multicolumn{4}{c}{$\geq \SI{50}{GeV}$} \\
\nJ & \multicolumn{2}{c|}{$\geq 1$} & \multicolumn{2}{c}{$\geq 2$} \\
$\mathcal{S}(\met)$ & $\mathbf{< 6}$ & $\geq 6$ & $\mathbf{< 6}$ & $\geq 6$ \\
\hline
\multirow{5}{*}{Other cuts} & \multicolumn{2}{c|}{$75<m_\text{SFOS}<\SI{105}{GeV}$} & \multicolumn{2}{c}{--} \\
 
& \multicolumn{2}{c|}{$m_{\ell\ell\ell}\notin [80,100] \gev$} & \multicolumn{2}{c}{--} \\
 
& \multicolumn{2}{c|}{--} & \multicolumn{2}{c}{$\boldsymbol{\mjj\geq \SI{350}{GeV}}$} \\
& \multicolumn{2}{c|}{--} & \multicolumn{2}{c}{$\pt\text{(jets)}\geq \SI{75}{GeV}$ for (sub)leading jets} \\
& \multicolumn{2}{c|}{--} & \multicolumn{2}{c}{$|m_{e^{\pm}e^{\pm}} - m_{Z}| \geq \SI{15}{GeV}$} \\
\hline \hline
Purity &  90\%  & 90\% &  45\%  & 55\%    \\
\hline \hline
\end{tabular}
\label{tab:Wh_CRVR}
\end{table}


The requirements on the numbers of leptons, $N_{\mathrm{Sig}}(\ell)$ and $N_{\mathrm{BL}}(\ell)$, number of jets, \nJ, and $\met$ applied in the CRs and VRs are similar to those applied in \SRhmttWh and \SRlmttWh, and are listed in~\cref{tab:Wh_CRVR}.  Values of the \met significance (\metsig) larger or smaller than six are used to distinguish the CRs from the VRs while keeping the VRs close to the SRs.
For \WZ-enriched regions, the third lepton satisfies the baseline lepton criteria \emph{without} fulfilling the signal lepton definition in order to maintain the orthogonality between CRs, VRs and SRs. In addition, the invariant mass of a pair of same-flavour opposite-sign leptons (SFOS), $m_\text{SFOS}$, is required to be within a window of $\pm15$~\gev\ around $m_{Z}$, and the invariant mass of the three leptons, \mlll, is required to be away from the $Z$ mass peak. Such criteria further improve the purity and suppress other backgrounds. The purity of the \WZ process in \CRWZWh and \VRWZWh is about 90\% with negligible contamination from signal.
 
To target the \WWSS process for the \WWSS-enriched regions \CRWWWh and \VRWWWh, two boosted jets with $\pt\geq \SI{75}{GeV}$ are required, while requiring $\mjj\geq \SI{350}{GeV}$ ensures orthogonality with respect to the SRs. To suppress the CF contribution, events are rejected if $|m_{e^{\pm}e^{\pm}} - m_{Z}| < \SI{15}{GeV}$. The final purity of \CRWWWh (\VRWWWh) is about 45\% (55\%) with a signal contamination of less than 3\% in both the CR and the VR.
The scale factors are $1.06^{+0.14}_{-0.08}$ and $1.00^{+0.25}_{-0.28}$ for the \WZ and \WWSS backgrounds, respectively, and are applied to these background events in the regions designed for the \Wh model. Both the statistical and systematic uncertainties, described in \cref{sec:systs}, are considered in the scale factors.
In~\cref{fig:Wh_CRVR}, good agreement between the observed data and the estimated backgrounds can be seen for \VRWZWh and \VRWWWh.
 
\begin{figure}[!tb]
\centering
\includegraphics[width=0.9\textwidth]{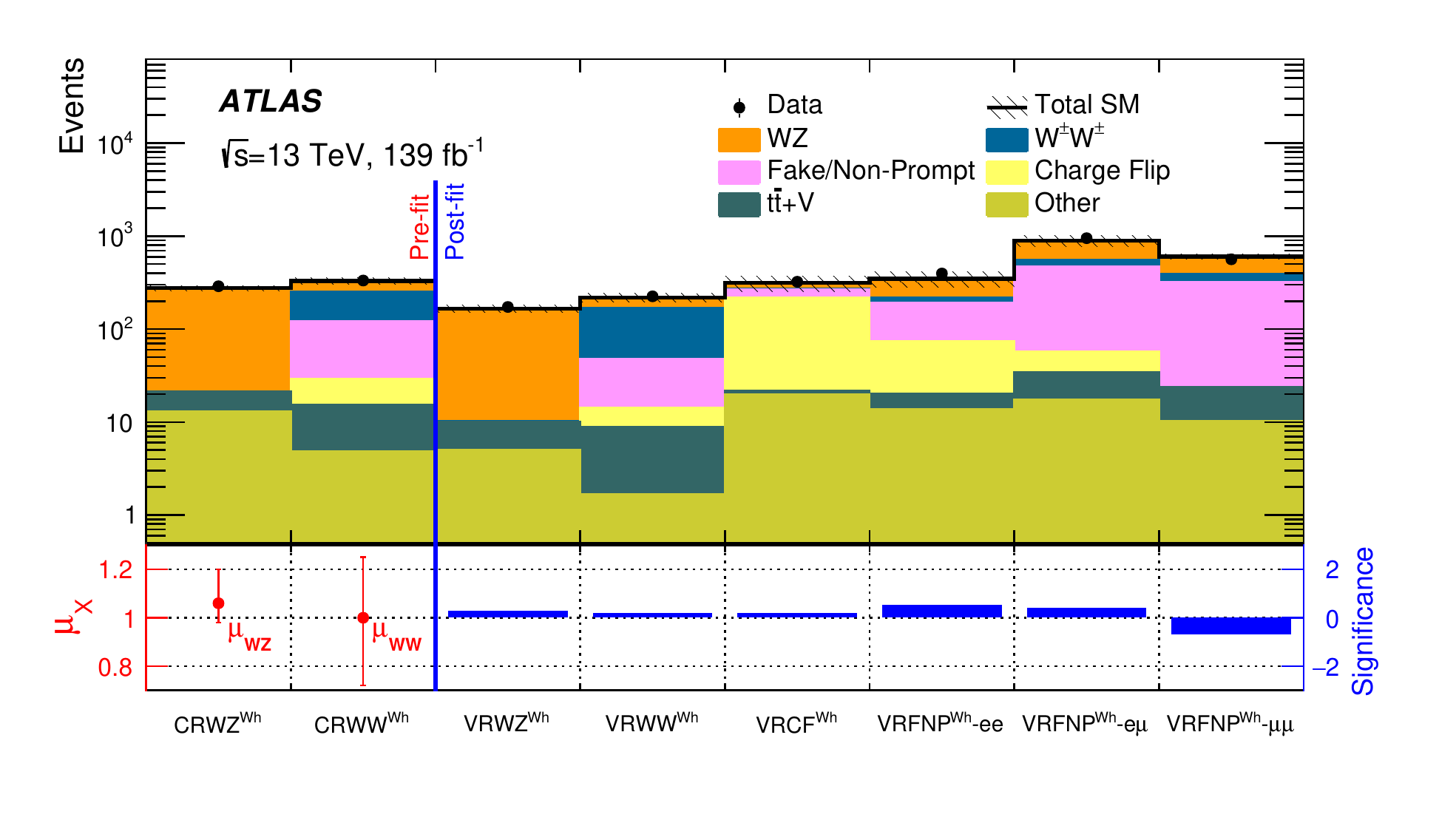}
\caption{Expected SM backgrounds and data yields in the \CRWZWh, \CRWWWh, \VRWZWh, \VRWWWh, \VRCFWh and \VRfakeWh designed for the \Wh model. The `Other' category contains the \ttHp, rare top, triboson, and other diboson processes with the SS final state. The error band includes the statistical, theoretical and experimental uncertainties. The bottom panel shows the obtained scale factors ($\mu_{\WZ}$, $\mu_{WW}$) in the CRs and the statistical significance~\cite{Cousins:2007bmb} of the discrepancy between the observed number of events and the SM expectation.}
\label{fig:Wh_CRVR}
\end{figure}
 
For SRs designed for the \WZ model and models of higgsino-like electroweakinos in RPV SUSY considered in this analysis, a general control region \CRWZtwoj for the \WZ process is defined in order to correct the cross section in a region of phase space with at least two jets, where imprecise modelling was observed in previous analyses~\cite{SUSY-2018-09}.
The validation regions for the \WZ process (\VRWZfourj and \VRWZfivej) and the \ttVp process (\VRttV), defined in \cref{tab:CRVR}, are designed to validate the estimates from the MC simulation of these processes. Large jet multiplicities are required in those validation regions in order to validate the modelling of those processes in the phase space where previous analyses observed the largest disagreements~\cite{SUSY-2018-09}.

\begin{table}[!tb]
\centering
\caption{Control region and validation region definitions for evaluating and validating the dominant irreducible backgrounds in SRs defined for the \WZ model and models of higgsino-like electroweakinos in (b)RPV SUSY. Requirements guaranteeing orthogonality with SRs are in boldface.}
\resizebox{\textwidth}{!}{
\begin{tabular}{c|cccc}
\hline \hline
& & & & \\ [1pt]
& \CRWZtwoj & \VRWZfourj & \VRWZfivej & \VRttV \\
& & & & \\ [1pt]
\hline \hline
$N_{\mathrm{BL}}(\ell)$ &  \multicolumn{3}{c}{= 3} & $\geq 2$ \\
$N_{\mathrm{Sig}}(\ell)$ &  \multicolumn{3}{c}{= 3} & $\geq 2$ \\
Charge$(\ell)$ &  \multicolumn{3}{c}{--} & same-sign \\
$\pt(\ell)$ &  \multicolumn{3}{c}{$\pt > \SI{20}{GeV}$ for (sub)leading leptons} & $\pt > \SI{30}{GeV}$ for SS pair leptons \\
\nb & \multicolumn{3}{c}{= 0} & $\geq 1$ \\
\nJ ($\pt\geq \SI{25}{GeV}$) & $\geq 2$ & $\geq 4$ & $\geq 5$ & $\geq 3$ with $\pt > \SI{40}{GeV}$ \\
\hline
\multirow{7}{*}{Other selections}
 
& $50 < \met < \SI{150}{GeV}$ & \multicolumn{2}{c}{$50 < \met < \SI{250}{GeV}$} & -- \\
 
& $\meff < \SI{1}{TeV}$ & \multicolumn{2}{c}{$\meff < \SI{1.5}{TeV}$} & -- \\
 
& $81<m_\text{SFOS}<101$~GeV & \multicolumn{2}{c}{$81<m_\text{SFOS}<101$~GeV} & -- \\
 
& \multicolumn{3}{c}{--} & $\Delta R (\ell_{1}, \mathrm{jet})_{\min}>1.1$ \\
 
& \multicolumn{3}{c}{--} & $\sum\pt^{\bjet}/\sum\pt^{\text{jet}}>0.4$ \\
 
& \multicolumn{3}{c}{--} & $\met/\meff > 0.1$ \\
 
& \multicolumn{4}{c}{\textbf{explicit veto on \SRhmttWZ \& \SRlmttWZ \& \SRtwolbrpv \& \SRthreelbrpv events}} \\
\hline
\multirow{4}{*}{\parbox{3cm}{Vetoing other possible BSM events}}
& \multicolumn{4}{c}{$\nb \geq 3$} \\
& \multicolumn{4}{c}{$\nb \geq 1$, $\nJ \geq 4$ ($\pt>\SI{50}{GeV}$), $\met > \SI{130}{GeV}$} \\
& \multicolumn{4}{c}{$\nb = 0$, $\nJ \geq 3$ ($\pt>\SI{50}{GeV}$), $\met > \SI{130}{GeV}$} \\
& \multicolumn{4}{c}{$\nb = 0$, $\nJ \geq 5$ ($\pt>\SI{50}{GeV}$)} \\
\hline \hline
Purity & 85\% & 84\% & 77\% & 62\% \\
\hline \hline
\end{tabular}
}
\label{tab:CRVR}
\end{table}


To define \WZ and bRPV CRs, requirements are placed on the number of signal leptons, $N_{\mathrm{Sig}}(\ell)$, the number of baseline leptons, $N_{\mathrm{BL}}(\ell)$, the number of jets, \nJ, and the number of \bjets , \nb. Additional requirements are set on \met, \meff, $m_\text{SFOS}$ and the presence of SS leptons.
A minimum angular separation between the leading lepton and the jets, $\Delta R (\ell_1, j)_{\min}$, is required in the validation regions targeting \ttVp events, as well as requirements on $\sum\pt^{\bjet}/\sum\pt^{\text{jet}}$.
The leading and subleading lepton \pt are required to be above \SI{20}{GeV}.
The events belonging to the SRs of the \WZ model and the bRPV model defined in~\cref{sec:selection} are vetoed. In addition, the selections given in \cref{tab:CRVR} are applied to ensure a more stringent rejection of possible bRPV and \textit{UDD} RPV signal events, as well as other SUSY signals with several ($b$-)jets and moderate \met in the final state. These vetoes help to reduce the expected signal contamination to a few percent. The purity of the target background process varies from a minimum of 62\% (\VRttV) to a maximum of 85\% (\CRWZtwoj).
 
The scale factor and its uncertainty are extracted from \CRWZtwoj and are found to be $0.88\pm0.30$. The estimated backgrounds and the observed data in \CRWZtwoj, \VRWZfourj, \VRWZfivej and \VRttV are shown in \cref{fig:CRVR}, where good agreement is observed.
 
\begin{figure}[!tb]
\centering
\includegraphics[width=0.9\textwidth]{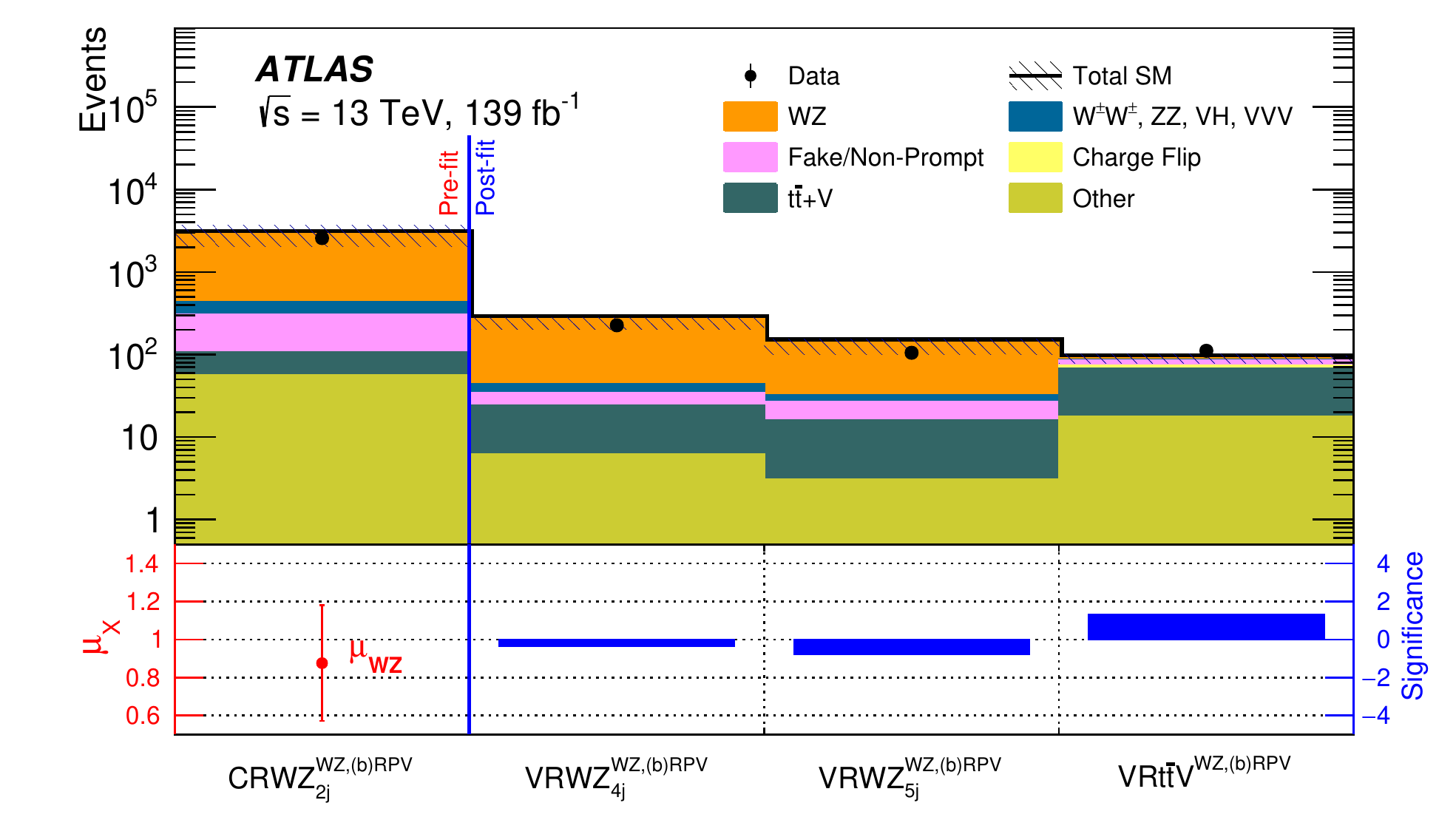}
\caption{Expected SM backgrounds and data yields in \CRWZtwoj, \VRWZfourj, \VRWZfivej and \VRttV designed for the \WZ model and models of higgsino-like electroweakinos in RPV SUSY. The `Other' category contains the \ttHp and rare top processes with the SS final state. The error band includes the statistical, theoretical and experimental uncertainties. The bottom panel shows the scale factor obtained from \CRWZtwoj ($\mu_{\WZ}$) and the statistical significance~\cite{Cousins:2007bmb} of the discrepancy between the observed number of events and the SM expectation.}
\label{fig:CRVR}
\end{figure}
 
The contributions of CF events are evaluated from reweighted data events with two opposite-sign leptons ($e^{\pm}e^{\mp}$, $e^{\pm}\mu^{\mp}$). The weight expresses the probability of one electron charge to be mismeasured and is a function of the electron CF rates.
This method largely improves the statistical accuracy by relying entirely on data to obtain the reweighting factors, thus eliminating uncertainties associated with MC simulations. An additional 25\% uncertainty stems from the choice of lepton selections, and was derived by comparing the nominal CF predictions with those obtained using BL leptons.
 
The CF rates are measured as a function of lepton $\pt$ and $|\eta|$ for simulated SM processes that contribute to the SRs due to CF.
They are multiplied by the scale factors obtained from a `tag and probe' method~\cite{EGAM-2018-01} to match the rates observed in data.
The nominal CF rates are no more than $\mathcal{O}(10^{-6})$ in the low-$\pt$ region, but reach $\mathcal{O}(1\%)$ in the higher $\pt$ and $|\eta|$ regions.
Systematic uncertainties are estimated from the statistical uncertainties of the measured CF rates and the uncertainties from the scale factors, leading to a 10\% to 40\% uncertainty in the predicted SR yields for the CF background.
 
The fake-factor method, the matrix method and the MC template method are used in this analysis to estimate the contributions of FNP events in the SRs.
Both the fake-factor method and the matrix method are purely data-driven methods, which are commonly employed in the ATLAS Collaboration~\cite{STDM-2011-24,ATL-PHYS-PUB-2010-005,TOPQ-2010-01} to estimate the FNP background in  dedicated regions. In this analysis, the fake-factor method is used to estimate the contribution of FNP events in the \Wh regions. Hence, the measurements of the values of the fake-factors are specifically tailored to reflect the FNP composition of the two-SS-lepton SRs of the \Wh model.
The implementation of the matrix method in this analysis is instead designed to be more universal, which enables it to estimate the FNP contribution in more complex regions. Therefore, it is used to evaluate the FNP events in SRs defined for the \WZ model and models of higgsino-like electroweakinos in RPV SUSY which have two SS leptons or three leptons, and $b$-vetoed or $b$-favoured channels.
Finally, the (semi-data-driven) MC template method~\cite{SUSY-2013-09} is used to validate specific matrix-method estimates  to ensure that the more universal matrix method is functioning well for the specific cases in this analysis.
 
The fake-factor method estimates the FNP events in a specific region by reweighting events passing the same selection except for inverted lepton identification and/or isolation requirements.
The reweighting factors, called `fake factors' (FFs), are measured separately for electrons and muons from data in FNP-enriched CRs (\CRfakeel and \CRfakemu) as functions of the lepton \pt and $|\eta|$.
The CRs listed in \cref{tab:Wh_CRVRFake} are designed to be as close as possible to \SRhmttWh and \SRlmttWh in order to share the same sources of FNP contributions as the target SRs. The measured FFs are around $0.1$ for both electrons and muons in most bins, but reach $0.3$ for some $\pt$ and $|\eta|$ bins.
The uncertainties of this method come principally from the measurement of the FFs which are propagated to the final estimate via the reweighting. In this analysis, the FF uncertainties coming from statistics, possible FNP contribution differences between the CRs and the targeted SRs, and prompt-lepton and CF background subtraction, amount to around 20\% of the final estimate in total.
Two validation regions \VRfakeWh and \VRCFWh, listed in \cref{tab:Wh_CRVRFake}, are defined in order to validate the data-driven methods applied to estimate the FNP and CF events in \SRhmttWh and \SRlmttWh. Good agreement between data and expectation is observed in the VRs, as shown in \cref{fig:Wh_CRVR}, thus validating the application of the above methods.

\begin{table}[!tb]
\centering
\caption{Definitions of the FNP-enriched control regions used to measure the FFs, and definitions of the validation regions used to  validate the estimates of the FNP and CF events in SRs defined for the \Wh model. Requirements guaranteeing orthogonality to SRs are in boldface.}
\begin{tabular}{c|cc|ccc|c}
\hline \hline
& & & \multicolumn{3}{c|}{} & \\ [1pt]
& \CRfakeel & \CRfakemu & \multicolumn{3}{c|}{\VRfakeWh} & \VRCFWh \\
& & & \multicolumn{3}{c|}{} & \\ [1pt]
& $e^{\pm}e^{\pm}$ & $\mu^{\pm}\mu^{\pm}$ & $e^{\pm}e^{\pm}$ & $e^{\pm}\mu^{\pm}$ & $\mu^{\pm}\mu^{\pm}$ & $e^{\pm}e^{\pm}$ \\
\hline \hline
$N_{\mathrm{BL}}(\ell)$ & \multicolumn{6}{c}{= 2} \\
Charge$(\ell)$ & \multicolumn{6}{c}{same-sign} \\
\hline
$N_{\mathrm{Sig}}(\ell)$ & \multicolumn{2}{c|}{= 1} & \multicolumn{4}{c}{= 2} \\
\hline
$\pt(\ell)$ & \multicolumn{6}{c}{$\geq \SI{25}{GeV}$} \\
\nJ & \multicolumn{6}{c}{$\geq 1$} \\
\hline
\nb & -- & = 1 & \multicolumn{4}{c}{= 0} \\
\met & $\mathbf{\in [30,50)}$ \textbf{GeV} & $\mathbf{<50~GeV}$ & \multicolumn{4}{c}{$\geq \SI{50}{GeV}$} \\
\hline
$|m_{\ell^{\pm}\ell^{\pm}} - m_{Z}|$ & $\geq \SI{15}{GeV}$ & -- & $\geq\SI{15}{GeV}$ & -- & -- & $<\SI{15}{GeV}$ \\
\hline
\mjj & \multicolumn{2}{c|}{--} & \multicolumn{4}{c}{$<\SI{350}{GeV}$} \\
\mttwo & \multicolumn{2}{c|}{--} & \multicolumn{4}{c}{$<\SI{80}{GeV}$} \\
\minmt & \multicolumn{2}{c|}{--} & \multicolumn{4}{c}{$<\SI{100}{GeV}$} \\
$\mathcal{S}(\met)$ & \multicolumn{2}{c|}{--} & \multicolumn{4}{c}{$\mathbf{<5}$} \\
\hline \hline
\end{tabular}
\label{tab:Wh_CRVRFake}
\end{table}


The matrix method involves the inversion of the matrix relating the numbers of observed baseline and signal leptons to the estimated numbers of real and FNP leptons via measured real ($\varepsilon$) and FNP ($\zeta$) lepton efficiencies; the implementation used in Ref.~\cite{SUSY-2018-09} is applied here.
The value of $\varepsilon$ is around 50\%--60\% (70\%) for electrons (muons) in the region of lepton $\pt$ around \SI{15}{GeV}, increasing to 98\% (99\%) for lepton $\pt>$ \SI{100}{GeV} (\SI{60}{GeV}).
The total uncertainty in $\varepsilon$ is 0.33\%--7\% (0.1\%--3\%) for electrons (muons) depending on the $(\pt, \eta)$ region.
The $\zeta$ probabilities are $\sim$10\%--20\% for both electrons and muons up to $\pt\sim\SI{45}{GeV}$, and increase to 30\%--40\% for $\pt>\SI{60}{GeV}$. They can be up to twice as large in events with two \btagged jets.
The effects of variations in the relative contributions of different sources of FNP leptons or in the overall event activity are considered as uncertainties of $\zeta$. For electrons (muons), the latter is 30\%--50\% (30\%--80\%), increasing with \pt.
The level of agreement between the data and the estimated background in a loose event preselection region requiring two SS leptons, $\met>\SI{50}{GeV}$ and at least one jet with $\pt>\SI{25}{GeV}$, in different lepton-flavour and \bjet-multiplicity combinations, as shown in~\cref{fig:MxM_vali}, indicates the universality of the matrix method in estimating the FNP lepton background in general cases.
Together with the level of agreement observed seen in~\cref{fig:CRVR}, this validates the estimation of the FNP background using the matrix method.
 
\begin{figure}[!tb]
\centering
\includegraphics[width=0.9\textwidth]{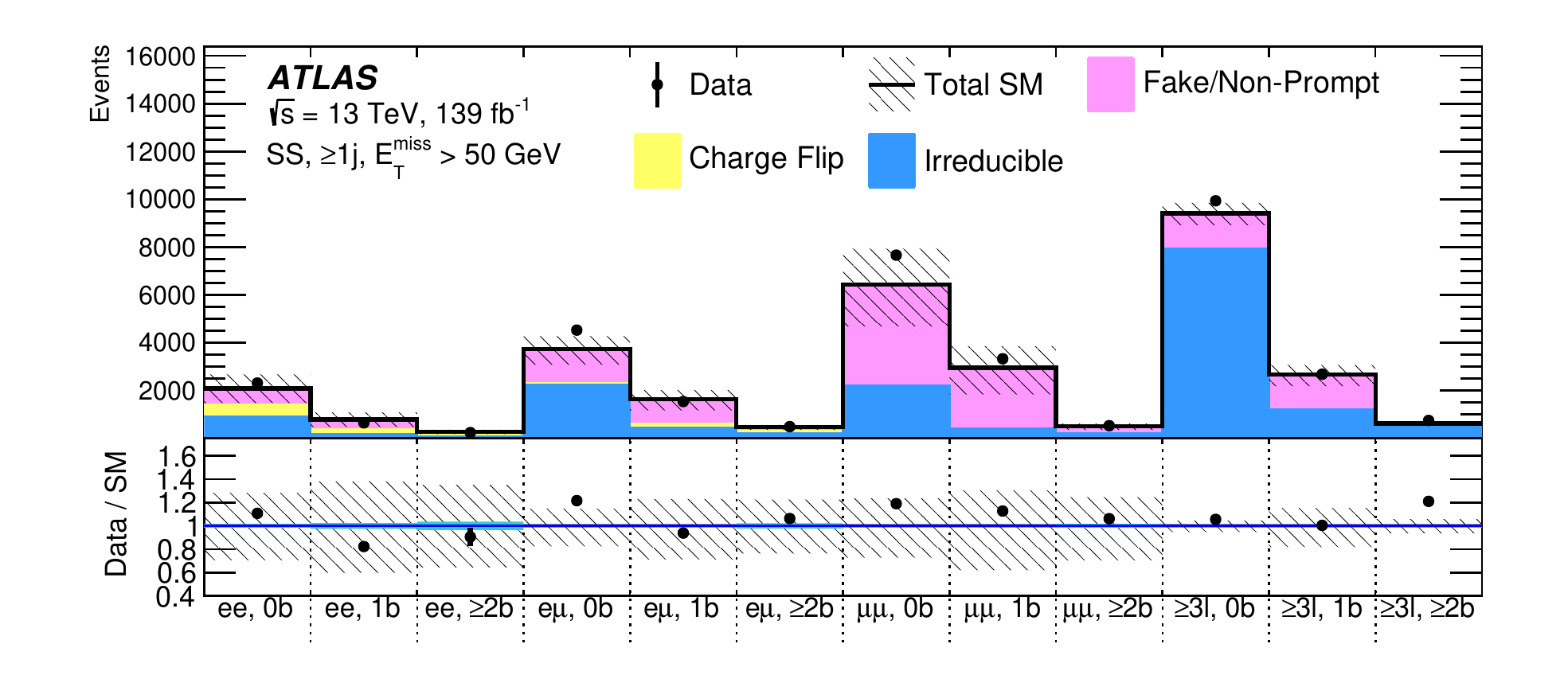}
\caption{Data event yields compared with the expected contributions from the irreducible and the reducible backgrounds after a loose preselection requiring SS leptons, $\met>\SI{50}{GeV}$ and at least one jet with $\pt>\SI{25}{GeV}$. The observed and predicted event yields are classified as a function of the number of leptons and their flavour, as well as the number of \bjets. The error bars only include the statistical uncertainty and the full uncertainties for the data-driven background estimates, in order to validate the matrix method itself. The bottom panel shows the ratio of the observed data to the predicted yields.}
\label{fig:MxM_vali}
\end{figure}
 
To further validate the estimation of the FNP and CF backgrounds, the MC template method is introduced to this analysis.
It relies on data-corrected CRs enriched in various sources of fake leptons and electron CF backgrounds to extrapolate the background predictions to the SRs. In this analysis, the scale factors are obtained for seven templates representing different types of backgrounds from six control regions, using discriminating variables such as \meff\ and lepton \pt. The uncertainties due to limited statistical precision and from the effects of the used discriminating variables are considered. The similarity of the \mttwo distributions obtained using the matrix method and the MC template method for SRs defined for the \WZ and bRPV models confirms the validity of the background estimation.


\FloatBarrier
\section{Systematic uncertainties}
\label{sec:systs}

\graphicspath{{figures/}}
 
Several sources of systematic uncertainty, besides the various statistical uncertainties, are considered in this analysis.
They are grouped into experimental uncertainties, theoretical uncertainties, uncertainties from the data-driven methods applied in this analysis, and normalisation and MC statistical uncertainties.
 
The experimental uncertainties encompass all possible differences between data and simulations in all analysis elements including the trigger, pile-up, and reconstructed objects.
A 1.7\% relative uncertainty in the luminosity~\cite{ATLAS-CONF-2019-021} is applied.
For leptons, uncertainties in the reconstruction efficiencies~\cite{EGAM-2018-01}, identification efficiencies~\cite{MUON-2018-03}, isolation efficiencies, energy scales~\cite{EGAM-2018-01} and resolutions, and trigger efficiencies are considered.
For jets, uncertainties in the jet vertex tagger~\cite{ATL-PHYS-PUB-2019-026} performance which affect the residual contamination from pile-up jets, uncertainties in the jet energy scale~\cite{PERF-2016-04} and jet energy resolution~\cite{JETM-2018-05}, and uncertainties in flavour tagging~\cite{FTAG-2018-01,FTAG-2020-08,ATLAS-CONF-2018-006}  are also considered.
The uncertainties associated with the objects used to compute the \met are propagated through the computation, and additional uncertainties in the scale and resolution of the contribution from low-momentum tracks not associated with the primary objects are also included~\cite{ATLAS-CONF-2018-023}.
These experimental uncertainties are correlated between the processes and regions that enter the simultaneous fit, including the signal models.
 
The theoretical uncertainties account for the uncertainties in modelling of the relevant SM and SUSY processes, including uncertainties in cross sections and due to the choice of scales, the PDF and the value of $\alphas$. Modelling uncertainties for backgrounds making subdominant contributions in the SRs are neglected except for cross-section uncertainties. If the background process is normalised to data, the associated uncertainties are applied instead of the total cross-section uncertainty. The uncertainties that affect the acceptance, such as the choice of scales and the PDF, are applied everywhere. The theoretical uncertainties vary from 10\% to 50\% in all regions defined in this analysis.
 
\begin{figure}[!tb]
\centering
\includegraphics[width=0.75\textwidth]{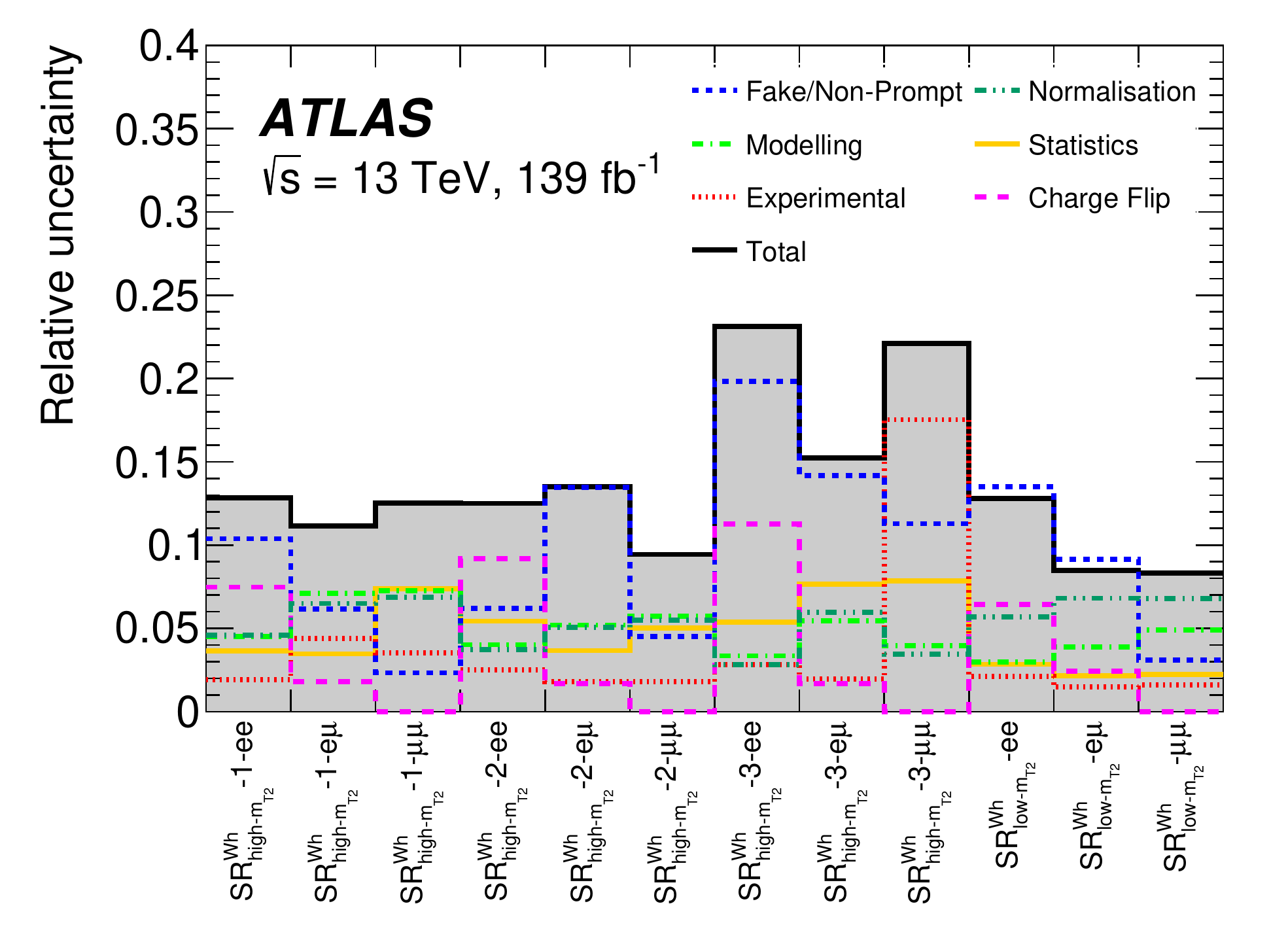}
\caption{Breakdown of the total systematic uncertainty in the background prediction for each of the SRs of the \Wh model. Total and individual uncertainties for different source categories are shown. The individual components can be correlated and therefore do not necessarily add up in quadrature to the total systematic uncertainty.}
\label{fig:Sys_SR_Wh}
\end{figure}
 
\begin{figure}[!tb]
\centering
\includegraphics[width=0.75\textwidth]{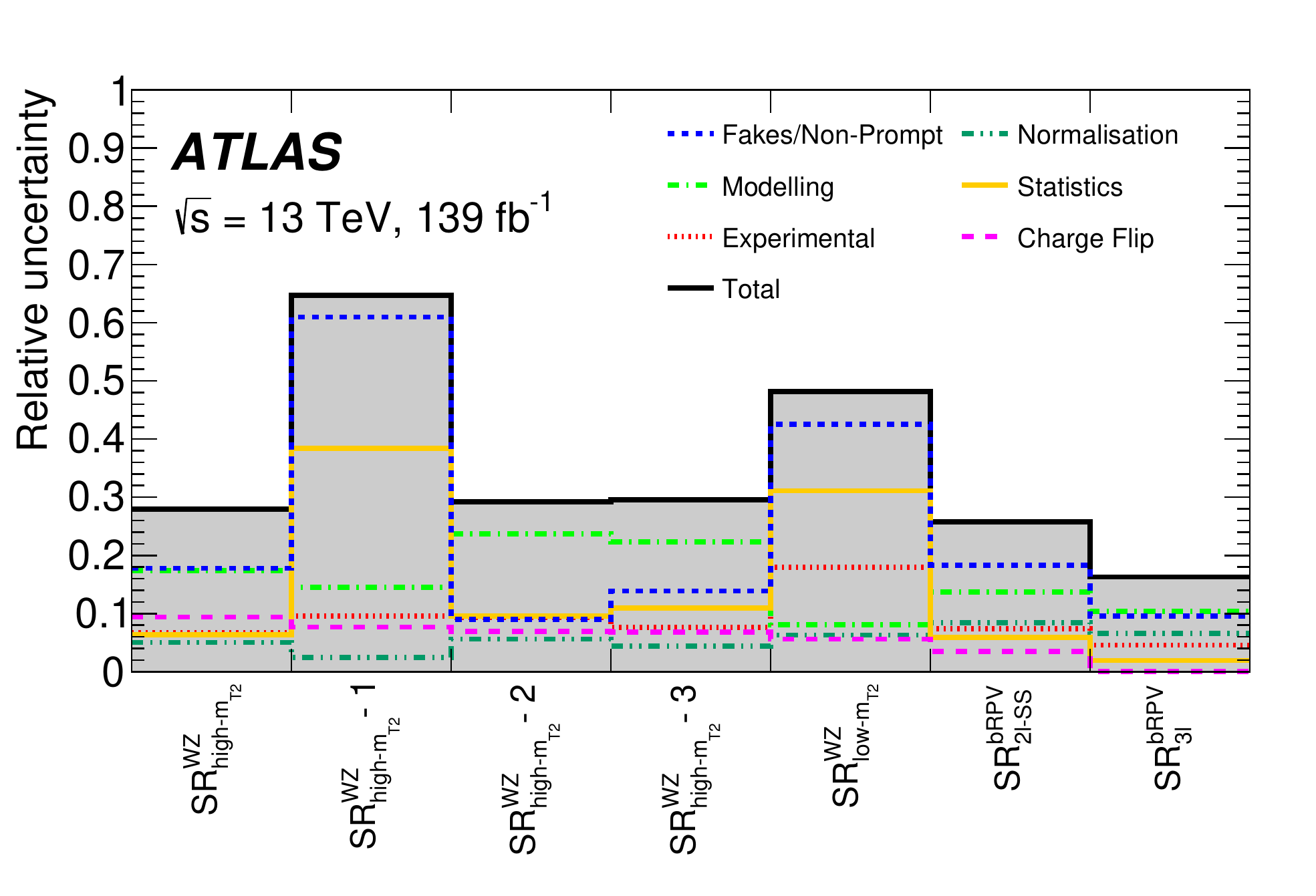}
\caption{Breakdown of the total systematic uncertainty in the background prediction for the each of the SRs of the \WZ and bRPV models. Total and individual uncertainties for different source categories are shown. The individual components can be correlated and therefore do not necessarily add up in quadrature to the total systematic uncertainty.}
\label{fig:Sys_SR_WZ_bRPV}
\end{figure}
 
The total uncertainty and the contributions from different sources are shown in
\cref{fig:Sys_SR_Wh,fig:Sys_SR_WZ_bRPV} for all the signal regions.
For regions designed for the \Wh model, the total uncertainties vary from 8\% to 25\%.
In some \Wh-model-specific SRs, the total uncertainty is less than the largest uncertainty contribution because of the large anti-correlation between the FNP-related uncertainties and the normalisation-related uncertainties.
The largest contribution comes from the estimation of the FNP background.
 
For SRs designed for the \WZ and bRPV models, total uncertainties vary from 30\% to 50\%, with the uncertainties from the estimations of FNP and CF events accounting for the largest contribution in most of the regions. They are larger than those of the \Wh model, as the correlations between the relevant systematic uncertainties were not constrained by the CRs during the fit.


\FloatBarrier
\section{Results}
\label{sec:result}

\graphicspath{{figures/}{figures/results/}}

The \met\ and $\mathcal{S}(\met)$ distributions for all events passing the \Wh\ SR requirements, except for the \met\ and $\mathcal{S}(\met)$ requirements themselves, are shown in \cref{fig:MET_unblindedSR1} and \cref{fig:METsig_unblindedSR2}, respectively. Data are compared with the expected SM background; each source is estimated as described in \cref{sec:background}. Separate distributions are provided for each SS-dilepton flavour: $e^{\pm}e^{\pm}$, $e^{\pm}\mu^{\pm}$ and $\mu^{\pm}\mu^{\pm}$. Fake and non-prompt leptons as well as the \WZ irreducible background dominate the events mimicking signal events, while the CF events are an important source of background in the $e^{\pm}e^{\pm}$ SRs, as observed in \cref{fig:Wh-ee-met} and \cref{fig:Wh-ee-sig}. The expected distributions for three representative signal mass points are also overlaid as indicated. Good agreement between the data and total expected SM background is observed.
 
\begin{figure}[!tb]
\centering
\subfloat[]{\includegraphics[width=0.48\textwidth]{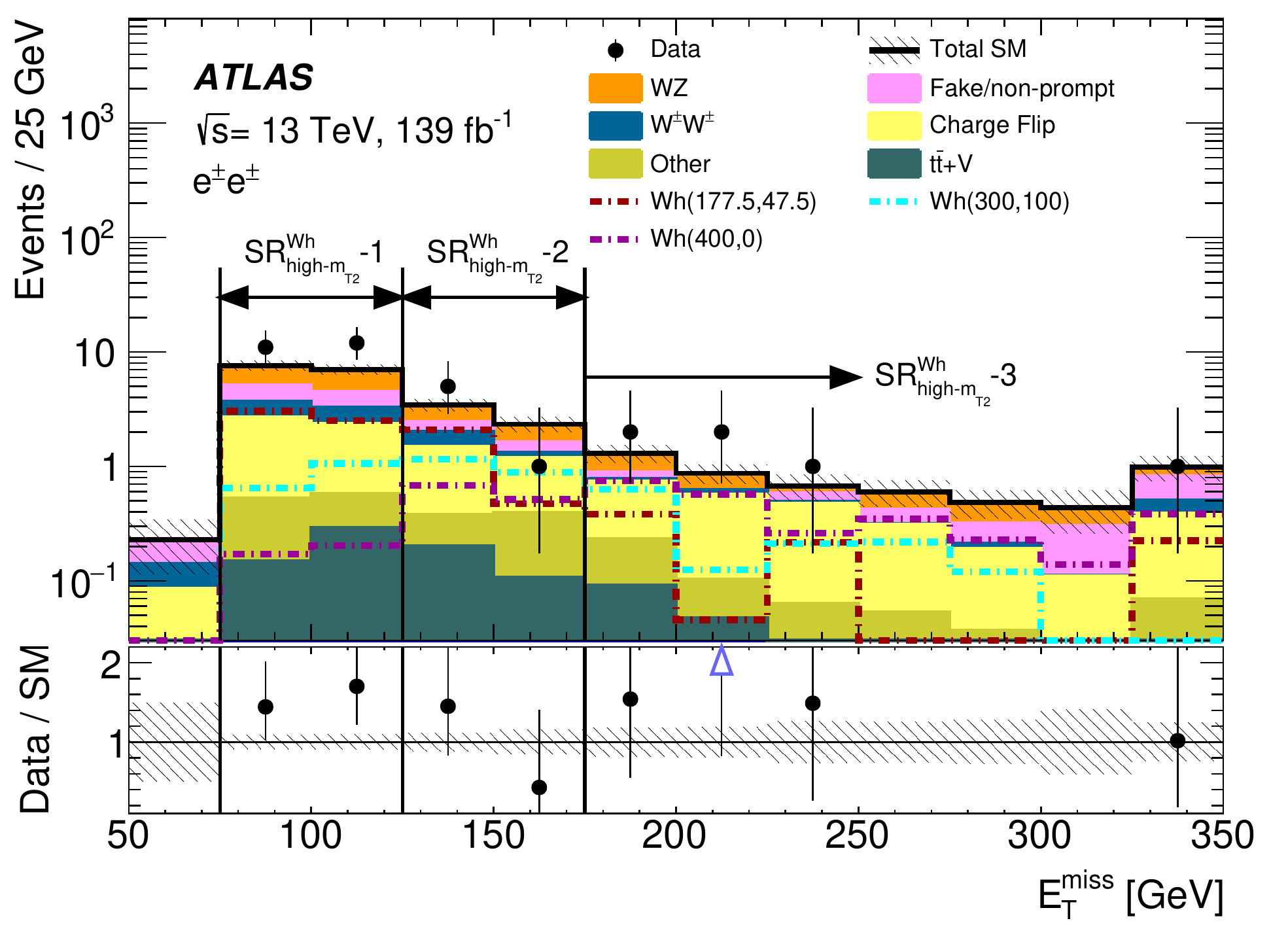}\label{fig:Wh-ee-met}}
\quad
\subfloat[]{\includegraphics[width=0.48\textwidth]{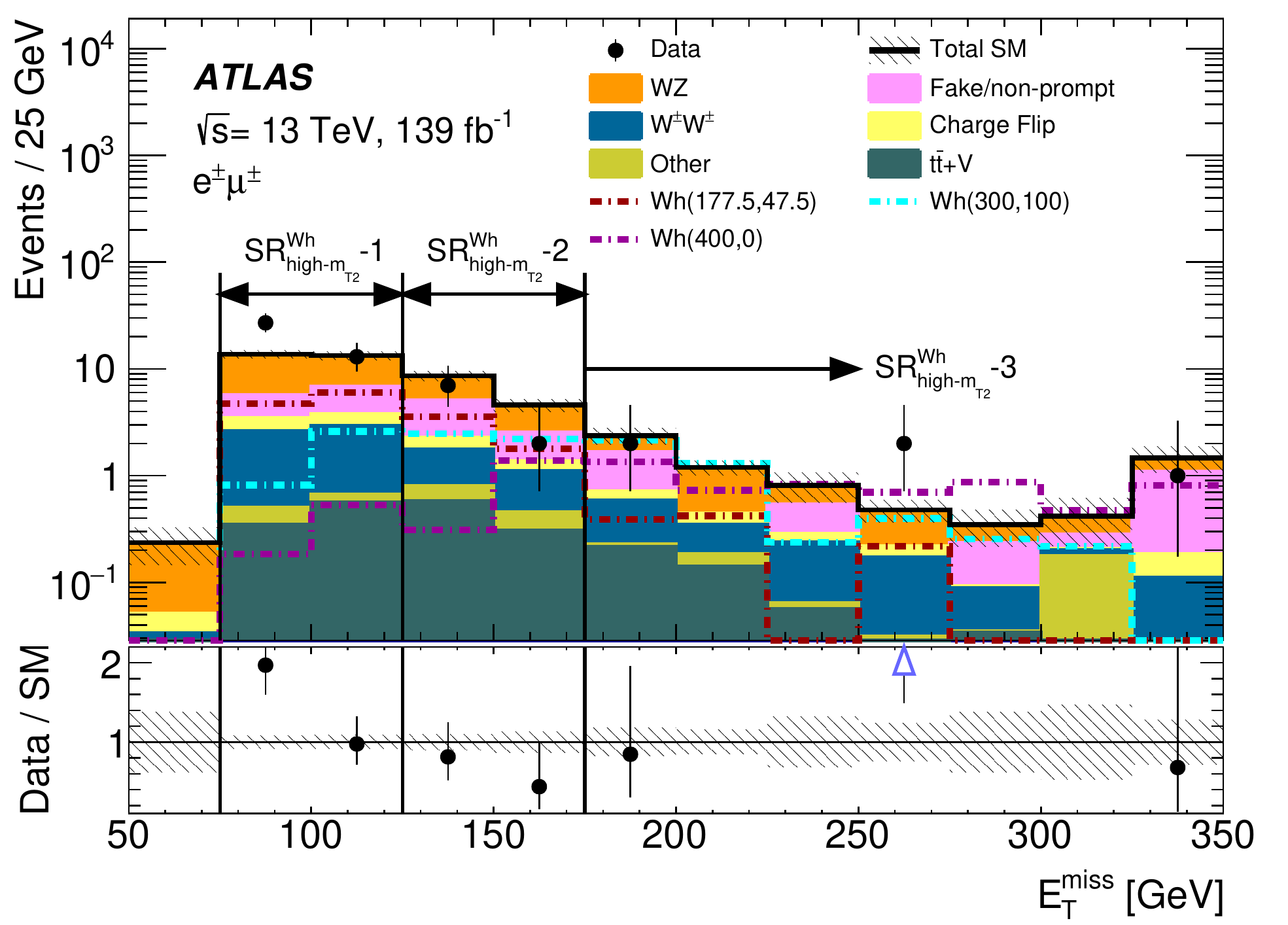}\label{fig:Wh-emu-met}}
\newline
\subfloat[]{\includegraphics[width=0.48\textwidth]{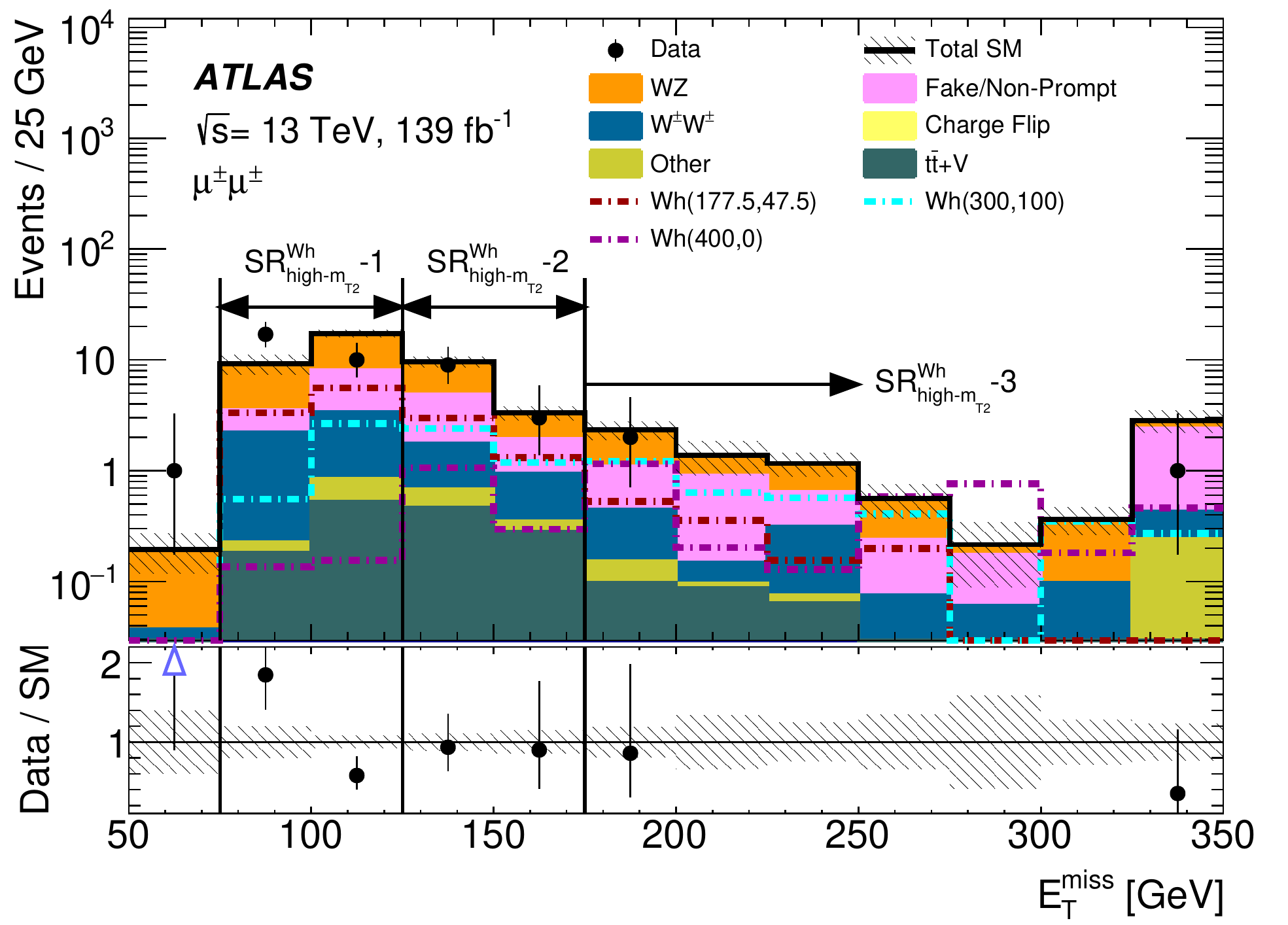}\label{fig:Wh-mumu-met}}
\caption{\met distributions after the background-only fit, showing the data and the post-fit expected background in all the flavour and \met bins of the \SRhmttWh region. The vertical black lines and the corresponding arrows indicate the cuts defining the three \met bins of the \SRhmttWh region: \SRhmttWh-1, \SRhmttWh-2, and \SRhmttWh-3.  The last bin includes overflow. The `Other' category contains the \ttHp, rare top, triboson, and other diboson processes with the SS final state. Distributions for three representative signal mass points of the \Wh model are overlaid. The bottom panel shows the ratio of the observed data to the predicted yields. The hatched bands indicate the combined theoretical, experimental, data-driven and MC statistical uncertainties.}\label{fig:MET_unblindedSR1}
\end{figure}
 
\begin{figure}[!tb]
\centering
\subfloat[]{\includegraphics[width=0.48\textwidth]{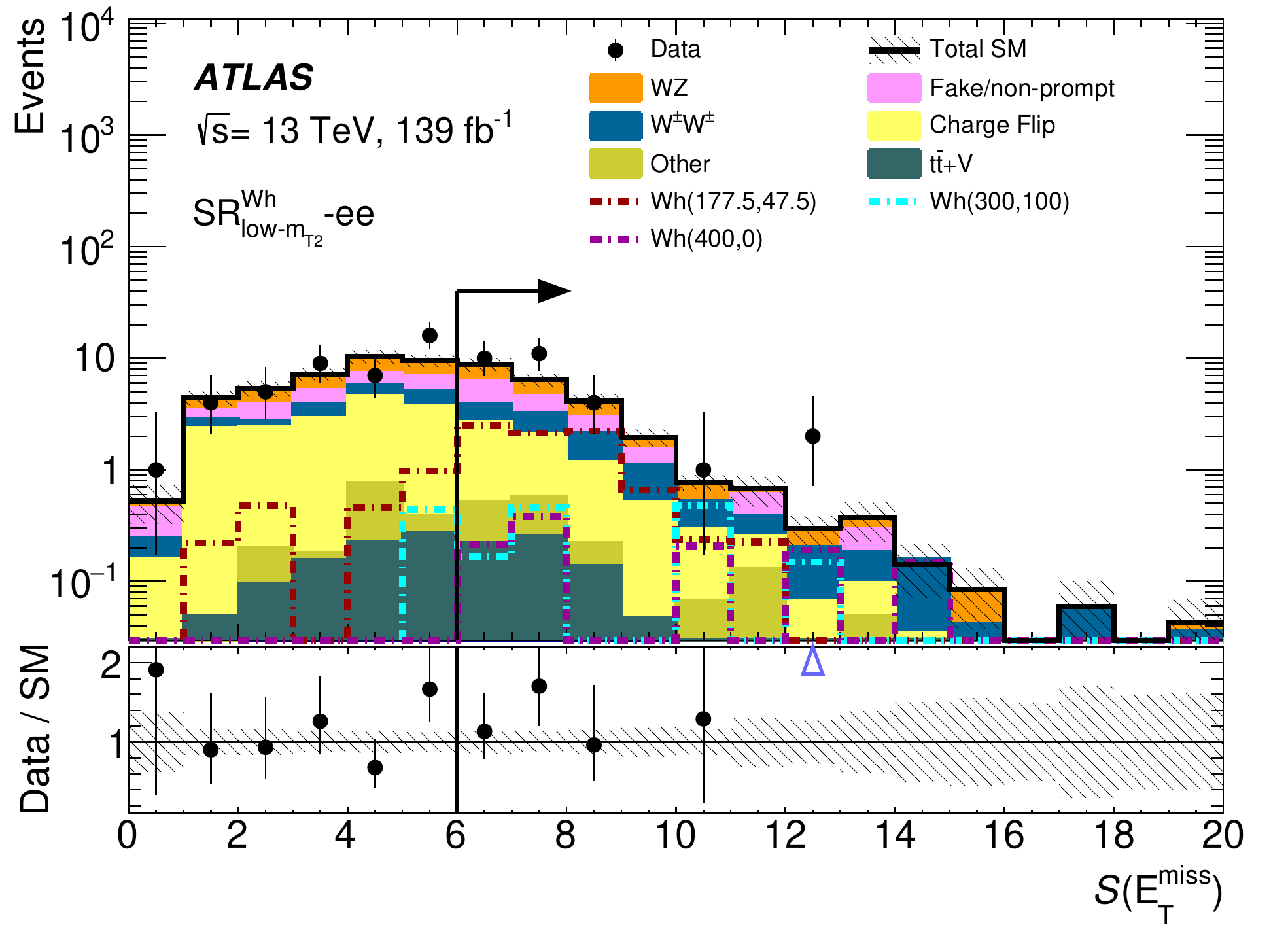}\label{fig:Wh-ee-sig}}
\quad
\subfloat[]{\includegraphics[width=0.48\textwidth]{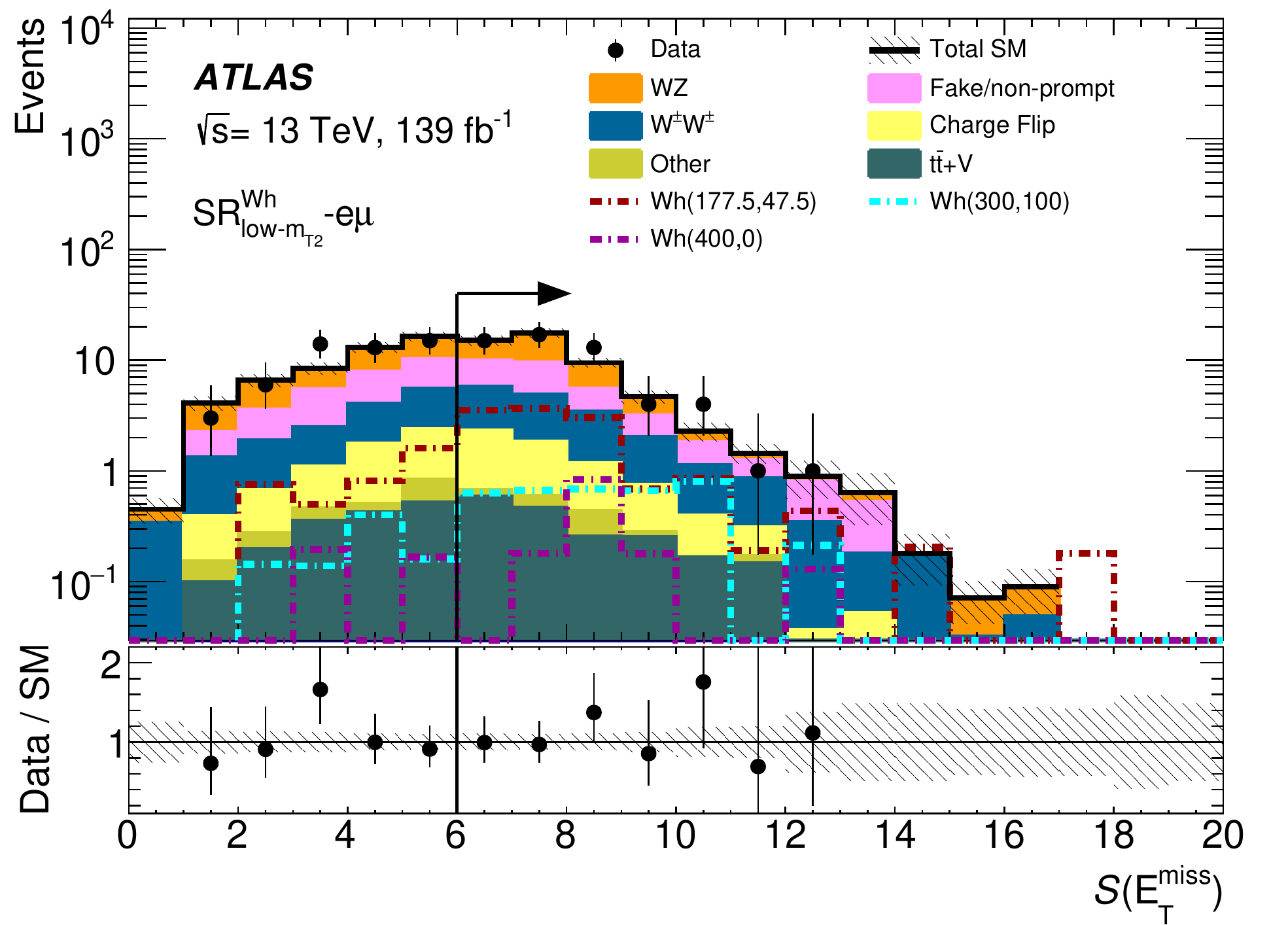}\label{fig:Wh-emu-sig}}
\newline
\subfloat[]{\includegraphics[width=0.48\textwidth]{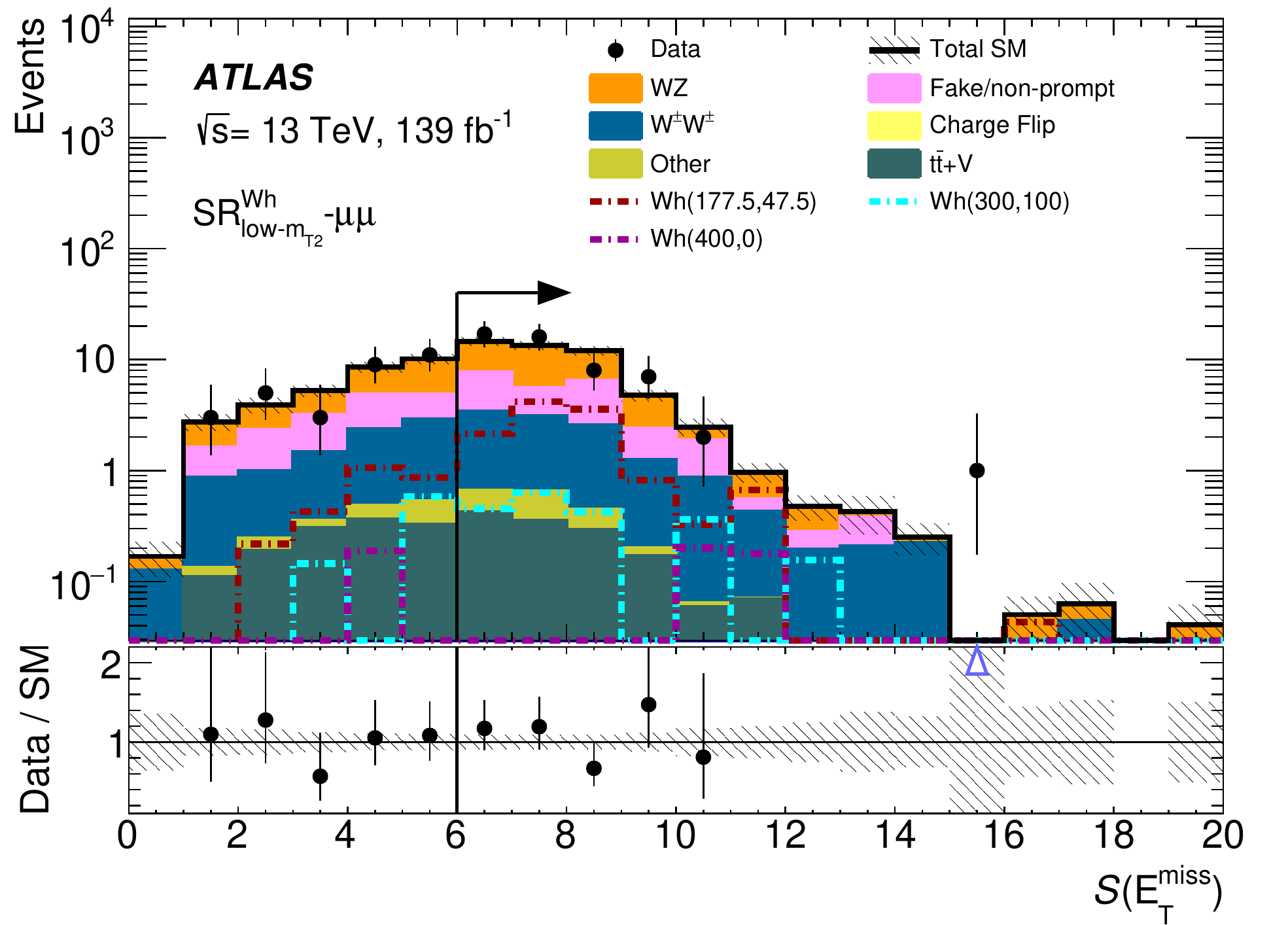}\label{fig:Wh-mumu-sig}}
\caption{$\mathcal{S}(\met)$ distributions after the background-only fit, showing the data and the post-fit expected background in all the flavour bins of the \SRlmttWh region. The vertical black line and the corresponding arrow indicates the cut defining the \SRlmttWh region. The last bin includes overflow. The `Other' category contains the \ttHp, rare top, triboson, and other diboson processes with the SS final state. Distributions for three representative signal mass points of the \Wh model are overlaid. The bottom panel shows the ratio of the observed data to the predicted yields. The hatched bands indicate the combined theoretical, experimental, data-driven and MC statistical uncertainties.}\label{fig:METsig_unblindedSR2}
\end{figure}
 
The observed number of events in each SR defined in \cref{sec:selection} for the \Wh model along with the background expectations and uncertainties are reported in \cref{fig:WhSS_pull_SR}. The observed data are compatible with the SM prediction, with a $-2.0\sigma$ data deficit observed in \SRhmttWh-3-$\mu\mu$. The largest excess of events is observed in \SRhmttWh-1-$ee$, with a significance of less than $2.0\sigma$.
 
\begin{figure}[!tb]
\centering
\includegraphics[width=0.8\textwidth]{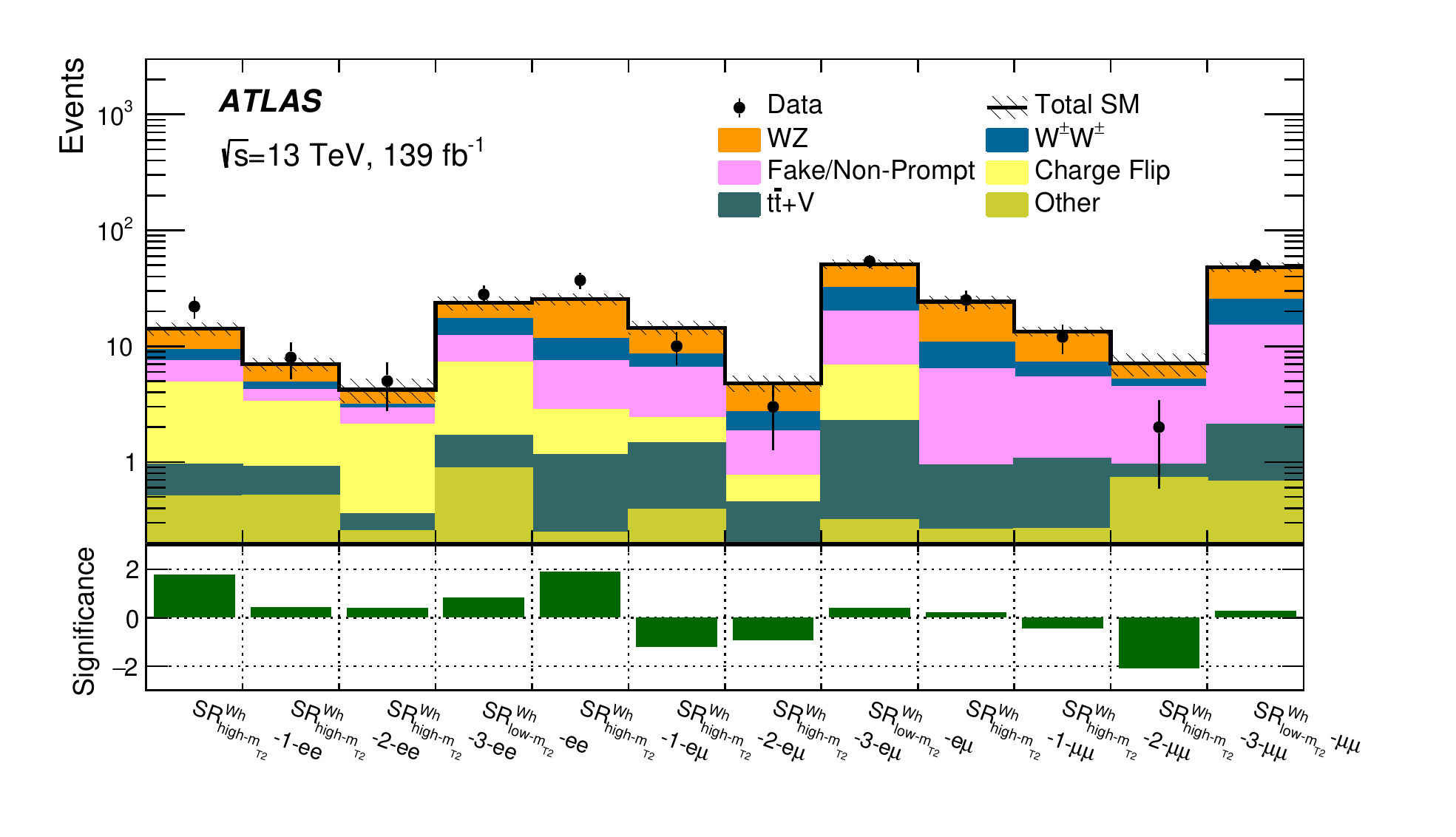}
\vspace{-1em}
\caption{Expected SM background and data yields in the SRs optimised for the \Wh\ model. The total uncertainties in the expected event yields are shown as the hashed bands. The SM prediction is taken from the background-only fit. The `Other' category contains the \ttHp, rare top, triboson, and other diboson processes with the SS final state. The bottom panel shows the statistical significance~\cite{Cousins:2007bmb} of the discrepancy between the observed number of events and the SM expectation.}
\label{fig:WhSS_pull_SR}
\end{figure}

The distributions of \mttwo in the SRs defined for the \WZ\ model and the bRPV models after applying all selection criteria apart from the \mttwo cut are shown in \crefrange{fig:WZ-high-mttwo}{fig:WZ-low-mttwo} and in  \crefrange{fig:bRPV-2lep-mttwo}{fig:bRPV-3lep-mttwo}, respectively. All considered sources of background are also plotted, estimated with the data-driven techniques detailed in \cref{sec:background}. The background is dominated by the SM \WZ process and the reducible background due to fake and non-prompt leptons. For comparison, representative signal mass points for winos/binos and higgsinos ($\tilde{H}$) are overlaid. The data distributions are in agreement with the background expectations.
 
\begin{figure}[!tb]
\centering
\subfloat[]{\includegraphics[width=0.45\textwidth]{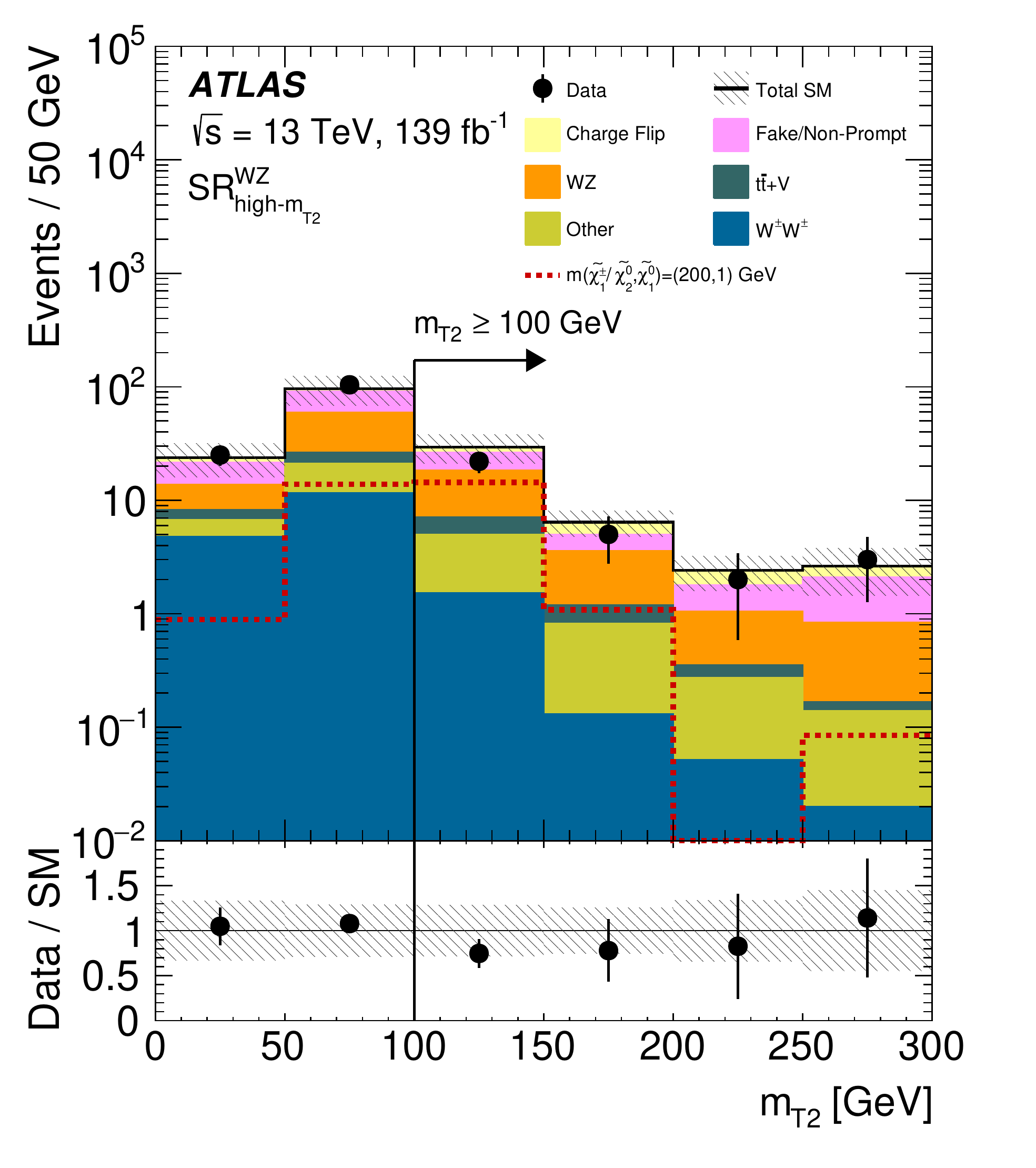}\label{fig:WZ-high-mttwo}}
\quad
\subfloat[]{\includegraphics[width=0.45\textwidth]{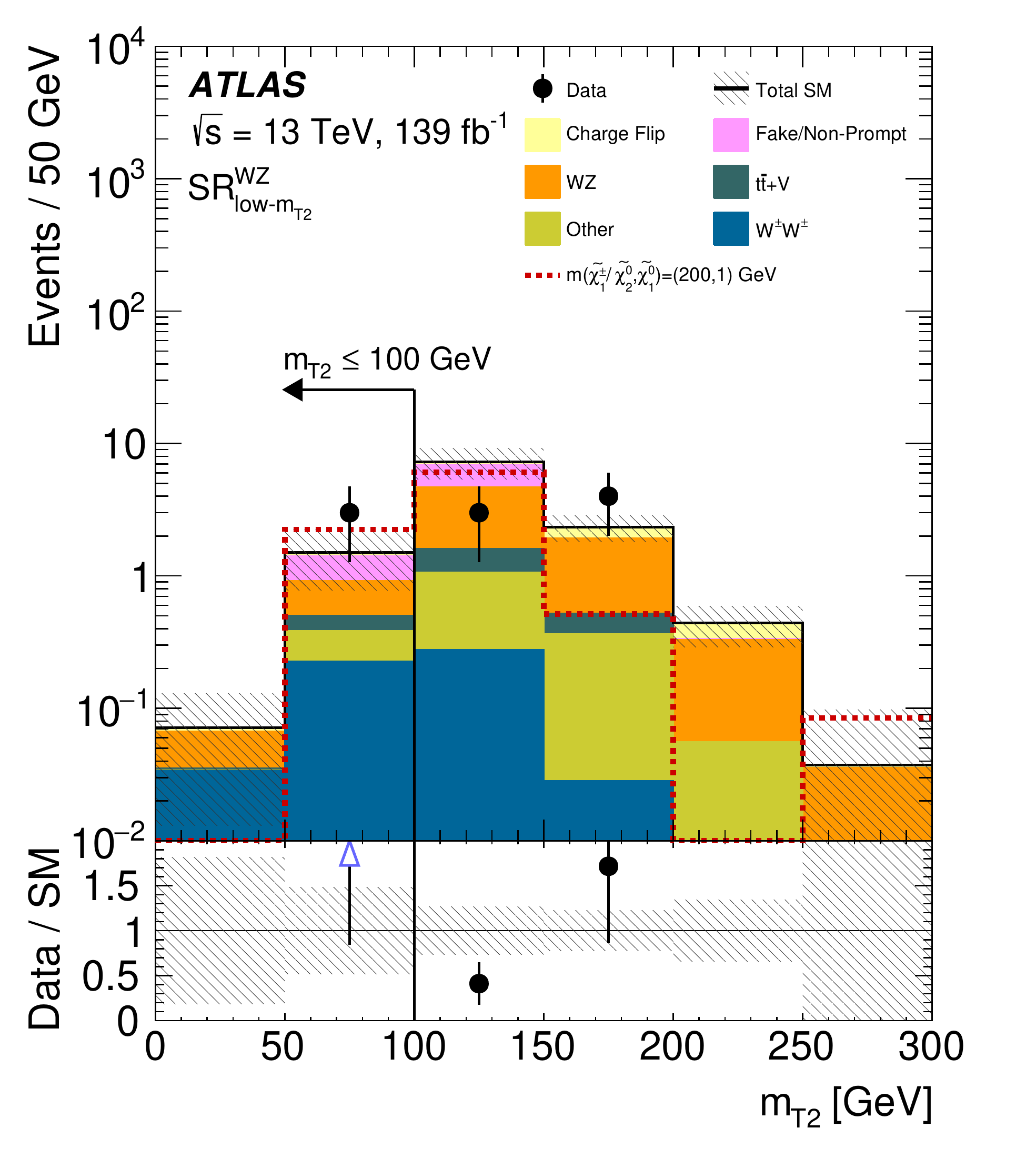}\label{fig:WZ-low-mttwo}}
\newline
\subfloat[]{\includegraphics[width=0.45\textwidth]{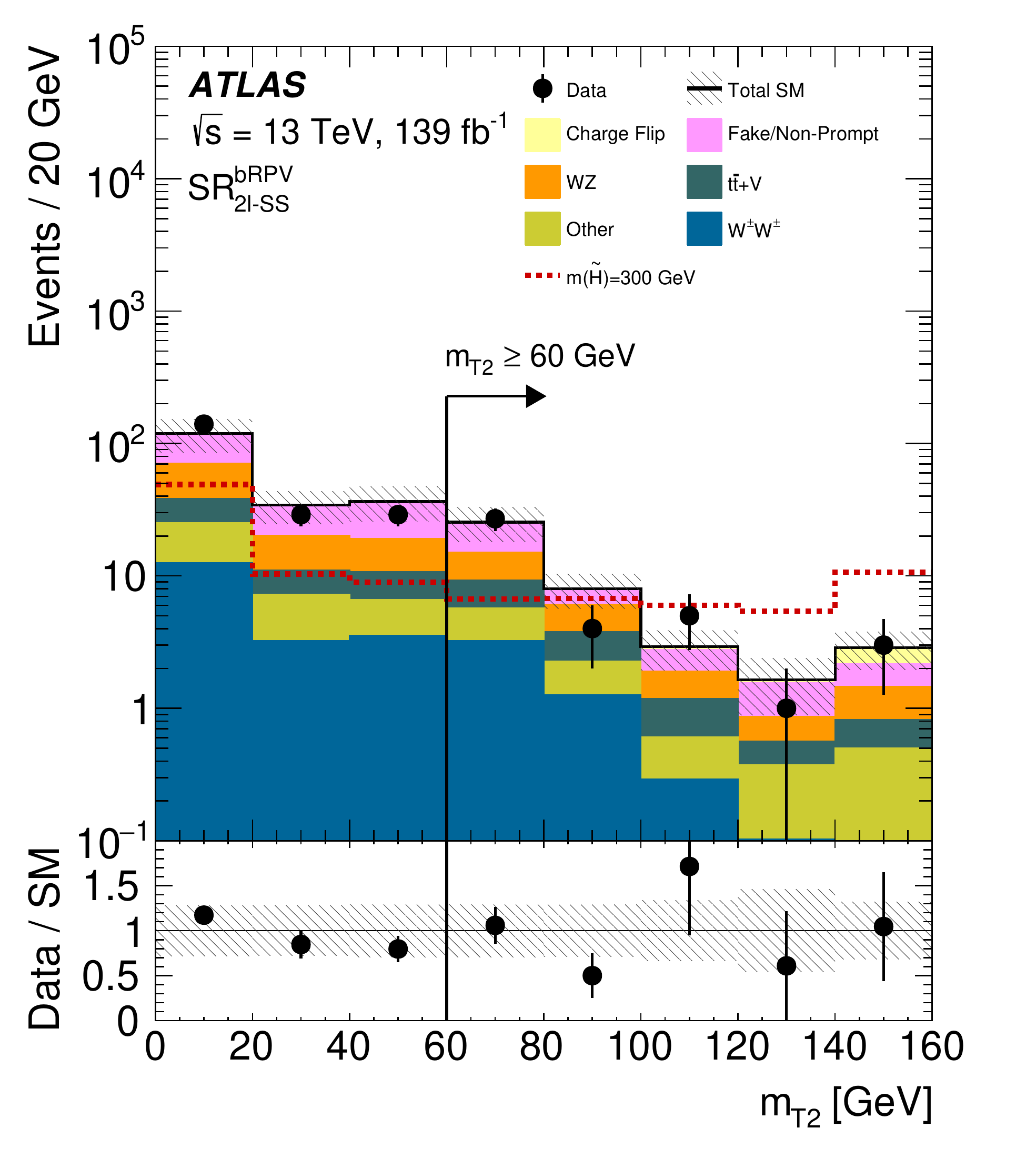}\label{fig:bRPV-2lep-mttwo}}
\quad
\subfloat[]{\includegraphics[width=0.45\textwidth]{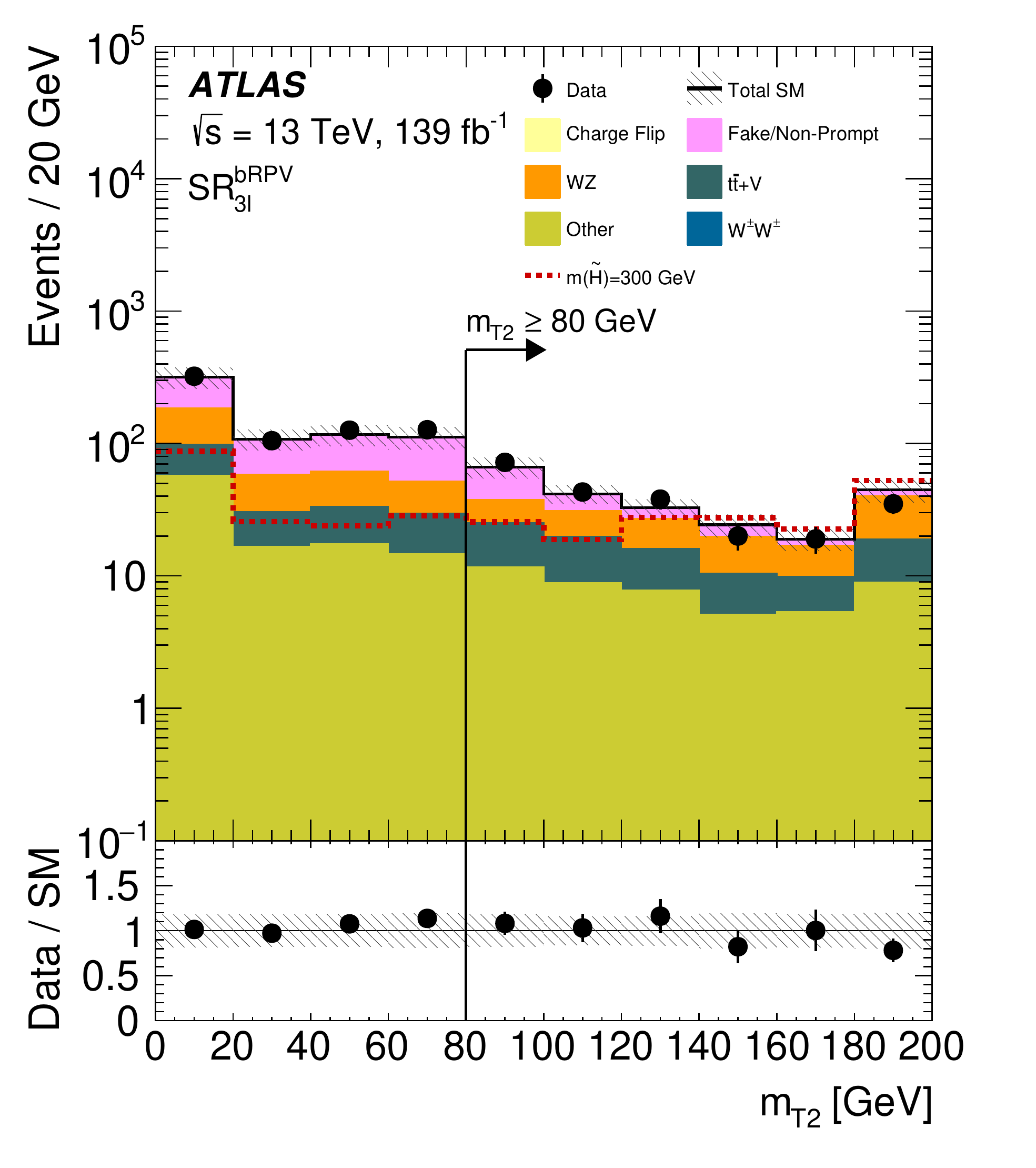}\label{fig:bRPV-3lep-mttwo}}
\caption{\mttwo distributions in SRs defined for the \WZ model (\protect\subref{fig:WZ-high-mttwo} and \protect\subref{fig:WZ-low-mttwo}) and the bRPV model (\protect\subref{fig:bRPV-2lep-mttwo} and \protect\subref{fig:bRPV-3lep-mttwo}). All SR selection criteria are satisfied except for that on \mttwo. The vertical black lines and the corresponding arrows indicate the cuts defining those regions. The matrix method is used for background estimation and the CF events are estimated via a data-driven method. The `Other' category contains the \ttHp, rare top, triboson, and other diboson processes with the SS final state. Uncertainties from theoretical, experimental, data-driven and MC statistical sources are all considered. The last bin includes overflow. Distributions for representative signal mass points are overlaid. The bottom panel shows the ratio of the observed data to the predicted yields.}\label{fig:Np1_MxM}
\end{figure}
 
In \cref{fig:SRyields}, the observed yields in each signal region defined in \cref{sec:selection} along with the background expectations and uncertainties are presented for the \WZ and bRPV models. The observed data are compatible with the SM prediction.
 
\begin{figure}[!tb]
\centering
\includegraphics[width=0.8\textwidth]{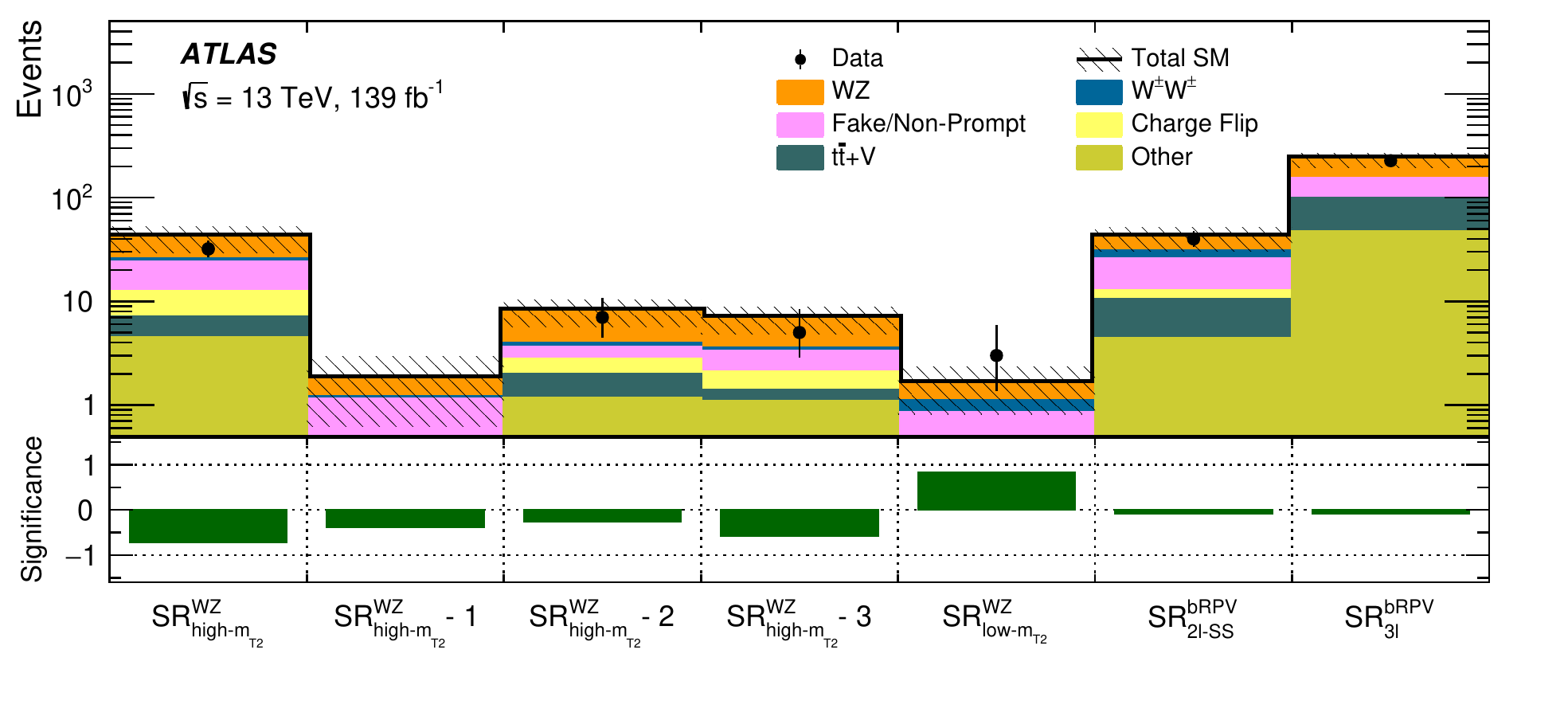}
\caption{Expected SM background and data yields in the SRs optimised for the \WZ\ and bRPV models. The SM prediction is taken from the background-only fit. The `Other' category contains the \ttHp, rare top, triboson, and other diboson processes with the SS final state. The total uncertainties in the expected event yields are shown as the hashed bands. The bottom panel shows the statistical significance~\cite{Cousins:2007bmb} of the discrepancy between the observed number of events and the SM expectation.}
\label{fig:SRyields}
\end{figure}
 
The results from the \textit{UDD} RPV model SRs are discussed in \cref{subsec:RPVUDD_results}. The data agree within  uncertainties with the SM expectation in the SRs designed for this model.

\FloatBarrier


\FloatBarrier
\section{Interpretation}
\label{sec:interpret}

\graphicspath{{figures/}{figures/interpretation/}}
 
Model-independent upper limits on the number of BSM events in each SR are derived using the \cls prescription~\cite{Junk:1999kv,Read:2002hq} and neglecting any possible signal contamination in the control regions. The \HISTFITTER~\cite{Baak:2014wma} framework is used for the statistical interpretation of the results. In order to quantify the probability for the background-only hypothesis to fluctuate to at least the observed number of events, a one-sided $p_0$-value is calculated using pseudo-experiments, where the profile likelihood ratio is used as a test statistic~\cite{Cowan:2010js} to exclude the signal-plus-background hypothesis. Normalisation to the integrated luminosity of the data sample allows an interpretation in terms of upper limits on the visible BSM cross section, defined as the product of the acceptance, reconstruction efficiency and production cross section.
 
The number of observed events and the background expectation in each SR are used to set an upper limit on the number of events from any BSM physics scenario without a sizable positive or negative interference in the CRs. The model-independent upper limits at 95\% confidence level (CL) on the visible cross section, \sigmavis,  for the \Wh, \WZ and bRPV signal regions are presented in \cref{table.results.exclxsec.pval.upperlimit}. Also listed are the 95\% CL upper limits on the number of signal events \sobs, as well as the expected 95\% CL upper limit on the number of signal events, \sexp. The last two columns indicate the \clb value and the discovery $p$-value, $p_0$ $(p(s)=0)$, with the corresponding Gaussian significance $Z$. The \clb value provides a measure of compatibility of the observed data with the 95\% CL signal strength hypothesis relative to fluctuations of the background, and  $p_0$ measures the compatibility of the observed data with the background-only (zero signal strength) hypothesis relative to fluctuations of the background. Larger values indicate greater relative compatibility.
 
\begin{table}[!b]
\caption{Model-independent statistical analysis for SRs (binned and inclusively) optimised for the \Wh, \WZ and bRPV models: 95\% CL upper limits on the visible cross section, \sigmavis, and on the number of signal events \sobs. The \sexp is the expected 95\% CL upper limit on the number of signal events, given the expected number (and $\pm1\sigma$ variations) of background events. The last two columns report the \clb  value for the background-only hypothesis, the one-sided $p_0$-value and the local significance $Z$ (the number of equivalent Gaussian standard deviations).}
\label{table.results.exclxsec.pval.upperlimit}
\centering
\renewcommand{\arraystretch}{1.2}
\small
\begin{tabular}{lccccc}
\noalign{\smallskip}\hline\noalign{\smallskip}{\textbf{Signal region}}                        & \sigmavis\ [fb]  &  \sobs  &  \sexp & \clb  & $p_0$~$(Z)$ \\
\noalign{\smallskip}\hline\noalign{\smallskip}
\SRhmttWh    & $0.28$ &  $39.3$ & \phantom{$_{0}$}$ { 33.9 }^{ +14.3 }_{ -10.0 }$ & $0.66$ & $ 0.34$~$(0.41)$ \\%
\SRhmttWh-1-$ee$ & $0.13$ &  $17.4$ & $ \phantom{0}{ 9.9 }^{ +4.4 }_{ -2.8 }$	& $0.94$ & $ 0.04$~$(1.72)$ \\%
\SRhmttWh-1-$e\mu$ & $0.17$ &  $23.6$ & $ { 12.9 }^{ +5.6 }_{ -3.6 }$	& $0.96$ & $ 0.03$~$(1.85)$ \\%
\SRhmttWh-1-$\mu\mu$ & $0.09$ &  $13.0$ & $ { 12.6 }^{ +5.4 }_{ -3.6 }$	& $0.55$ & $ 0.45$~$(0.14)$ \\%
\SRhmttWh-2-$ee$ & $0.06$ &  \phantom{0}$7.8$  & \phantom{0}$ { 7.2 }^{ +3.1 }_{ -2.2 }$	& $0.63$ & $ 0.36$~$(0.36)$ \\%
\SRhmttWh-2-$e\mu$ & $0.05$ &  \phantom{0}$6.8$  & \phantom{0}$ { 9.5 }^{ +4.0 }_{ -2.7 }$	& $0.16$ & $ 0.50$~$(0.00)$ \\%
\SRhmttWh-2-$\mu\mu$ & $0.07$ &  \phantom{0}$9.6$  & \phantom{0}$ { 7.7 }^{ +0.6 }_{ -0.2 }$	& $0.64$ & $ 0.50$~$(0.00)$ \\%
\SRhmttWh-3-$ee$ & $0.05$ &  \phantom{0}$6.9$  & \phantom{0}$ { 6.1 }^{ +3.0 }_{ -1.6 }$	& $0.61$ & $ 0.37$~$(0.33)$ \\%
\SRhmttWh-3-$e\mu$ & $0.03$ &  \phantom{0}$4.8$  & \phantom{0}$ { 6.1 }^{ +3.0 }_{ -1.6 }$	& $0.24$ & $ 0.50$~$(0.00)$ \\%
\SRhmttWh-3-$\mu\mu$ & $0.03$ &  \phantom{0}$4.3$  & \phantom{0}$ { 6.9 }^{ +3.0 }_{ -2.0 }$	& $0.06$ & $ 0.50$~$(0.00)$ \\%
\SRlmttWh    & $0.24$ &  $33.0$ & \phantom{$^{+0}$}$ { 29.5 }^{ +11.7 }_{ -8.8 }$\phantom{$_{-}$} & $0.63$ & $ 0.33$~$(0.43)$ \\%
\SRlmttWh-$ee$ & $0.12$ &  $16.2$ & $ { 12.6 }^{ +5.4 }_{ -3.6 }$	& $0.76$ & $ 0.23$~$(0.76)$ \\%
\SRlmttWh-$e\mu$ & $0.14$ &  $19.9$ & $ { 17.6 }^{ +7.4 }_{ -5.1 }$	& $0.63$ & $ 0.36$~$(0.35)$ \\%
\SRlmttWh-$\mu\mu$ & $0.13$ &  $18.2$ & $ { 17.0 }^{ +7.0 }_{ -4.9 }$	& $0.59$ & $ 0.41$~$(0.22)$ \\%
\SRhmttWZ     & $0.13$ &  $18.7$ & $ { 24.4 }^{ +6.8 }_{ -5.0 }$ & $0.12$ & $ 0.50$~$(0.00)$ \\%
\SRhmttWZ-1   & $0.01$ &  \phantom{0}$1.7$ & \phantom{0}$ { 3.6 }^{ +1.3 }_{ -0.6 }$ & $0.02$ & $ 0.45$~$(0.12)$ \\%
\SRhmttWZ-2   & $0.05$ &  \phantom{0}$7.4$ & \phantom{0}$ { 8.3 }^{ +3.2 }_{ -2.2 }$ & $0.34$ & $ 0.50$~$(0.00)$ \\%
\SRhmttWZ-3   & $0.04$ &  \phantom{0}$5.2$ & \phantom{0}$ { 7.3 }^{ +2.7 }_{ -1.8 }$ & $0.11$ & $ 0.50$~$(0.00)$ \\%
\SRlmttWZ    & $0.04$ &  \phantom{0}$5.9$ & \phantom{0}$ { 4.4 }^{ +1.8 }_{ -0.8 }$ & $0.81$ & $ 0.22$~$(0.76)$ \\%
\SRtwolbrpv       & $0.16$ &  $22.6$ & $ { 25.8 }^{ +7.9 }_{ -5.8 }$ & $0.29$ & $ 0.50$~$(0.00)$ \\%
\SRthreelbrpv    & $0.44$ &  $61.4$ & \phantom{$_{0}$}$ { 93.0 }^{ +56.0 }_{ -20.3 }$ & $0.02$ & $ 0.50$~$(0.00)$ \\
\noalign{\smallskip}\hline\noalign{\smallskip}
\end{tabular}
\end{table}
 
For \SRhmttWZ, \SRtwolbrpv and \SRthreelbrpv, $p_0$ is capped at $0.5$ since the predictions exceed the data. In all other SRs the significances are low, with no appreciable excess observed over the expected background.  The most stringent observed limit is from \SRlmttWZ, where visible cross sections larger than \SI{0.04}{fb} are excluded. Model-independent limits are also provided in \cref{app:RPVUDD} (\cref{table.results.exclxsec.pval.upperlimit.nonWhSRs}) for the \textit{UDD} RPV SRs.
 
\FloatBarrier
 
Model-dependent exclusion limits were extracted by performing hypothesis tests on the background-only hypothesis and the signal-plus-background hypothesis using the \HISTFITTER\ package. Both fits were carried out simultaneously in all SRs of each model and for each benchmark point together with its uncertainty. Since the signal contamination in the CRs is low, it was not accounted for in the fit. All SRs corresponding to a model are statistically combined. Following the \cls prescription, the $p$-values of the signal-plus-background hypothesis are tested against those of the background-only hypothesis to extract the corresponding \cls values for each point. A signal point is considered excluded at 95\% CL when such values fall below a 5\% threshold.
 
The resulting expected and observed exclusion limits for the \Wh\ model are shown in \cref{fig:Wh-exclusion}. The large $\pm 1\sigma$ uncertainty band around the expected limit is almost entirely dominated by the statistical uncertainty of the MC simulated signals. The observed bounds are stronger than the expected ones due to the deficit of data compared to the SM background expectation seen in \SRhmttWh-3-$\mu\mu$, as shown in \cref{fig:WhSS_pull_SR}. This SR features the strongest sensitivity in the region with high chargino--LSP mass splittings, as it combines the highest \met threshold and the greatest expected number of events and purity that characterises the dimuon channel.
However, this discrepancy falls within a $2\sigma$ fluctuation of the expected limit.
 
\begin{figure}[!tb]
\centering
\subfloat[]{\includegraphics[width=0.5\textwidth]{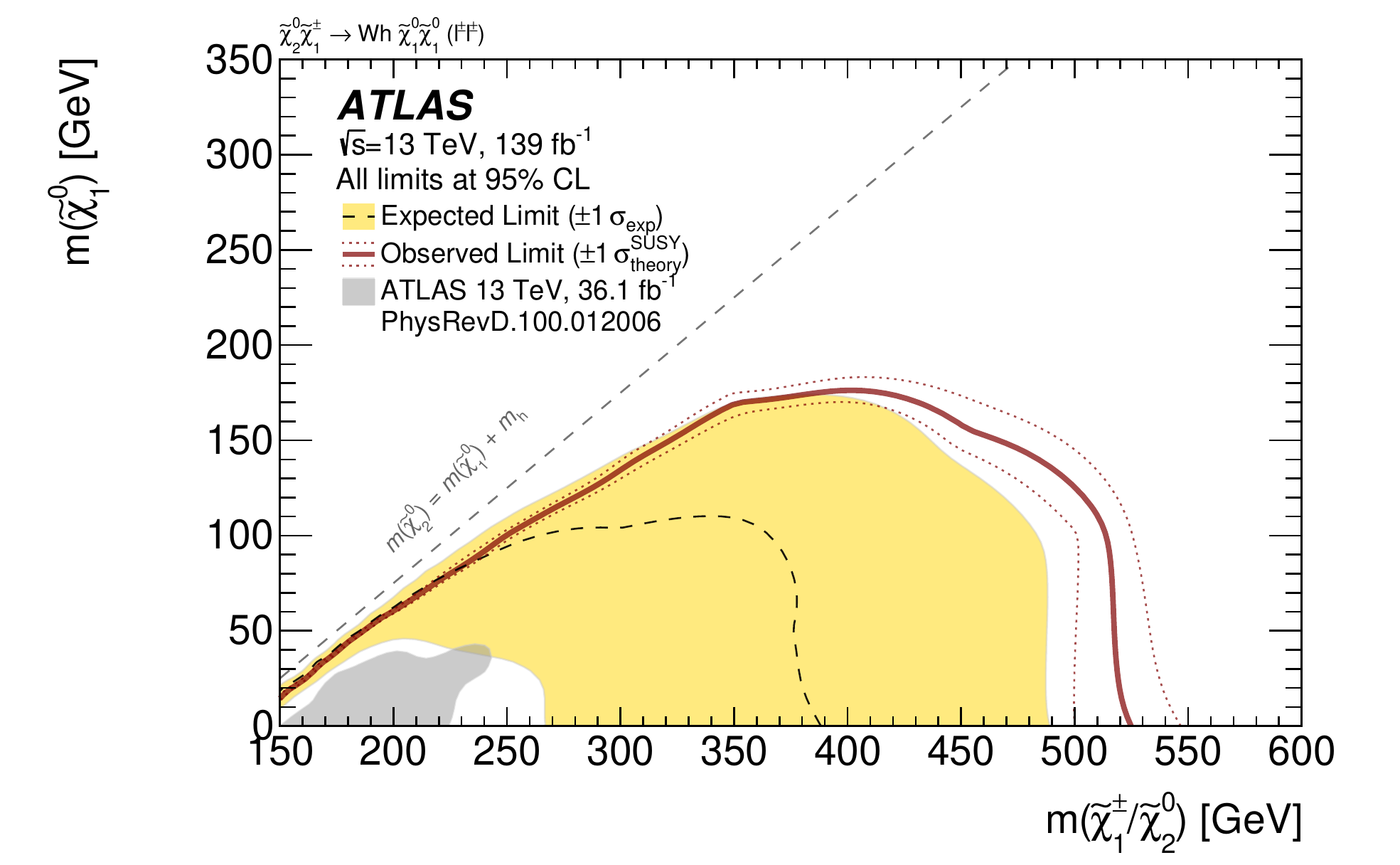}\label{fig:Wh-exclusion}}
\hfill
\subfloat[]{\includegraphics[width=0.5\textwidth]{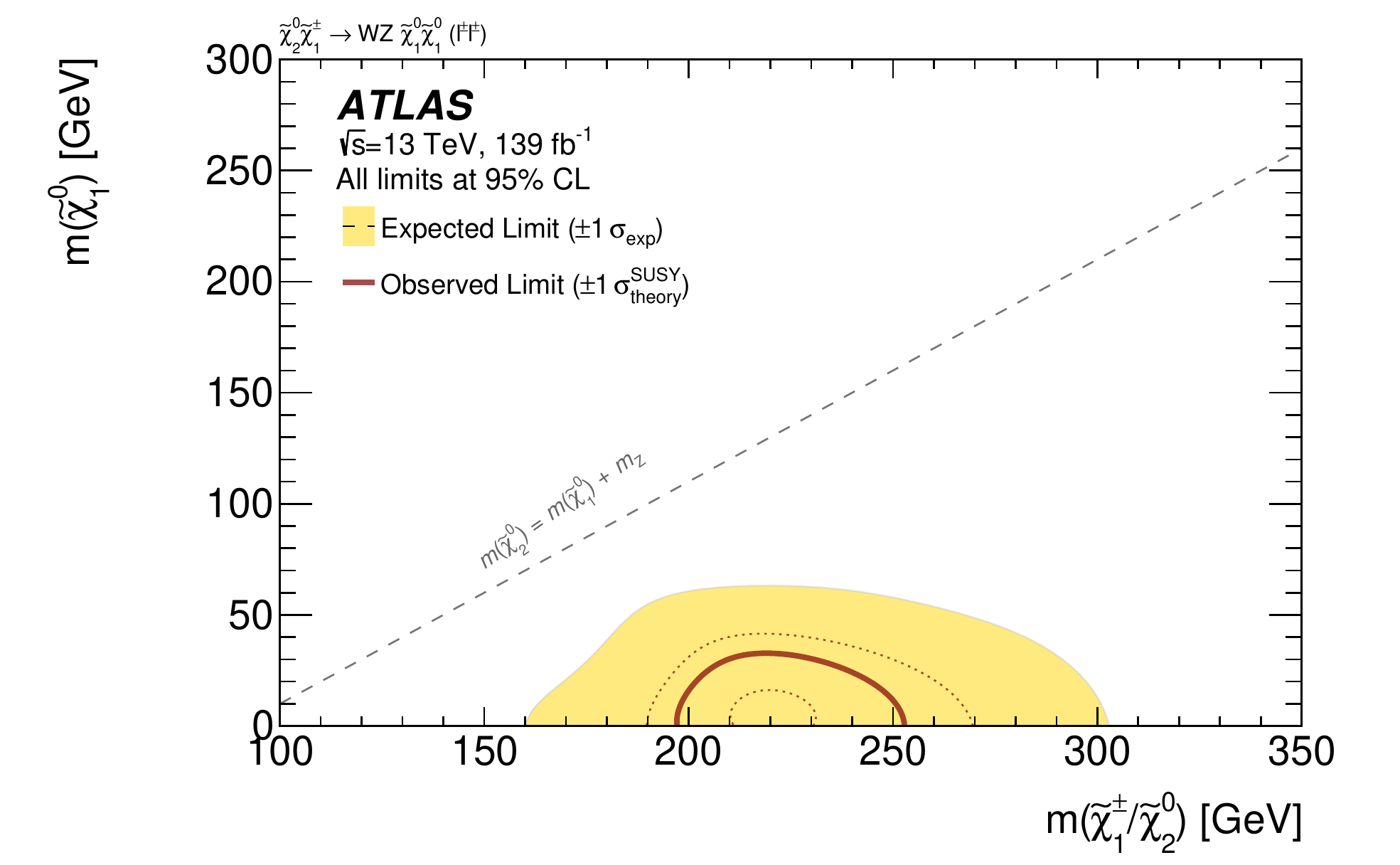}\label{fig:WZ-exclusion-2d}}
\caption{Exclusion limits at 95\% CL for the \protect\subref{fig:Wh-exclusion} \Wh-mediated and the \protect\subref{fig:WZ-exclusion-2d} \WZ-mediated simplified model of wino \conentwo\ production. Observed (solid) and expected (dashed) limits on \chinoonepm/\ninotwo\ and \ninoone\ masses are indicated. The red dotted lines around the observed limit reflect the theoretical variation due to the signal cross-section uncertainty. The band around the expected limits expresses the $\pm1\sigma$ variation due to all uncertainties except theoretical uncertainties in the signal cross section. The grey region in \protect\subref{fig:Wh-exclusion} denotes the observed limits obtained in a previous search in  the same channel with 36.1~\ifb\ of data~\cite{SUSY-2017-01}.}
\label{fig:WhWZ-exclusion}
\end{figure}
 
In the \Wh model, \chinoonepm/\ninotwo masses are excluded up to about \SI{525}{\gev} for a massless \ninoone. On the other hand, the exclusion for \ninoone masses reaches about \SI{180}{\gev} for $m(\chinoonepm/\ninotwo) \simeq \SI{400}{\gev}$. The comparison with the observed exclusion limits from the previous \SI{36.1}{\ifb} search~\cite{SUSY-2017-01} in the same channel demonstrates that the current analysis has a greatly improved reach.
 
The observed and expected exclusion limits for the \WZ\ model are shown in \cref{fig:WZ-exclusion-2d}, where two orthogonal SRs, \SRhmttWZ and \SRlmttWZ, are statistically combined. The deficit of data events compared to the SM expectation in \SRhmttWZ, seen in \cref{fig:SRyields}, leads to the observed limits being more stringent than the expected ones, yet within the $1\sigma$ band around the latter. The uncertainty in the expected exclusion limit is dominated by the FNP background determination, as observed in \cref{fig:Sys_SR_WZ_bRPV}.
For a massless $m(\ninoone)$, \chinoonepm/\ninotwo\ masses in the interval 200--250~\gev\ are excluded.
This is the first analysis in ATLAS with sensitivity to the \WZ\ model in the two-SS-lepton channel. Previous analyses, assuming a nearly massless \ninoone, excluded \chinoonepm/\ninotwo\ masses of up to \SI{640}{GeV} by selecting three-lepton events~\cite{SUSY-2019-09}, while the search with boosted hadronically decaying bosons was sensitive to higher masses, excluding a mass range of 440--960~\gev~\cite{SUSY-2018-41}.
 
The expected and observed production cross-section upper limits for light higgsinos in the bRPV model can be seen in \cref{fig:bRPV-exclusion} with the statistical combination of two orthogonal SRs, namely \SRtwolbrpv and \SRthreelbrpv. By comparing the observed upper limits on the cross section with the theoretical expected cross section, higgsino \ninoone/\ninotwo/\chinoonepm\ masses smaller than \SI{440}{GeV} are excluded when assuming inclusive higgsino production and allowing all predicted sparticle decay modes. These are the first experimental constraints on bRPV models with degenerate higgsino masses.
 
\begin{figure}[!tb]
\centering
\includegraphics[width=0.7\textwidth]{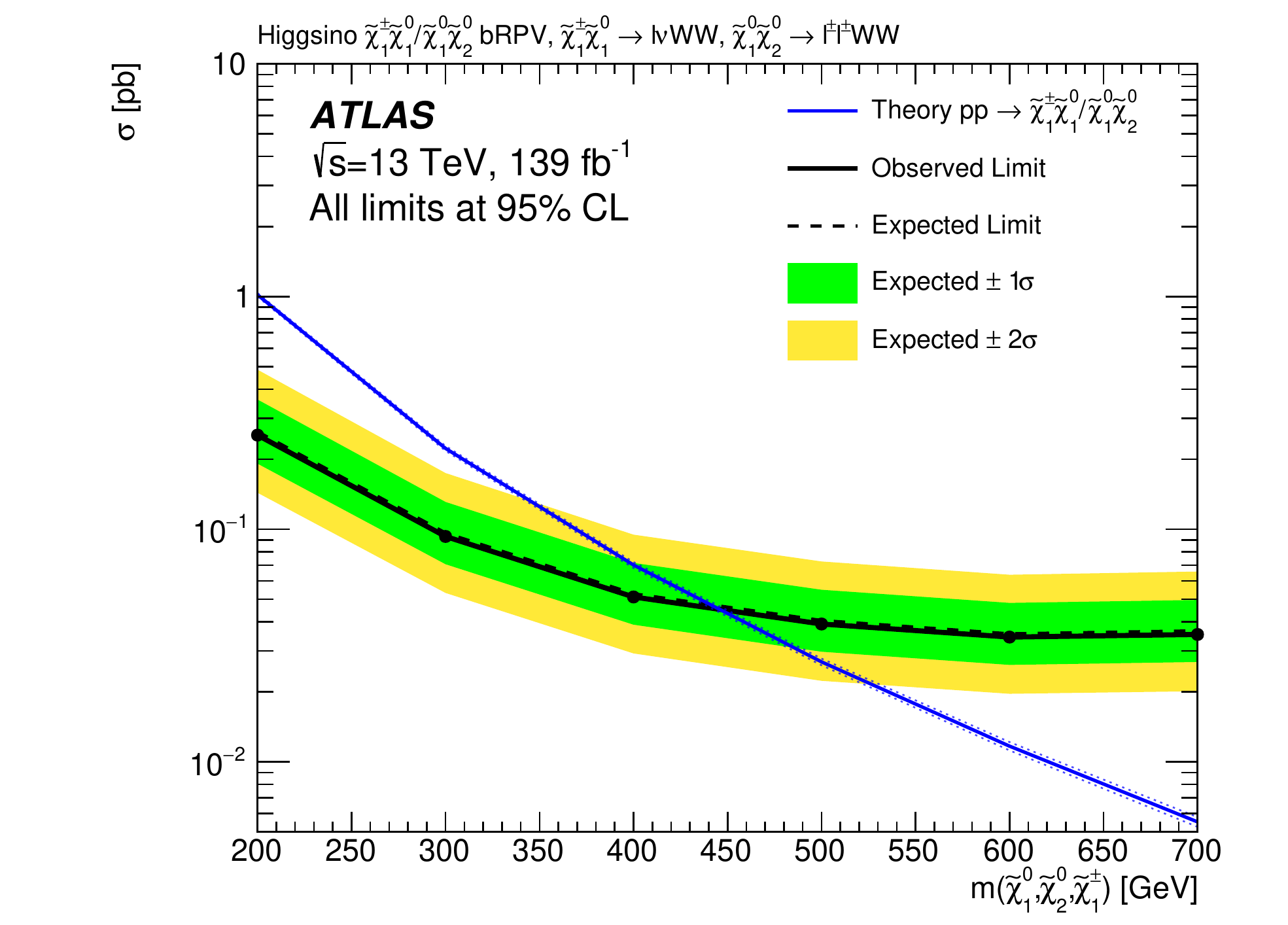}
\caption{Observed (black solid line) and expected (black dashed line) 95\% CL exclusion limits as a function of higgsino  \ninoone/\ninotwo/\chinoonepm\ mass in the bilinear RPV model. The green (yellow) contours of the band around the expected limit are the  $\pm1\sigma$ ($\pm2\sigma$) variations including all uncertainties. The prediction for the theoretical production cross section is also shown (blue solid line) with its uncertainty (blue dotted lines).}
\label{fig:bRPV-exclusion}
\end{figure}


\FloatBarrier
\section{Conclusion}
\label{sec:conclusion}

 
This paper presents a search for directly produced electroweak gauginos and higgsinos in events with two electrons or muons of the same charge or three leptons based on a 139~\ifb\ sample of $\sqs = 13$~\tev\ proton--proton collisions collected by the ATLAS experiment at the LHC from 2015 to 2018. Events are categorised according to the number of jets, the number of \bjets, the missing transverse momentum, the effective mass and other relevant observables, substantially improving the sensitivity to specific \Rp-conserving and \Rp-violating scenarios. No significant excess over the expected background is observed. Observed 95\% CL limits are placed on the visible cross section in the defined signal regions and constraints are set on the parameters of the simplified topologies and complete models considered. In a wino--bino  \Wh-mediated model, NLSP masses of up to 525~\gev\ are excluded for a massless lightest neutralino, a considerable improvement on previous limits of 240~\gev\ and 300~\gev\ set by ATLAS~\cite{SUSY-2017-01} with a 36.1~\ifb\ data set and CMS~\cite{CMS-SUS-19-012} with a 137~\ifb\ sample, respectively. The analogous excluded \chinoonepm/\ninotwo\ mass range for the \WZ\ topology is between 200~\gev\ and 250~\gev\ in a channel probed for the first time by ATLAS in the two-same-sign-lepton final state. In a natural RPV model with bilinear terms, never explored before in electroweak SUSY production, mass-degenerate higgsinos \ninoone/\ninotwo/\chinoonepm\ lighter than 440~\gev\ are excluded. Model-independent upper bounds on the visible cross section as low as 40~ab are set in signal regions inspired by an \Rp-violating scenario with a baryon-number-violating term. Search regions orthogonal to those in other ATLAS analyses were deployed in all considered models, allowing better future statistical combinations with other channels.


\section*{Acknowledgements}


We thank CERN for the very successful operation of the LHC, as well as the
support staff from our institutions without whom ATLAS could not be
operated efficiently.
 
We acknowledge the support of
ANPCyT, Argentina;
YerPhI, Armenia;
ARC, Australia;
BMWFW and FWF, Austria;
ANAS, Azerbaijan;
CNPq and FAPESP, Brazil;
NSERC, NRC and CFI, Canada;
CERN;
ANID, Chile;
CAS, MOST and NSFC, China;
Minciencias, Colombia;
MEYS CR, Czech Republic;
DNRF and DNSRC, Denmark;
IN2P3-CNRS and CEA-DRF/IRFU, France;
SRNSFG, Georgia;
BMBF, HGF and MPG, Germany;
GSRI, Greece;
RGC and Hong Kong SAR, China;
ISF and Benoziyo Center, Israel;
INFN, Italy;
MEXT and JSPS, Japan;
CNRST, Morocco;
NWO, Netherlands;
RCN, Norway;
MEiN, Poland;
FCT, Portugal;
MNE/IFA, Romania;
MESTD, Serbia;
MSSR, Slovakia;
ARRS and MIZ\v{S}, Slovenia;
DSI/NRF, South Africa;
MICINN, Spain;
SRC and Wallenberg Foundation, Sweden;
SERI, SNSF and Cantons of Bern and Geneva, Switzerland;
MOST, Taiwan;
TENMAK, T\"urkiye;
STFC, United Kingdom;
DOE and NSF, United States of America.
In addition, individual groups and members have received support from
BCKDF, CANARIE, Compute Canada and CRC, Canada;
PRIMUS 21/SCI/017 and UNCE SCI/013, Czech Republic;
COST, ERC, ERDF, Horizon 2020, ICSC-NextGenerationEU and Marie Sk{\l}odowska-Curie Actions, European Union;
Investissements d'Avenir Labex, Investissements d'Avenir Idex and ANR, France;
DFG and AvH Foundation, Germany;
Herakleitos, Thales and Aristeia programmes co-financed by EU-ESF and the Greek NSRF, Greece;
BSF-NSF and MINERVA, Israel;
Norwegian Financial Mechanism 2014-2021, Norway;
NCN and NAWA, Poland;
La Caixa Banking Foundation, CERCA Programme Generalitat de Catalunya and PROMETEO and GenT Programmes Generalitat Valenciana, Spain;
G\"{o}ran Gustafssons Stiftelse, Sweden;
The Royal Society and Leverhulme Trust, United Kingdom.
 
The crucial computing support from all WLCG partners is acknowledged gratefully, in particular from CERN, the ATLAS Tier-1 facilities at TRIUMF (Canada), NDGF (Denmark, Norway, Sweden), CC-IN2P3 (France), KIT/GridKA (Germany), INFN-CNAF (Italy), NL-T1 (Netherlands), PIC (Spain), ASGC (Taiwan), RAL (UK) and BNL (USA), the Tier-2 facilities worldwide and large non-WLCG resource providers. Major contributors of computing resources are listed in Ref.~\cite{ATL-SOFT-PUB-2023-001}.


\clearpage
\appendix
\part*{Appendix}
\addcontentsline{toc}{part}{Appendix}

\graphicspath{{figures/}{figures/appendix/}{figures/appendix/N-1}}
 
\section{RPV analysis with \textit{UDD} terms}
\label{app:RPVUDD}
 
\subsection{Signal regions}
\label{subsec:RPVUDD_SR}
 
The signal regions designed to maximise the sensitivity to signals in this model are listed in~\cref{tab:SR_RPV}.
Orthogonal signal regions are defined with the number of \bjets.
In each case, the signal regions are split according to the jet multiplicity targeting different higgsino mass ranges.
 
Within each SR, variables such as the sum of the jets' \pt (\sumjet), the sum of the \bjets' \pt divided by the sum of the jets' \pt (\fracbjet), the minimum angular distance between the leading lepton and jets (\DRlj) and the angular distance between the two SS leptons are used to maximise the sensitivity to the target signal, based on a series of dedicated optimisation studies.

\begin{table}[!htb]
\centering
\caption{Signal region definitions designed for the \textit{UDD} RPV model. The variables are defined in the text.}
\resizebox{\textwidth}{!}{
\begin{tabular}{c|c|c|c|c|c|c|c|c}
\hline \hline
& \multicolumn{2}{c|}{\SRonebjrpv} & \multicolumn{3}{c|}{\SRtwobjrpv} & \multicolumn{3}{c}{\SRthreebjrpv} \\
& L & M & L & M & H & L & M & H \\
\hline \hline
 
$N_{\mathrm{BL}}(\ell)$ & \multicolumn{8}{c}{= 2} \\
 
$N_{\mathrm{Sig}}(\ell)$ & \multicolumn{8}{c}{= 2} \\
 
Charge$(\ell)$ & \multicolumn{8}{c}{same-sign} \\
 
$\pt(\ell)$ & \multicolumn{8}{c}{$> 25~\gev$} \\
 
\nJ	($\pt > 25~\gev$) & \multicolumn{8}{c}{$\geq 1$} \\
\hline \hline
\nb & \multicolumn{2}{c|}{= 1} & \multicolumn{3}{c|}{= 2} &  \multicolumn{3}{c}{$\geq$ 3} \\
$\sum \pt(\ell)$ & \multicolumn{2}{c|}{$\geq \SI{100}{GeV}$} & \multicolumn{3}{c|}{--} & \multicolumn{3}{c}{--} \\
\met & $\geq \SI{100}{GeV}$ & $\geq \SI{50}{GeV}$ & \multicolumn{3}{c|}{$\geq \SI{80}{GeV}$} & \multicolumn{3}{c}{$\geq \SI{20}{GeV}$} \\
\hline \hline
\nJ ($\pt > 25~\gev$) & $\leq 2$ & = 2 or = 3 & $\leq 3$ & =3 or = 4 & $\geq 5$ and $\leq 6$ & $\leq 3$ & $\leq 3$ & $\leq 6$ \\
\fracbjet & $\geq 0.7$ & $\geq 0.45$ & $\geq 0.9$ & $\geq 0.75$ & -- & $\geq 0.8$ & $\geq 0.8$ & $\geq 0.5$ \\
\sumjet & $\geq \SI{120}{GeV}$ & $\geq \SI{400}{GeV}$ & $\geq \SI{300}{GeV}$ & $\geq \SI{420}{GeV}$ & $\geq \SI{420}{GeV}$ & -- & -- & $\geq \SI{350}{GeV}$ \\
\DRlj & $\leq 1.2$ & $\leq 1.0$ & $\leq 1.0$ &  $\leq 1.0$ &  $\leq 1.0$ &  $\leq 1.5$ & -- & $\leq 1.0$ \\
$\Delta R (\ell^{\pm}, \ell^{\pm})$ & $\geq 2.0$ & $\geq 2.5$ & $\geq 2.5$ & $\geq 2.5$ & $\geq 2.0$ & $\geq 2.0$ & -- & $\geq 2.0$ \\
\hline \hline
\end{tabular}
}
\label{tab:SR_RPV}
\end{table}


\subsection{Background estimation and systematic uncertainties}
\label{subsec:RPVUDD_bkg}
 
The background composition is similar to that in the SRs described in \cref{sec:selection}, but with \ttVp as the dominant irreducible background in the above SRs because of the \bjet requirement.
 
The background estimation strategy is the same as that used for the regions designed for the \WZ and bRPV models, described in detail in~\cref{sec:background}.
The irreducible backgrounds are estimated through MC simulation, after applying data-driven scale factors for the \WZ background events with at least two jets obtained from \CRWZtwoj.
The CF events are estimated via the data-driven method described in \cref{sec:background}. The FNP events are estimated from the data by applying the matrix method, after it was validated by comparing its estimates with those of the MC template method.
 
\Cref{fig:Sys_RPVUDD} shows the uncertainties' contributions in the signal regions designed for this model. The uncertainties vary from 20\% to 50\% depending on the regions. The largest contribution comes from the data-driven methods applied.
 
\begin{figure}[!tb]
\centering
\includegraphics[width=0.9\textwidth]{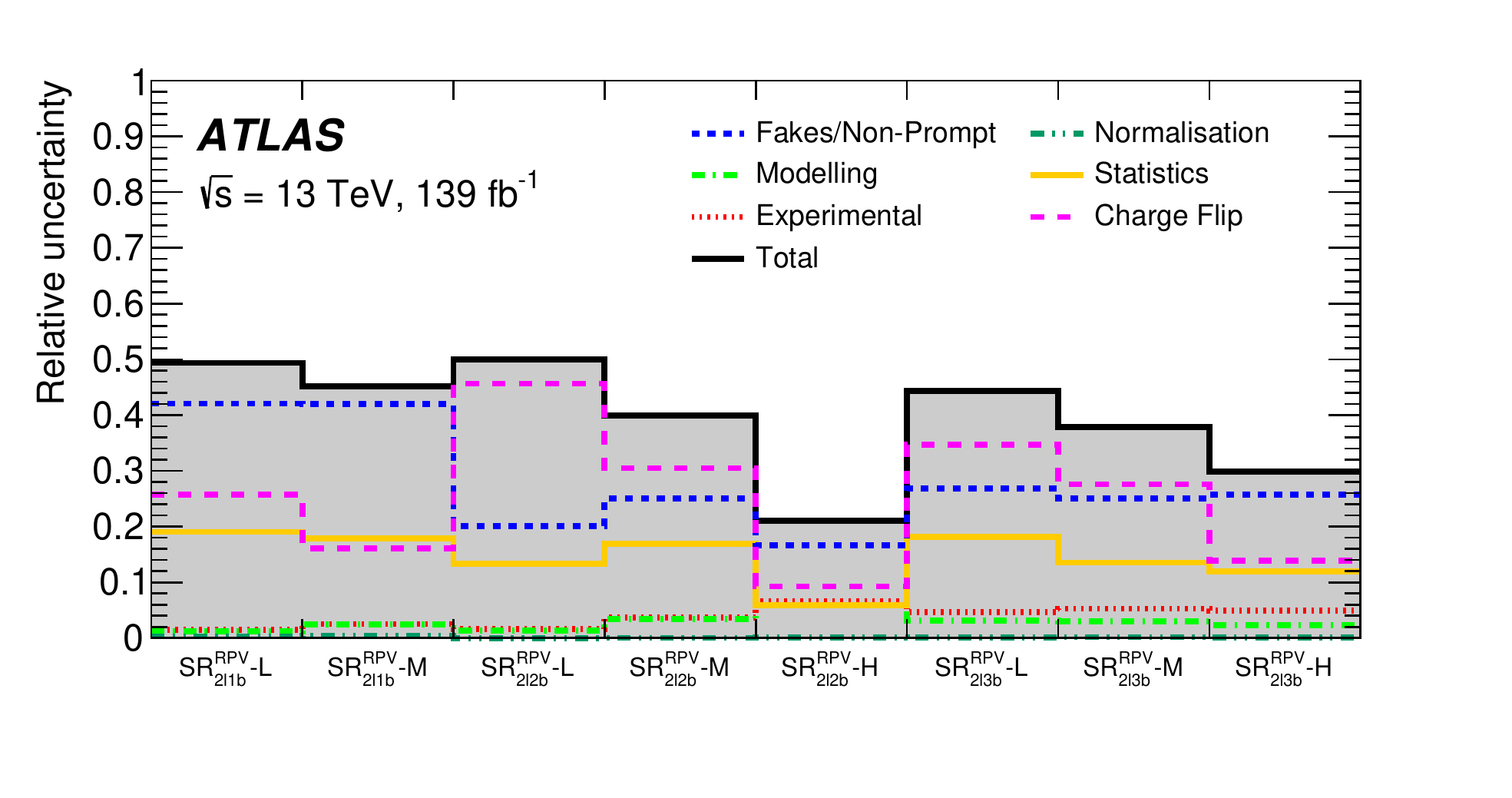}
\caption{Relative contributions from experimental and theoretical uncertainties in SRs defined for the \textit{UDD} RPV model. The individual components can be correlated and therefore do not necessarily add up in quadrature to the total systematic uncertainty.}
\label{fig:Sys_RPVUDD}
\end{figure}
 
\subsection{Results}
\label{subsec:RPVUDD_results}
 
The \fracbjet distributions for the data and background sources are presented for a subset of the SRs in \cref{fig:RPVUDD_N-1_DD}. All selection criteria defined in \cref{tab:SR_RPV} are applied apart from the one on \fracbjet, which is indicated in the graphs by a vertical line and an arrow. The data and total background expectation are in agreement, considering the involved uncertainties.
 
\begin{figure}[!tb]
\centering
\subfloat[]{\includegraphics[width=0.33\textwidth]{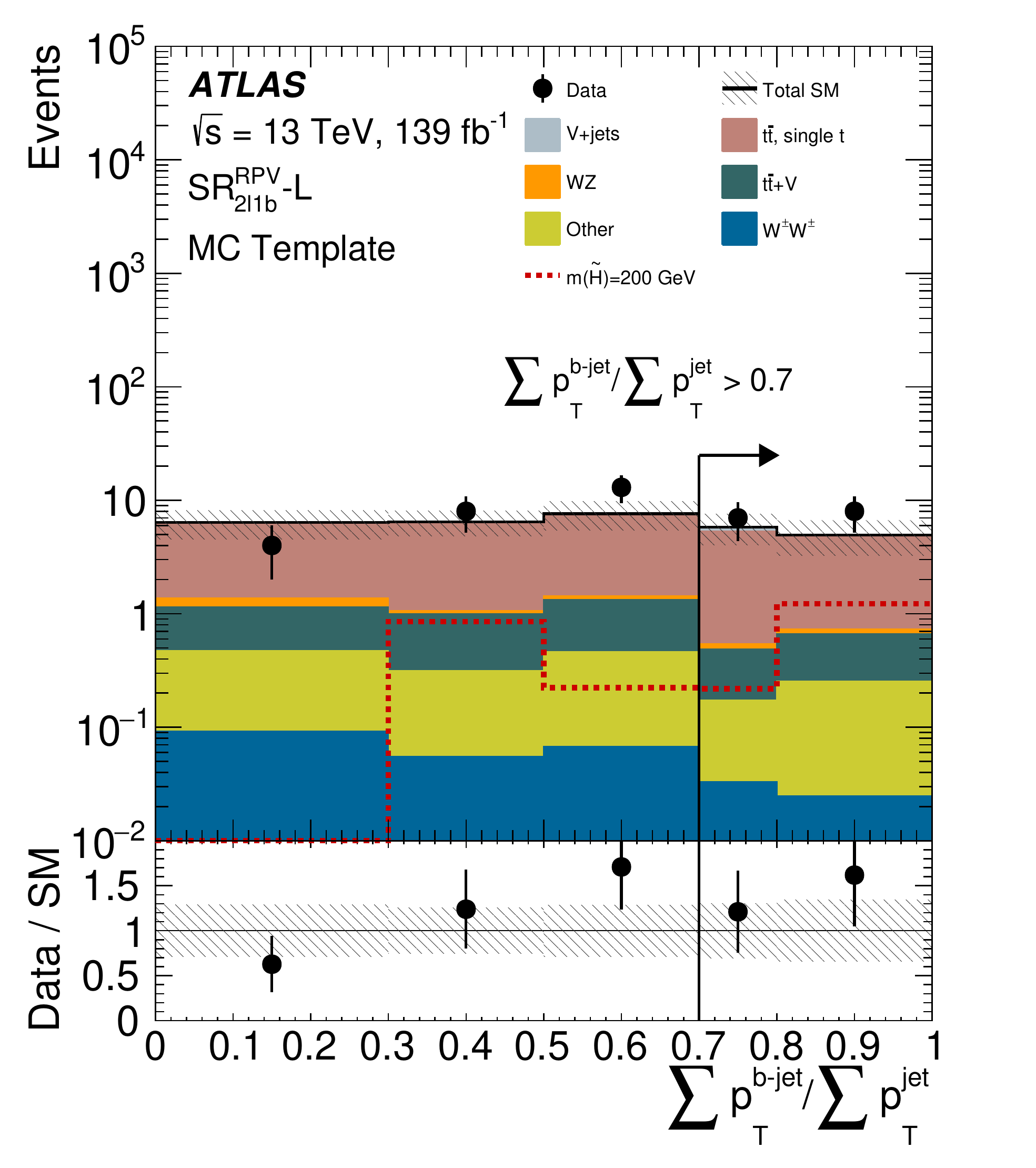}}
\hfill
\subfloat[]{\includegraphics[width=0.33\textwidth]{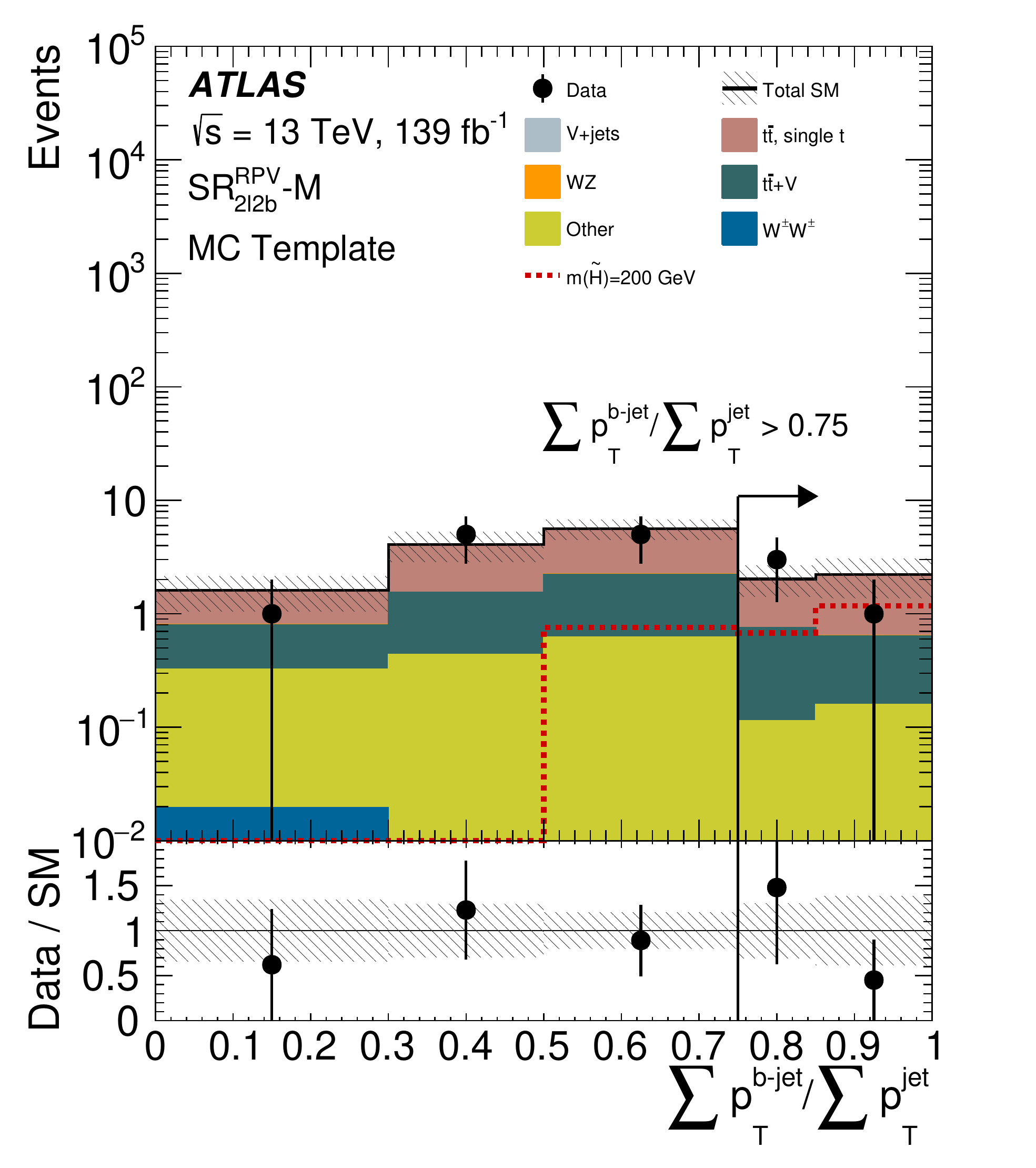}}
\hfill
\subfloat[]{\includegraphics[width=0.33\textwidth]{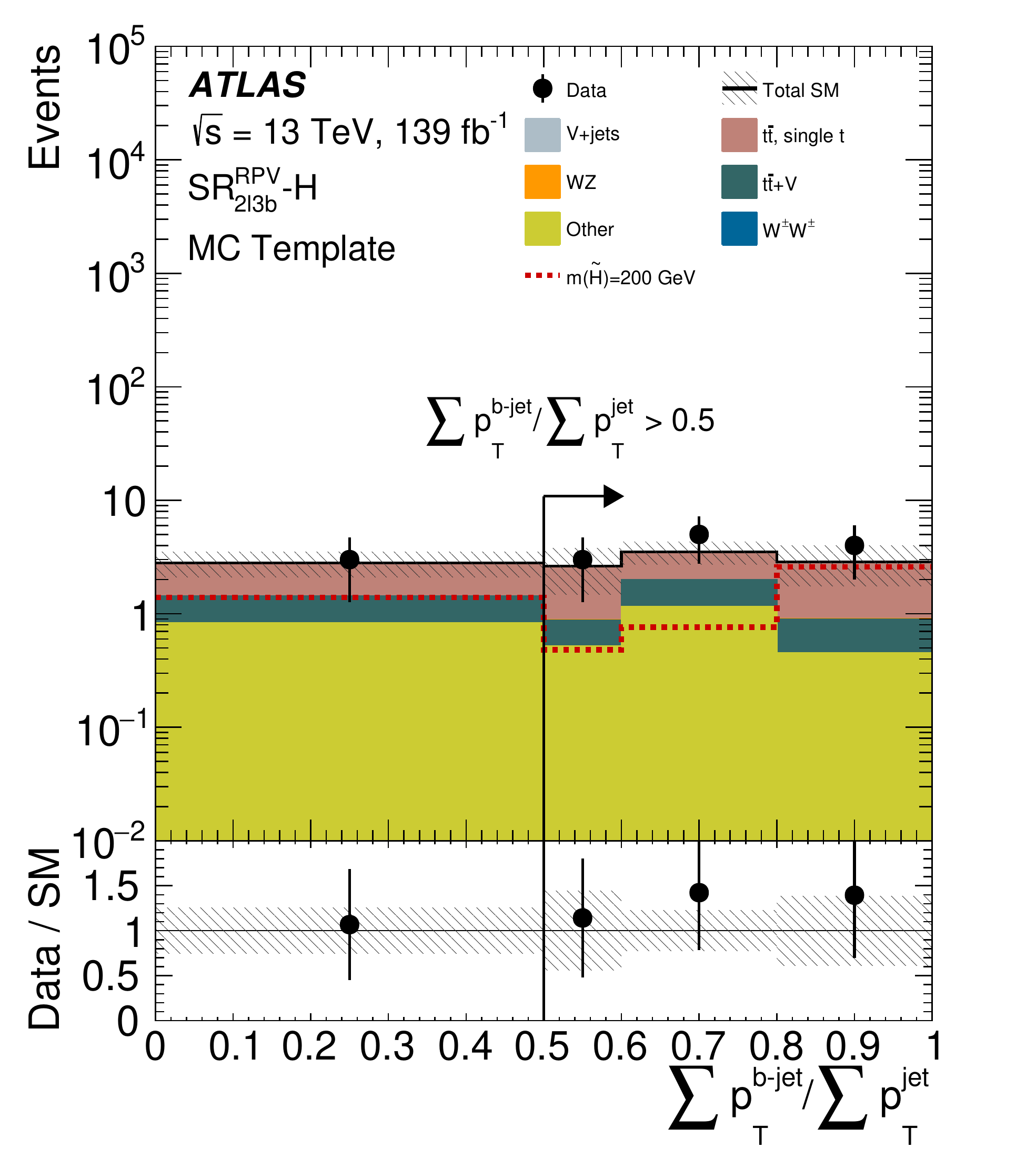}}
\caption{\fracbjet distributions of the data and the expected background in some SRs defined for the \textit{UDD} RPV model with data-driven methods applied. All uncertainties are considered. The vertical black lines and the corresponding arrows indicate the cuts defining those regions. The last bin includes overflow. The `Other' category contains the \ttHp, rare top, triboson, and other diboson processes with the SS final state. Distributions for a representative signal mass point are overlaid. The bottom panel shows the ratio of the observed data to the predicted yields.}
\label{fig:RPVUDD_N-1_DD}
\end{figure}
 
A comparison between the data and background yields for all the SRs defined for the \textit{UDD} RPV model is shown in \cref{fig:Rpvyields}. The observed and expected numbers of events are compatible in all SRs, with the largest excess being ${\sim}1\sigma$, observed in \SRthreebjrpv-H. Following the procedure described in \cref{sec:interpret}, these results are used to set model-independent upper limits as low as 40~ab on BSM production cross sections, as listed in \cref{table.results.exclxsec.pval.upperlimit.nonWhSRs}.
 
\begin{figure}[h!]
\begin{center}
\includegraphics[width=0.8\textwidth]{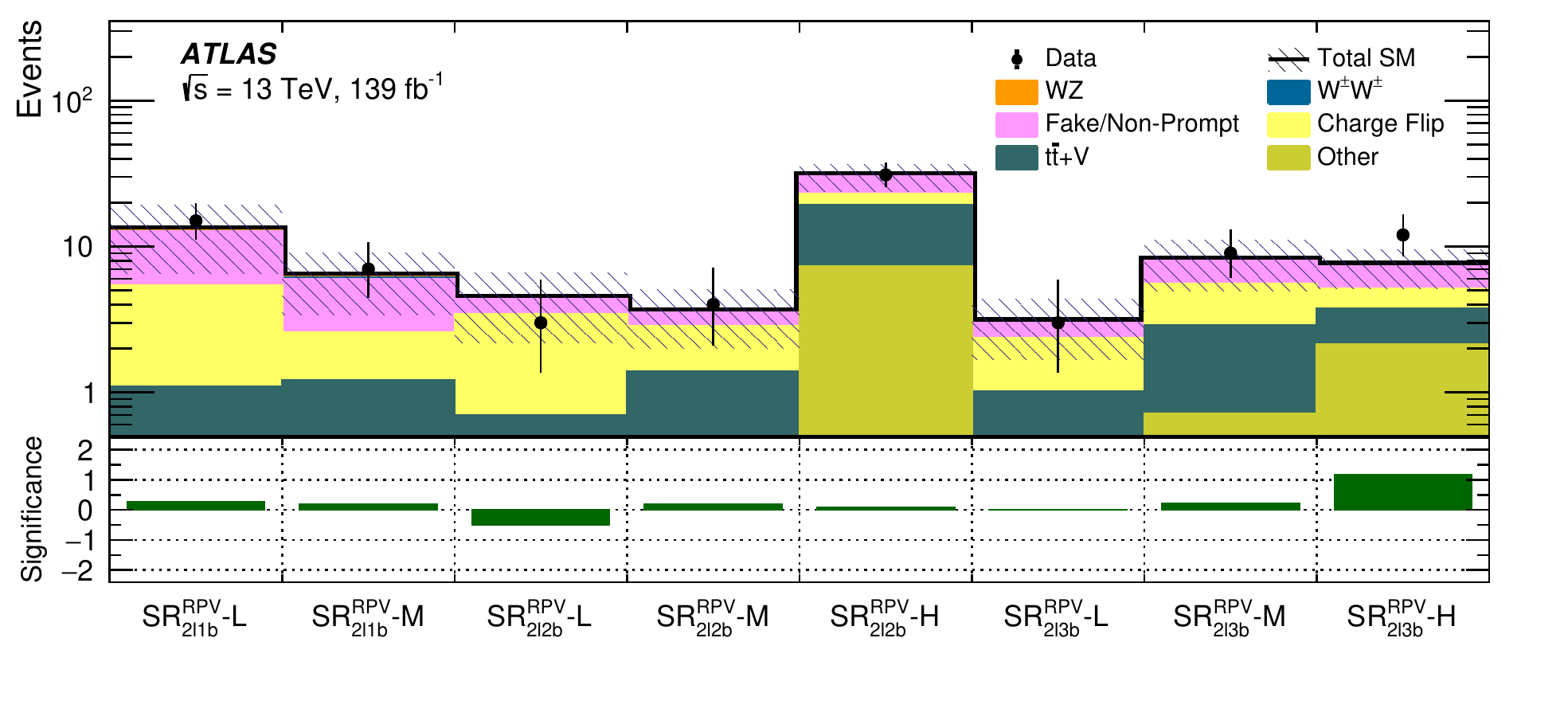}
\end{center}
\vspace{-2em}
\caption{Expected SM background and data yields in the SRs optimised for the \textit{UDD} RPV model. The SM prediction is taken from the background-only fit. The `Other' category contains the \ttHp, rare top, triboson, and other diboson processes with the SS final state. The total uncertainties in the expected event yields are shown as the hashed bands. The bottom panel shows the statistical significance~\cite{Cousins:2007bmb} of the discrepancy between the observed number of events and the SM expectation.}
\label{fig:Rpvyields}
\end{figure}
 
\begin{table}[!tb]
\caption{Model-independent statistical analysis for SRs optimised for the \textit{UDD} RPV models: the 95\% CL upper limit on the visible cross section times efficiency (\sigmavis), the observed number of signal events (\sobs), and the signal events given the expected number of background events (\sexp, $\pm1\sigma$ variations of the expected number). The last two columns report the \clb  value for the background-only hypothesis, the one-sided $p_0$-value and the local significance $Z$ (the number of equivalent Gaussian standard deviations).}
\label{table.results.exclxsec.pval.upperlimit.nonWhSRs}
\centering
\renewcommand{\arraystretch}{1.2}
\small
\begin{tabular}{lccccc}
\noalign{\smallskip}\hline\noalign{\smallskip}{\textbf{Signal channel}}                        & \sigmavis\ [fb]  &  \sobs  &  \sexp & \clb  & $p_0$~$(Z)$ \\
\noalign{\smallskip}\hline\noalign{\smallskip}
\SRonebjrpv-L    & $0.13$ &  $17.5$ & $ { 15.1 }^{ +4.8 }_{ -3.7 }$\phantom{0} & $0.69$ & $ 0.38$~$(0.32)$ \\%
\SRonebjrpv-M    & $0.07$ &  $10.1$ & $ { 8.9 }^{ +3.1 }_{ -1.7 }$ & $0.66$ & $ 0.46$~$(0.11)$ \\%
\SRtwobjrpv-L    & $0.04$ &  \phantom{0}$6.1$ & $ { 6.2 }^{ +2.4 }_{ -1.1 }$ & $0.48$ & $ 0.50$~$(0.00)$ \\%
\SRtwobjrpv-M    & $0.05$ &  \phantom{0}$6.8$ & $ { 6.0 }^{ +2.3 }_{ -1.2 }$ & $0.65$ & $ 0.38$~$(0.30)$ \\%
\SRtwobjrpv-H    & $0.15$ &  $20.7$ & $ { 18.6 }^{+6.0 }_{ -4.3 }$\phantom{0} & $0.64$ & $ 0.41$~$(0.22)$ \\%
\SRthreebjrpv-L    & $0.04$ &  \phantom{0}$6.1$ & $ { 5.7 }^{ +1.9 }_{ -1.0 }$ & $0.61$ & $ 0.50$~$(0.00)$ \\%
\SRthreebjrpv-M    & $0.08$ &  $11.5$ & $ { 9.7 }^{ +3.2 }_{ -1.8 }$ & $0.70$ & $ 0.35$~$(0.37)$ \\%
\SRthreebjrpv-H    & $0.10$ &  $13.5$ & $ { 8.6 }^{ +3.2 }_{ -2.5 }$ & $0.92$ & $ 0.10$~$(1.31)$ \\%
\noalign{\smallskip}\hline\noalign{\smallskip}
\end{tabular}
\end{table}
 
\Cref{fig:RPV-exclusion} shows the expected upper limits for the higgsino \textit{UDD} RPV model. All combinations of orthogonal SRs which target the same mass point are considered, including \SRrpv-L (the statistical combination of \SRonebjrpv-L, \SRtwobjrpv-L, \SRthreebjrpv-L), \SRrpv-M (the statistical combination of \SRonebjrpv-M, \SRtwobjrpv-M, \SRthreebjrpv-M) and \SRrpv-H (the statistical combination of \SRtwobjrpv-H, \SRthreebjrpv-H). Among  these combinations, the one providing the strongest expected limit is chosen for each $\tilde{\chi}^0_{1,2}$ mass point.
 
\begin{figure}[!tb]
\centering
\includegraphics[width=0.7\textwidth]{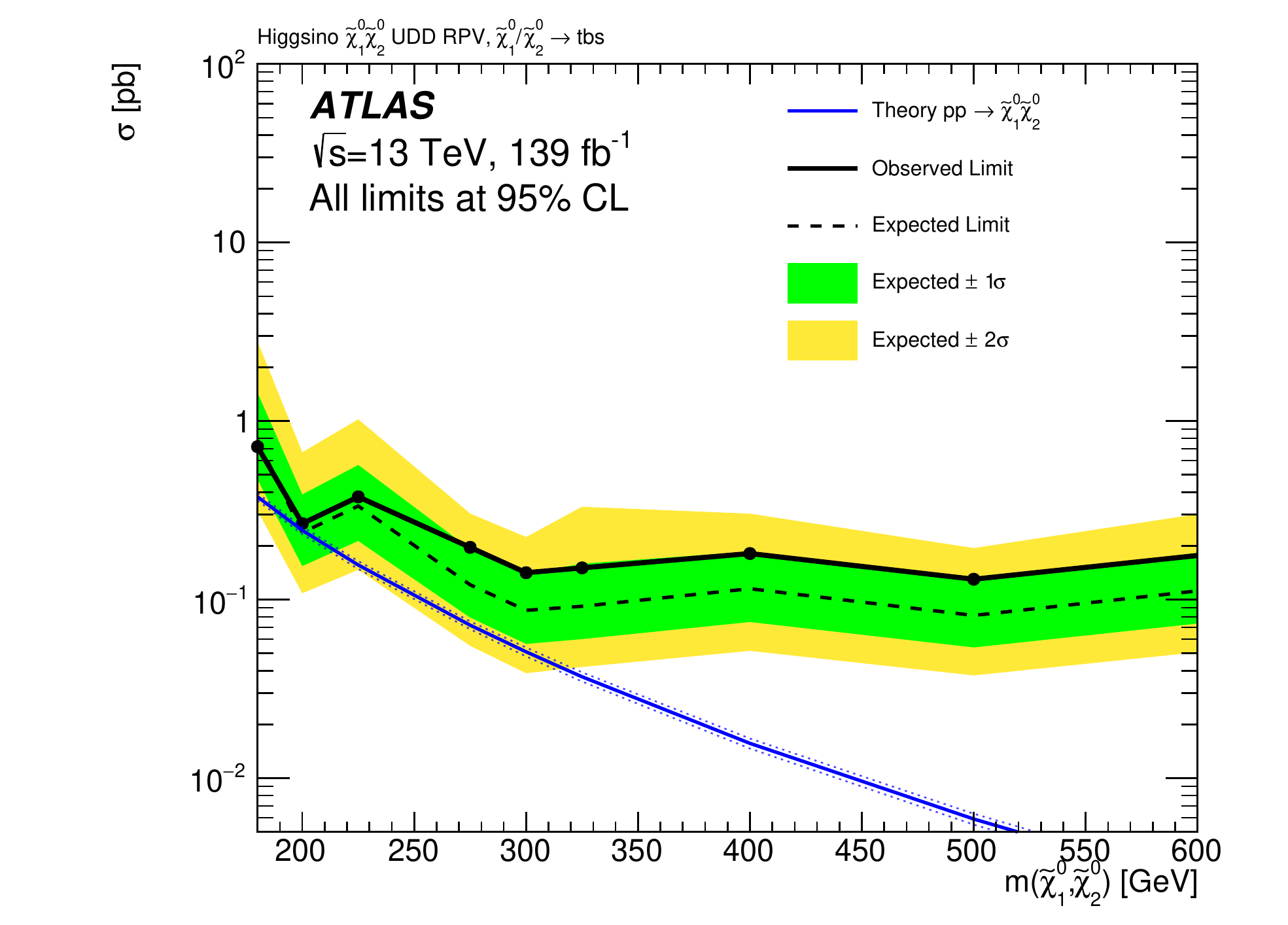}
\caption{Observed (black solid line) and expected (black dashed line) 95\% CL exclusion limits as a function of higgsino \ninoone/\ninotwo\ mass in the \textit{UDD} RPV model. The green (yellow) contours of the band around the expected limit are the  $\pm1\sigma$ ($\pm2\sigma$) variations including all uncertainties. The prediction for the theoretical production cross section is also shown (blue solid line) with its uncertainty (blue dotted lines).}
\label{fig:RPV-exclusion}
\end{figure}
 
A higgsino-like \ninoone/\ninotwo\ mass of \SI{200}{GeV} is excluded in this analysis, considering \nonentwo\ production only. This value was excluded by a previous ATLAS search~\cite{SUSY-2019-04} based on the selection of events with one lepton, but also using \conenone\ and \conentwo\ production with  $\chinoonepm\to bbs$.


\FloatBarrier
\printbibliography
 
\clearpage
 
\input{atlas_authlist}

\end{document}

%% file: atlas_authlist.tex
 
\begin{flushleft}
\hypersetup{urlcolor=black}
{\Large The ATLAS Collaboration}

\bigskip

\AtlasOrcid[0000-0002-6665-4934]{G.~Aad}$^\textrm{\scriptsize 102}$,
\AtlasOrcid[0000-0002-5888-2734]{B.~Abbott}$^\textrm{\scriptsize 120}$,
\AtlasOrcid[0000-0002-7248-3203]{D.C.~Abbott}$^\textrm{\scriptsize 103}$,
\AtlasOrcid[0000-0002-1002-1652]{K.~Abeling}$^\textrm{\scriptsize 55}$,
\AtlasOrcid[0000-0002-8496-9294]{S.H.~Abidi}$^\textrm{\scriptsize 29}$,
\AtlasOrcid[0000-0002-9987-2292]{A.~Aboulhorma}$^\textrm{\scriptsize 35e}$,
\AtlasOrcid[0000-0001-5329-6640]{H.~Abramowicz}$^\textrm{\scriptsize 151}$,
\AtlasOrcid[0000-0002-1599-2896]{H.~Abreu}$^\textrm{\scriptsize 150}$,
\AtlasOrcid[0000-0003-0403-3697]{Y.~Abulaiti}$^\textrm{\scriptsize 117}$,
\AtlasOrcid[0000-0003-0762-7204]{A.C.~Abusleme~Hoffman}$^\textrm{\scriptsize 137a}$
% ...
\bigskip
\\

% ...
$^{29}$Physics Department, Brookhaven National Laboratory, Upton NY; United States of America.\\
% ...
$^{35}$$^{(a)}$Facult\'e des Sciences Ain Chock, R\'eseau Universitaire de Physique des Hautes Energies - Universit\'e Hassan II, Casablanca;$^{(b)}$Facult\'{e} des Sciences, Universit\'{e} Ibn-Tofail, K\'{e}nitra;$^{(c)}$Facult\'e des Sciences Semlalia, Universit\'e Cadi Ayyad, LPHEA-Marrakech;$^{(d)}$LPMR, Facult\'e des Sciences, Universit\'e Mohamed Premier, Oujda;$^{(e)}$Facult\'e des sciences, Universit\'e Mohammed V, Rabat;$^{(f)}$Institute of Applied Physics, Mohammed VI Polytechnic University, Ben Guerir; Morocco.\\
% ...
$^{55}$II. Physikalisches Institut, Georg-August-Universit\"{a}t G\"ottingen, G\"ottingen; Germany.\\
% ...
$^{102}$CPPM, Aix-Marseille Universit\'e, CNRS/IN2P3, Marseille; France.\\
$^{103}$Department of Physics, University of Massachusetts, Amherst MA; United States of America.\\
% ...
$^{117}$Department of Physics, New York University, New York NY; United States of America.\\
% ...
$^{120}$Homer L. Dodge Department of Physics and Astronomy, University of Oklahoma, Norman OK; United States of America.\\
% ...
$^{137}$$^{(a)}$Departamento de F\'isica, Pontificia Universidad Cat\'olica de Chile, Santiago;$^{(b)}$Millennium Institute for Subatomic physics at high energy frontier (SAPHIR), Santiago;$^{(c)}$Instituto de Investigaci\'on Multidisciplinario en Ciencia y Tecnolog\'ia, y Departamento de F\'isica, Universidad de La Serena;$^{(d)}$Universidad Andres Bello, Department of Physics, Santiago;$^{(e)}$Instituto de Alta Investigaci\'on, Universidad de Tarapac\'a, Arica;$^{(f)}$Departamento de F\'isica, Universidad T\'ecnica Federico Santa Mar\'ia, Valpara\'iso; Chile.\\
% ...
$^{150}$Department of Physics, Technion, Israel Institute of Technology, Haifa; Israel.\\
$^{151}$Raymond and Beverly Sackler School of Physics and Astronomy, Tel Aviv University, Tel Aviv; Israel.\\
% ...
$^{a}$ Also Affiliated with an institute covered by a cooperation agreement with CERN.\\
$^{b}$ Also at An-Najah National University, Nablus; Palestine.\\
$^{c}$ Also at Borough of Manhattan Community College, City University of New York, New York NY; United States of America.\\
$^{d}$ Also at Bruno Kessler Foundation, Trento; Italy.\\
$^{e}$ Also at Center for High Energy Physics, Peking University; China.\\
$^{f}$ Also at Center for Interdisciplinary Research and Innovation (CIRI-AUTH), Thessaloniki; Greece.\\
% ...

\end{flushleft}
